\documentclass[superscriptaddress,reprint,showpacs,preprintnumbers,nofootinbib, amsmath,amssymb,aps,prd,floatfix,]{revtex4-1}
\pdfoutput=1
\usepackage{graphicx,amssymb,amsmath,xcolor}
\usepackage{soul}

\newcommand{\be}{\begin{equation}}
\newcommand{\ee}{\end{equation}}

\usepackage{lineno}

\begin{document}

\title{Implementation of the SuSAv2-MEC 1p1h and 2p2h models in GENIE and analysis of nuclear effects in T2K measurements}

\author{S. Dolan}
\affiliation{IN2P3-CNRS, Laboratoire Leprince-Ringuet, Palaiseau 91120, France}
\affiliation{DPhP, IRFU, CEA Saclay, 91191 Gif-sur-Yvette, France}
\affiliation{CERN, European Organization for Nuclear Research, Geneva, Switzerland}

\author{G.D.~Megias}
\affiliation{IN2P3-CNRS, Laboratoire Leprince-Ringuet, Palaiseau 91120, France}
\affiliation{DPhP, IRFU, CEA Saclay, 91191 Gif-sur-Yvette, France}
\affiliation{University of Tokyo, Institute for Cosmic Ray Research, Research Center for Cosmic Neutrinos, Kashiwa, Japan}

\author{S. Bolognesi}
\affiliation{DPhP, IRFU, CEA Saclay, 91191 Gif-sur-Yvette, France}

\date{\today}
\begin{abstract}

We first present the implementation and validation of the SuSAv2-MEC 1p1h and 2p2h models in the GENIE neutrino-nucleus interaction event generator and a comparison of the subsequent predictions to measurements of lepton and hadron kinematics from the T2K experiment. These predictions are also compared to those of other available models in GENIE. We further compare  semi-inclusive predictions of the implemented 1p1h model to those of the microscopic model on which SuSAv2 is based - Relativistic Mean Field (RMF) - to begin to test the validity of widely-used `factorisation' assumptions employed by generators to predict hadron kinematics from inclusive input models. The results highlight that a more precise treatment of hadron kinematics in generators is essential in order to attain the few-\% level uncertainty on neutrino interactions necessary for the next generation of accelerator-based long-baseline neutrino oscillation experiments.

\end{abstract}


\maketitle

\section{Introduction}\label{intro}
The modelling of neutrino-nucleus interactions in the one-to-few GeV region is one of the most complicated issues facing current long-baseline neutrino oscillation measurements (T2K, NOVA) and is expected to be one of the limiting factors for the sensitivity of the future experiments such as DUNE and T2HK~\cite{nustec2017}. A key systematic uncertainty arises from the description of multi-nucleon correlations in the initial state which may induce 2-particle-2-hole (2p2h) final states. It is particularly important to understand the size of the 2p2h interaction cross section compared to the single-body contributions (1p1h) as a poor modelling of this leads to a direct bias on the reconstruction of neutrino energy and therefore must be covered with large systematic uncertainties in current oscillation analyses~\cite{Abe:2017uxa,NOvA:2018gge}. Various models~\cite{Nieves:2011pp,Nieves:2011yp,Martini:2010ex,Martini:2012mec,Mosel:MEC,Meucci,Giusti:2016t2k,Rocco:2015cil,Lovato:2015qka,Martini:low,Ivanov19,Butkevich:2018hll} have been developed to describe such 1p1h and 2p2h processes. In this paper we focus on the SuSAv2 models~\cite{Gonzalez-Jimenez:2014eqa,Megias:2014qva,Megias:2016lke,Megias:2016fjk}. 

The SuSAv2 1p1h model, originally based on the superscaling phenomenon~\cite{super,super1,Chiara1,Amaro:2004bs} shown by electron-nucleus scattering data, has recently been improved through the inclusion of Relativistic Mean Field theory effects~\cite{Caballero:2006wi,Caballero:2005sj,Caballero:2007tz,Meucci:2009}. This model has proven its validity to describe the nuclear dynamics observed in electron-nucleus reactions while taking into account the experimentally-observed enhancement of the transverse scaling function, compared with its longitudinal counterpart, as a genuine relativistic effect together with a careful treatment of the final-state interactions (FSI) between the outgoing nucleon and the residual nucleus. For the description of the 2p2h-MEC (meson exchange current) contributions the model makes use of the fully relativistic calculations from~\cite{Simo:2016ikv} which allows for a proper separation of neutron-proton and proton-proton pairs in the final state via the analysis of the direct-exchange interference terms~\cite{RuizSimo:2016ikw}. The combined SuSAv2-MEC model, covering the 1p1h and 2p2h channels, has been shown to be capable of reproducing the nuclear dynamics and superscaling properties observed in ($e,e'$) reactions~\cite{super,super1,Megias:2019jlab}, which serves as a stringent test for nuclear models, whilst also providing an accurate description of existing neutrino data~\cite{Megias:2016lke,Megias:2016fjk,Megias:2017cuh,Megias:2019mnv,Megias:2019jlab}. 
To date, SuSAv2-MEC is the only fully relativistic model that can be extended without approximations to the full-energy range of interest for present and future neutrino experiments. In this paper we present the implementation of SuSAv2-MEC 1p1h and 2p2h contributions in the GENIEv3 Monte Carlo neutrino interaction simulation~\cite{Andreopoulos:2009rq, Andreopoulos:2015wxa} and use it to better characterise nuclear effects in T2K neutrino scattering cross-section measurements.

Such implementations of the neutrino-nucleus interaction models in event generators is crucial for a variety of reasons. Firstly, a proper modelling of neutrino interactions in the simulation of oscillation experiments is needed in order to perform a correct extrapolation of the near detector constraints to the far detector in the analyses aimed at measuring the neutrino oscillation parameters. This argument is evident for experiments which use (or are planning to use) different detector technologies and nuclear targets at near and far detectors. Even in the case of two detectors exploiting the same technology and targets, such an extrapolation is not straightforward because of the different acceptance of the two detectors, due to different size (and possibly different selections). But beyond such issues, the most complex systematic in the near-to-far extrapolation actually comes from effects which are independent on the detector technology. Due to the neutrino oscillations, the neutrino energy distribution is different in the near and far sites, therefore the cross-section must be evaluated at different energies. Moreover, usually the near detector constrains only the product of neutrino flux and cross-section, which each extrapolates to the far detector differently. The disentangling of the two is based on a simulation (and tuning) of the flux and of the neutrino interactions. 

The implementation of the neutrino interaction models in generators is essential to perform a proper comparison of such models with some of the most recent cross-section measurements. Indeed, in order to provide the most model-independent unbiased results possible, experiments prefer to measure cross sections of interaction topologies (e.g. charged-current with zero pions in the final state, CC0$\pi$) rather than measuring the physical interaction processes (e.g. 1p1h), thereby avoiding correcting the data for effects due to hadronic Final State Interactions (FSI) inside the target nucleus which can cause nuclear emission or absorption. Consequently, FSI effects must be added to the models in order to compare to the data. For example, a measurement of zero-pion final states contain, in addition to the bulk of 1p1h processes, further contributions from 2p2h processes and from resonant interactions with subsequent pion absorption by FSI. The latter contribution is very difficult to describe in a pure microscopic model but is included in neutrino interaction event generators~\cite{Andreopoulos:2009rq,Hayato:2009zz,Buss:2011mx, Golan:2012rfa, Golan:2012wx, Juszczak:2005zs}. Furthermore, recent experimental results focus on multidimensional and/or `semi-inclusive' measurements, where the outgoing lepton and some hadron(s) are detected in coincidence  (e.g. lepton and highest momentum proton kinematics in measurements of CC0$\pi$ interactions), whilst many of the available models are only able to calculate `inclusive' cross sections, which integrate over all possible hadronic final states (i.e. they are able to directly predict outgoing lepton kinematics but can say nothing about outgoing hadrons)~\cite{Moreno:2014kia}. For these models, it is only by their implementation in event generators that they can be used to predict the semi-inclusive final states. While such an approach relies on substantial approximations, which will be discussed in Sec.~\ref{implementation}, it is still the only option available today for the majority of models and, more importantly, is the technique used in neutrino oscillation measurements. Therefore, the implementation of more sophisticated neutrino interaction models, such as SuSAv2, even in such a very approximated approach, is important in order to improve the predictions for the oscillation measurements. The comparison to cross-section measurements is then crucial in order to estimate the systematic uncertainties induced by the usage of such approximated approaches in measurements of neutrino oscillations.

In this manuscript we present, in Sec.~\ref{implementation}, the implementation of the SuSAv2 models in the GENIE event generator, alongside a discussion of all the approximations involved. The comparison of the SuSAv2 2p2h implementation with other 2p2h implementations is then shown in Sec.~\ref{results2p2h}. In Sec.~\ref{results-ssinc}, the RMF, SuSAv2 and SuSAv2-GENIE models are compared for the 1p1h channel at T2K kinematics together with a dedicated analysis of T2K CC0$\pi$ measurements with a restriction on the outgoing nucleon momentum, allowing a first test of some of the key approximations built into neutrino interaction event generators. The SuSAv2-MEC implementation is also tested against other recent T2K semi-inclusive measurements (CC0$\pi$ with and without protons) in Sec.~\ref{results-sinc} where an analysis of the single-transverse variables is also shown. In the Appendices~\ref{appendix-inc}-\ref{appendix-fact}, we test the SuSAv2-MEC implementation for T2K CC0$\pi$ inclusive data and compare the SuSAv2-MEC implementation with the Valencia one in both inclusive and semi-inclusive reactions. Finally, we present our conclusions in Sec.~\ref{conc}.

Throughout this manuscript all GENIE predictions are made using GENIE version R-3\_00\_02~\cite{genie302Git} which serves as the base model for the implementations presented here~\footnote{The latest R-3\_00\_06 update contains some fixes to the Valenica 1p1h model implementation as well as to the Delta-resonance decay simulation with respect to the version used in this manuscript (R-3\_00\_02). These changes will not affect the new SuSA model implementations at all. Although it is not expected that the changes will dramatically affect any distribution shown here, it should nonetheless be noted that the GENIE predictions we show are not from the latest version.}. The full SuSAv2 implementation is expected to be in the next public GENIE release (R-3\_02\_00) but a preliminary version of the code can be found in~\cite{geniesusaGit}.

\section{Implementation in GENIE and the factorisation approach}\label{implementation}

The GENIE event generator generally simulates 1p1h and 2p2h neutrino-nucleus interactions using Monte-Carlo methods to produce events at a rate which is proportional to their modelled cross section. For the newly implemented SuSAv2 1p1h and 2p2h interactions this is calculated as a double differential \textit{inclusive} cross-section in momentum and energy transfer to the hadronic system ($q_3$, $q_0$) by contracting a generic leptonic tensor with the hadron tensors taken from the theoretical model. The current implementation directly uses the SuSAv2 hadron tensor for the 1p1h predictions and uses the tensor produced with the original theoretical model for 2p2h~\cite{RuizSimo:2016ikw} \textit{before} it is parameterised within SuSAv2-MEC~\cite{Megias:2014qva,Megias:2016lke,Megias:2016fjk}. Although the parameterised model is known mostly reproduce the original calculation, at very high or low neutrino energy there are some discrepancies. The use of the hadron tensor from the pre-parameterised microscopic model ensures an almost direct reproduction of its predictions at all kinematics. The input hadron tensors are finely binned (5 MeV in the energy and momentum transfer of the interaction) and are evaluated using an interpolation method similar to the one described in~\cite{Schwehr2017}. The validation of these implementations is discussed in Appendix~\ref{appendix-inc}.

For any inclusive cross-section model implementation, the generation of the outgoing hadronic state is (as with most model implementations in neutrino interaction event generators) largely factorised from the rest of the procedure. For SuSAv2 2p2h interactions this is mostly based on the methods already employed by the implementation for the Valencia group's 2p2h model in GENIE~\cite{Gran:2013kda,Schwehr2017} (which also uses a global Fermi gas initial-state model). The initial state nucleon momenta are chosen by independently sampling from a Fermi gas nuclear model (as was used in the theoretical model to produce the inclusive cross section prediction) before combining the two nucleons into a single `cluster'. The energy of this cluster is then reduced to account for a simple constant removal energy for each nucleon. The probability of the initial nucleons being a neutron-proton or neutron-neutron (or proton-proton in the case of incoming anti-neutrinos) pair is chosen based on the kinematics of the selected inclusive interaction using the SuSAv2 2p2h theoretical model~\cite{Simo:2016ikv,RuizSimo:2016ikw}. The four-momentum transfer from the inclusive interaction is then given to the cluster and the nucleon content is changed appropriately (for example for incoming neutrinos a neutron is turned into a proton) before the cluster is decayed isotropically to two nucleons in its centre of mass frame. The two outgoing nucleons are then separately propagated through a semiclassical FSI model\footnote{In this manuscript FSI is described using GENIE's `hN' semiclassical cascade model, rather than the `hA' empirical model which is known to have issues in all GENIEs up to the current GENIE version (R-3\_00\_06)~\cite{Harewood:2019rzy}.}, simulating re-interactions inside the nuclear medium thereby altering the nucleon's kinematics and potentially stimulating additional nuclear emission (of hadrons or further nucleons) and absorption. 

The hadron kinematics for the 1p1h model are generated with a similar methodology. However, here the removal energy of the nucleon is chosen based on a momentum-transfer dependent SuSAv2 analysis~\cite{Megias:2017PhD,Megias:2016ee,Gonzalez-Jimenez:2014eqa}, which represents a first step away from factorisation by correlating the hadronic initial state with the interaction kinematics. The global Fermi-gas used for the 2p2h case is also replaced with a local Fermi gas in the current version of the model implementation. Future work will aim to replace this with a RMF spectral function. The 1p1h case also demands more thought to keep the outgoing nucleon on-shell. To do this the momentum transfer to the nucleon is altered to satisfy its dispersion relation. Momentum is then conserved by giving the appropriate amount to the nuclear remnant.

This implementation scheme produces almost identical \textit{inclusive} predictions to the input model used to calculate the hadron tensor. However, the ability of this implementation to give reasonable semi-inclusive or exclusive predictions, as with almost all current model implementations in neutrino interaction event generators, clearly relies on several approximations and has a lot of unconstrained inputs (for example the spectral function used, the treatment of removal energy, the FSI model or how much four-momentum is given to the nuclear remnant). Primarily, instead of computing a fully exclusive cross-section in terms of all the particles in the final state (which would require 16 hadron tensor components to be parameterised~\cite{Moreno:2014kia}), an inclusive cross section is modelled properly by SuSAv2 as a function of muon kinematics only and the nucleon part is added {\itshape a posteriori}. Here the primary interaction is factorised from both FSI and the sampling of the nucleon spectral function (i.e. both are evaluated independently of an interaction's momentum and energy transfer). Moreover the energy transfer predicted from the inclusive interaction is initially given entirely to the target nucleon(s) and none to the nuclear remnant (the impulse approximation). So, while the model is expected to describe the lepton kinematics well, there is no guarantee that the final state proton kinematics and the proton-muon correlations are properly modelled by such an approach. Despite this, as previously explained, the simulation of semi-inclusive final states is necessary in the data analysis aiming at the measurement of neutrino oscillation parameters. As two obvious examples, we cite the correction for neutron in the neutrino energy reconstruction with the calorimetric approach used by the NO$\nu$A experiment~\cite{nova} and the subtraction of proton background from neutrino interactions in the antineutrino dominated beam. Recent measurements of cross sections as a function of the outgoing muon and proton kinematics and their correlations offer the opportunity to compare such approximated simulations to data, as will be discussed in the next section.

\section{Comparison of SuSAv2 2p2h with other models}\label{results2p2h}

The SuSAv2 2p2h model is based on quite different theoretical assumptions than the other  models available in GENIE: the Valencia model~\cite{Nieves:2011yp,Gran:2013kda} and GENIE's own `empirical' model~\cite{Katori:2013eoa}. The latter is not directly based on any microscopic calculation but is widely used by the $\mu$BooNE~\cite{Adams:2018sgn} and NO$\nu$A~\cite{NOvA:2018gge} experiments. It places a smooth contribution in the `dip' area of invariant-mass phase space (between the 1p1h and resonant peaks) amounting to around 45\% of the strength of the default GENIE RFG 1p1h model. SuSAv2 2p2h and the Valencia model are both based on the same fundamental RFG-based 2p2h microscopic calculation~\cite{DePace:2003}, but are different implementations of it. A particular difference stems from the treatment of the $\Delta$-resonance propagator. 
SuSAv2 2p2h implements only the real part of the $\Delta$-resonance propagator in the 2p2h pion-exchange diagrams in order to avoid double counting of possible effects related to $\Delta$-excitation effects in both 2p2h channel and the inelastic regime, while the Valencia model implements only partially the real part and partially the imaginary part, including also higher energy resonance exchange ($\rho$). The treatment of the $\Delta$-resonance propagator in the SuSAv2 2p2h model follows refs.~\cite{DePace:2003,Simo:2016ikv}, which are also used by other groups~\cite{Rocco,DePace:2004,Butkevich,Dekker,Alberico,VanOrden}, and can be viewed as an empirical approach that provides very good agreement with ($e,e'$) scattering data~\cite{Megias:2016lke,Megias:2019jlab}. Nevertheless, one could argue that contributions from the imaginary part of the $\Delta$ propagator in a 2p2h RFG approach do not lead to real pions in the final state. Indeed, the treatment of the $\Delta$-excitation effects is still an open question to be addressed by theoretical models as possible double-counting effects between the 2p2h channels and the inelastic regime could be considered in the analysis, depending on how the inelastic response is modelled and how the medium modification of the $\Delta$ decay width is treated in both 2p2h and pion-production regimes. More dedicated analyses of the $\Delta$ propagator will be addressed in further works although some preliminary results have been shown in~\cite{Megias:2019mnv} where overall no large effects are expected for T2K and MINERvA CC0$\pi$ inclusive measurements ($\lesssim 10\%$, mainly at large $q_0$ for a given $q_3$ value). Therefore, the inclusion of SuSAv2 2p2h in GENIE provides a complementary addition which, crucially, has been carefully validated using electron scattering data. 

\begin{figure*}[!hp]
\begin{center}
\includegraphics[scale=0.425, angle=0]{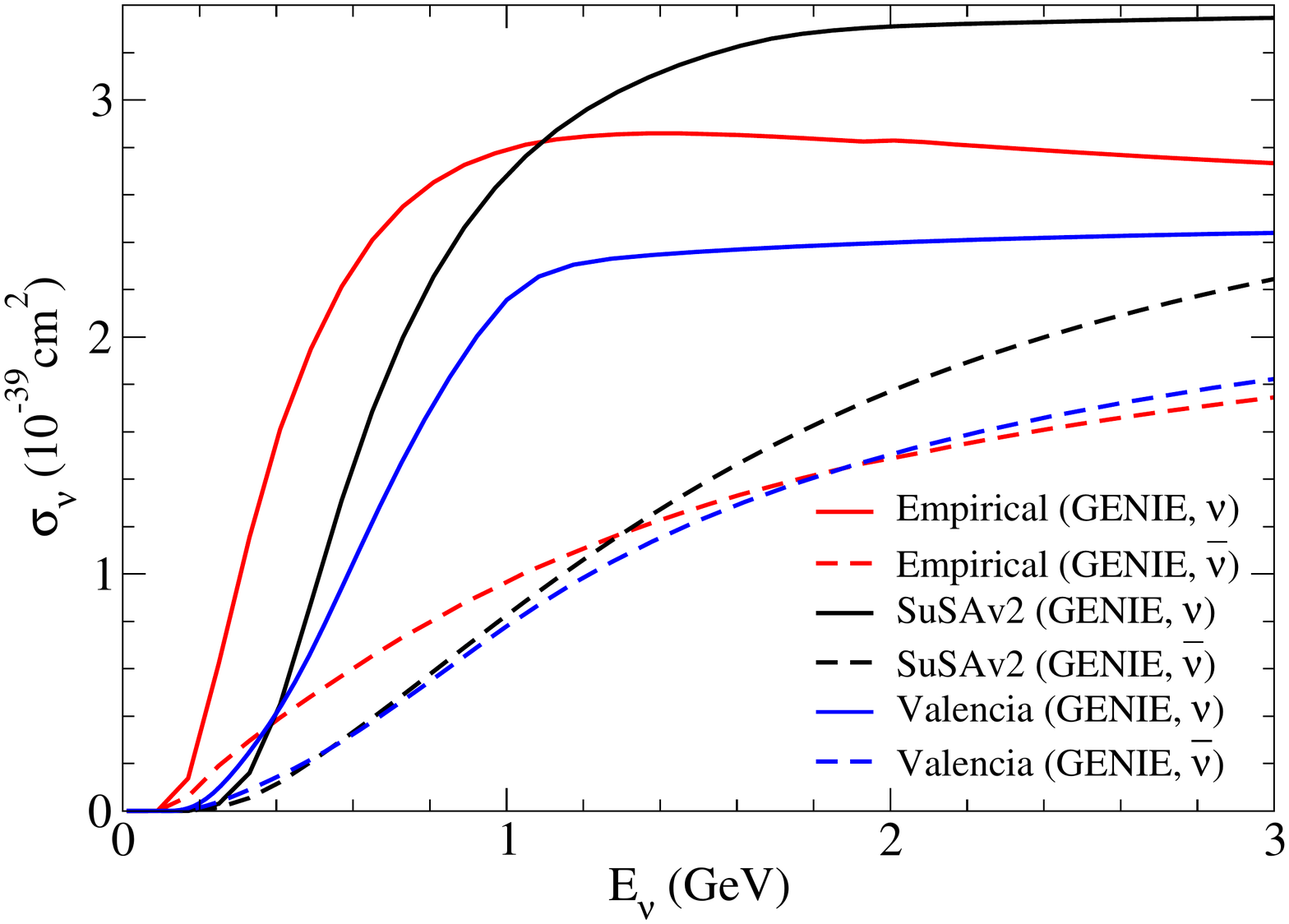}
\end{center}
\caption{Total 2p2h cross section for muon-neutrino and anti-neutrino interactions on Carbon for the ``Empirical'' model available in GENIE, the SuSAv2 2p2h model discussed in this manuscript and the Valencia model as implemented in GENIE~\cite{Nieves:2011yp,Gran:2013kda}.}
\label{fig:2p2hEnu}
\end{figure*}

The dependence of the SuSAv2 2p2h, Valencia and empirical 2p2h neutrino and antineutrino cross sections with the incoming neutrino energy is shown in Fig.~\ref{fig:2p2hEnu}. It can be seen that all the models differ substantially in both normalisation and shape. At higher energy part of the difference between the SuSAv2 and Valencia models stems from the fact that the latter is only available up to 1.2 GeV of momentum transfer but there are also substantial differences at lower energy as well. This different behavior is due to fundamental differences in the nuclear response functions encoded in the hadron tensors. Indeed, while the only hadron tensor element with explicit energy dependence is the V-A interference term ($W_3$ in the Valencia model notation in~\cite{ValverdePhD}), all of the hadron tensor terms have an implicit dependency on the energy because of the integration limits on $q_3$,$q_0$. For a detailed view of the energy dependence of the various hadron tensors in SuSAv2 model, see~\cite{Megias:2016fjk,Megias:2017PhD}.

\begin{figure*}[!hp]
\begin{center}
\includegraphics[width=0.49\linewidth]{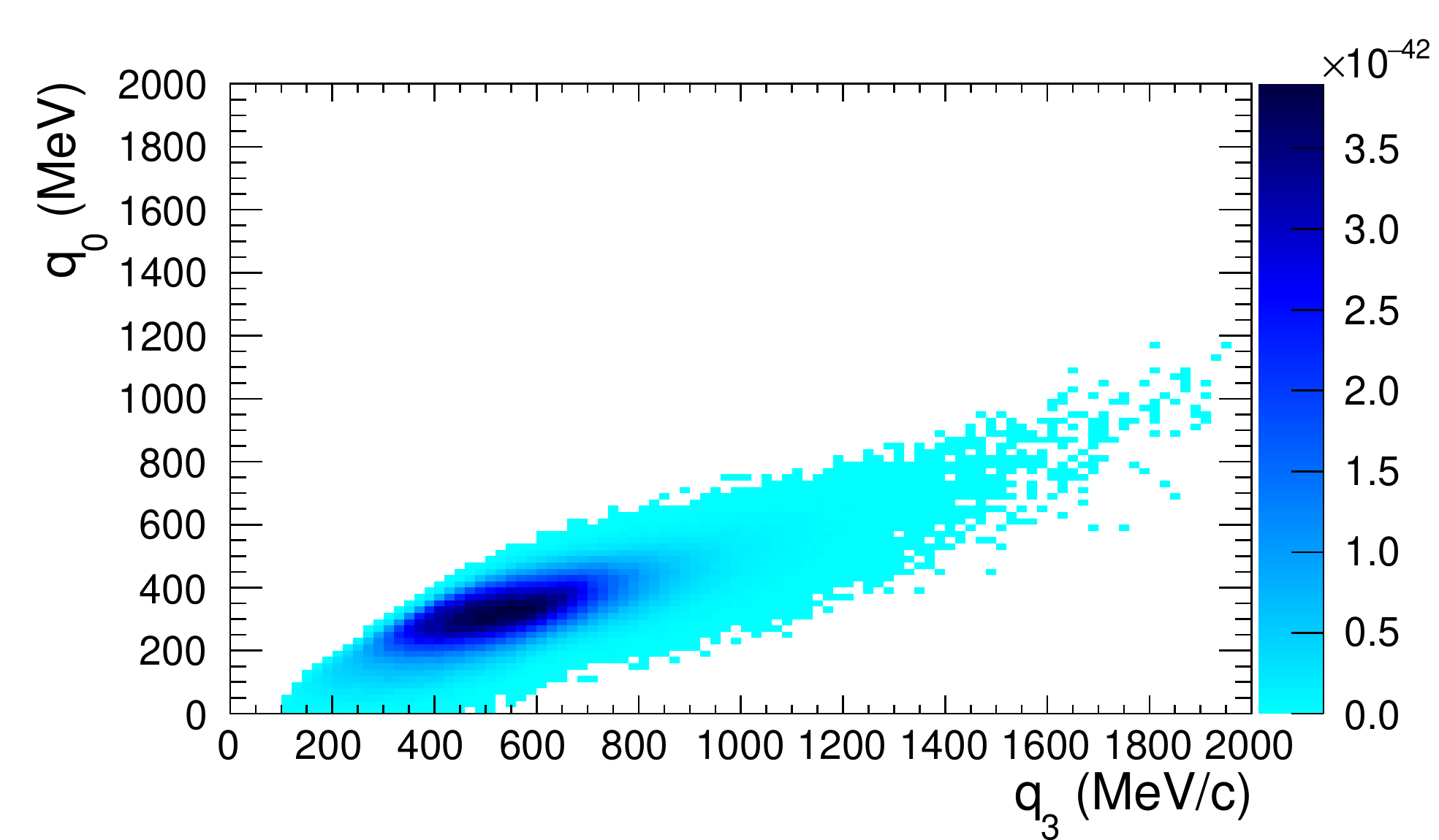}
\includegraphics[width=0.49\linewidth]{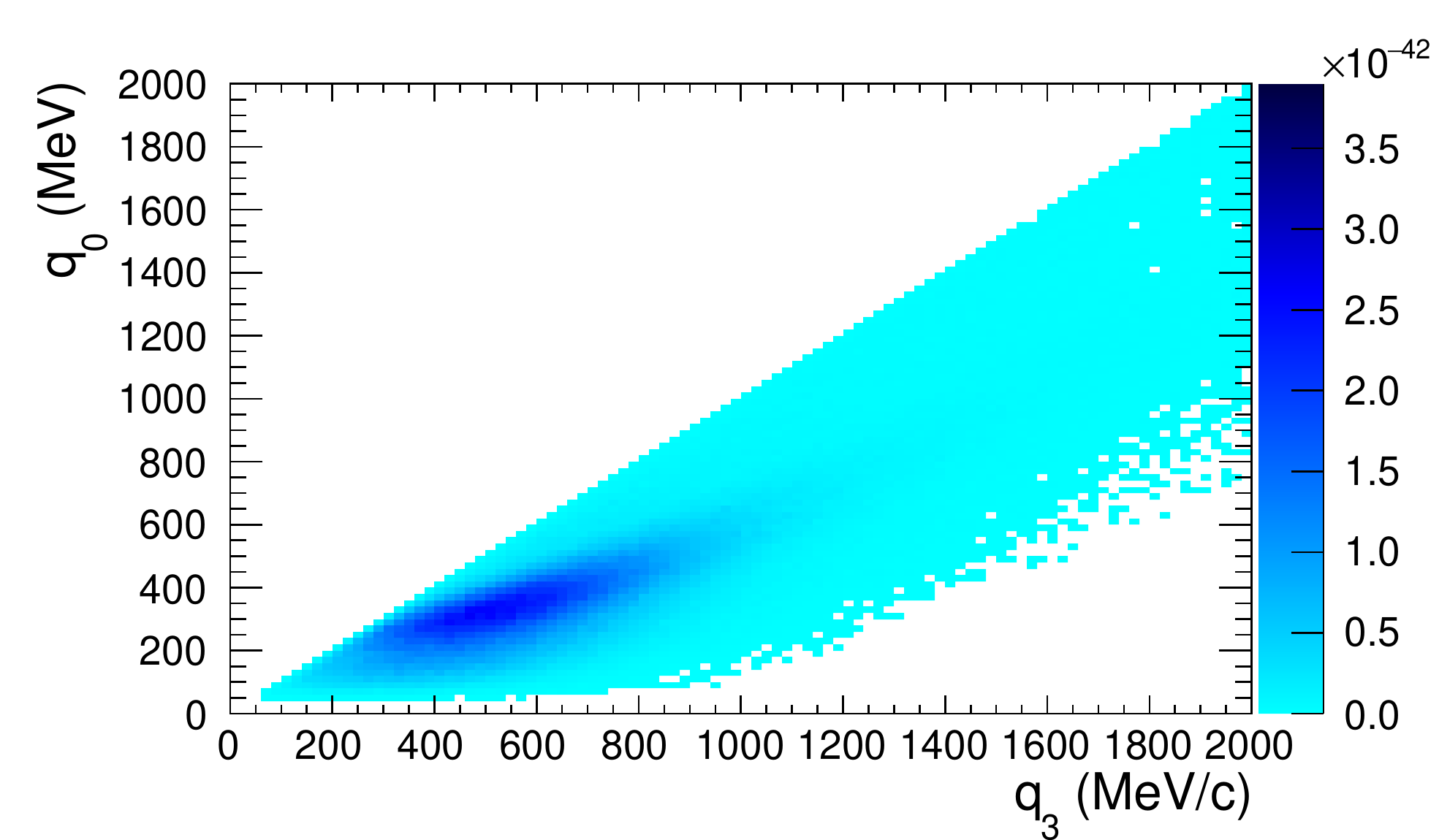}
\includegraphics[width=0.49\linewidth]{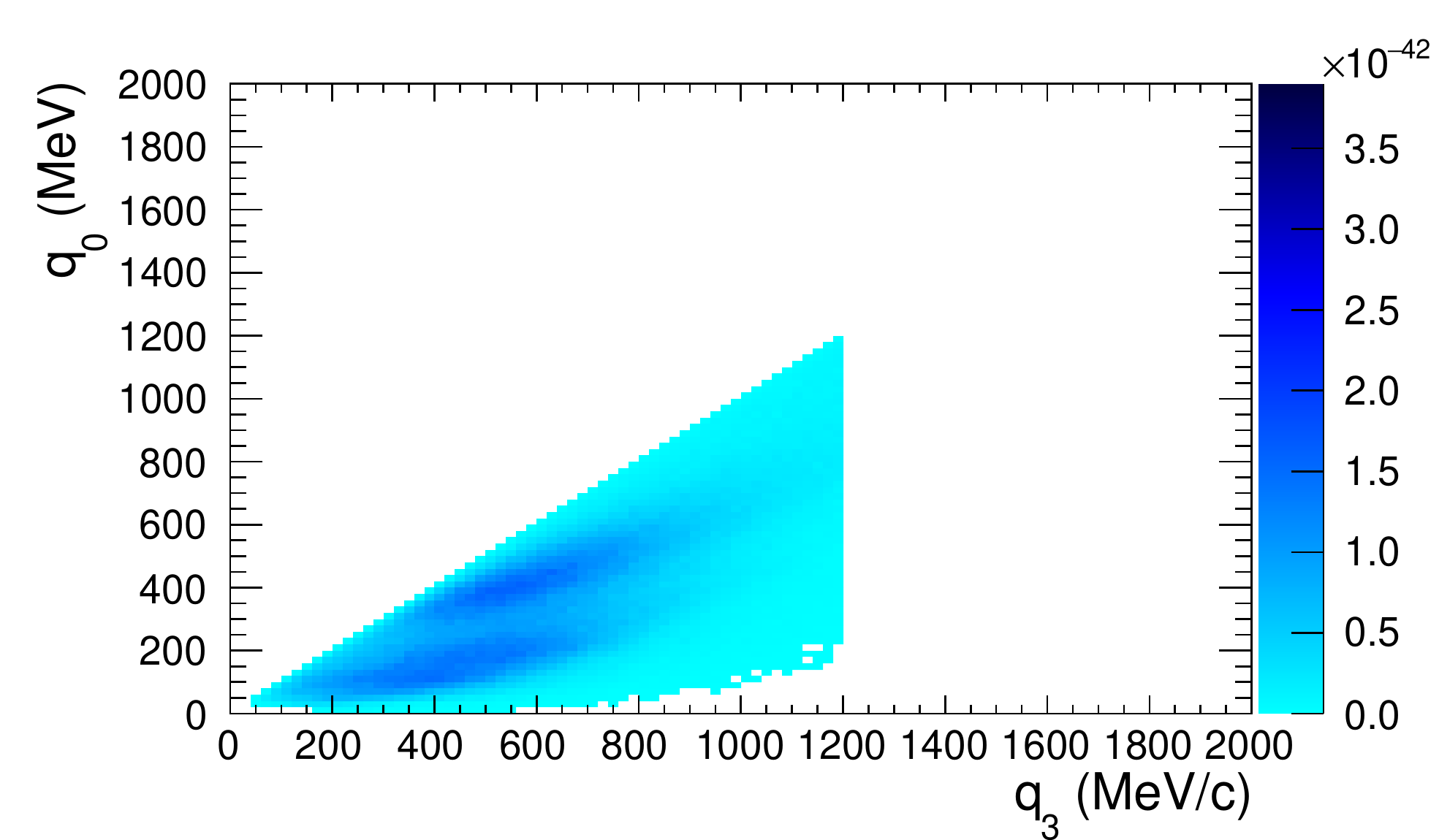}
\end{center}
\caption{2p2h cross-sections (in units of cm$^2$) in bins of energy ($q_0$) and momentum ($q_3$) transfer  for muon-neutrino interactions on Carbon for different models. The top left is the GENIE ``Empirical'' model, the top right is the implemented SuSAv2 2p2h prediction and the lower plot is the GENIE implementation of the the Valencia model (where the 1.2 GeV cut off in the model discussed in the text is clear).}
\label{fig:2p2hq0q3}
\end{figure*}

\begin{figure*}[!hp]
\begin{center}
\includegraphics[width=0.32\linewidth]{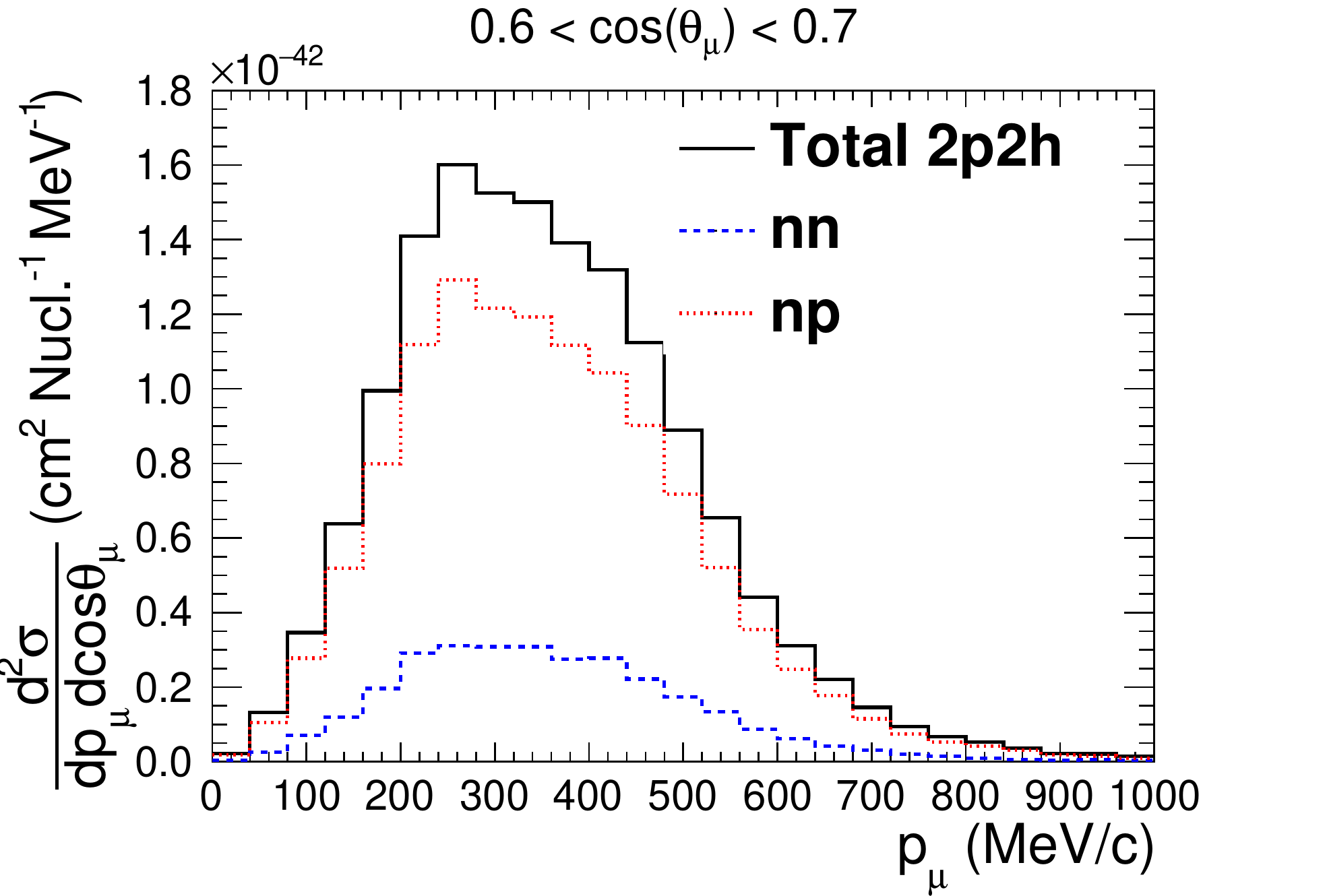}
\includegraphics[width=0.32\linewidth]{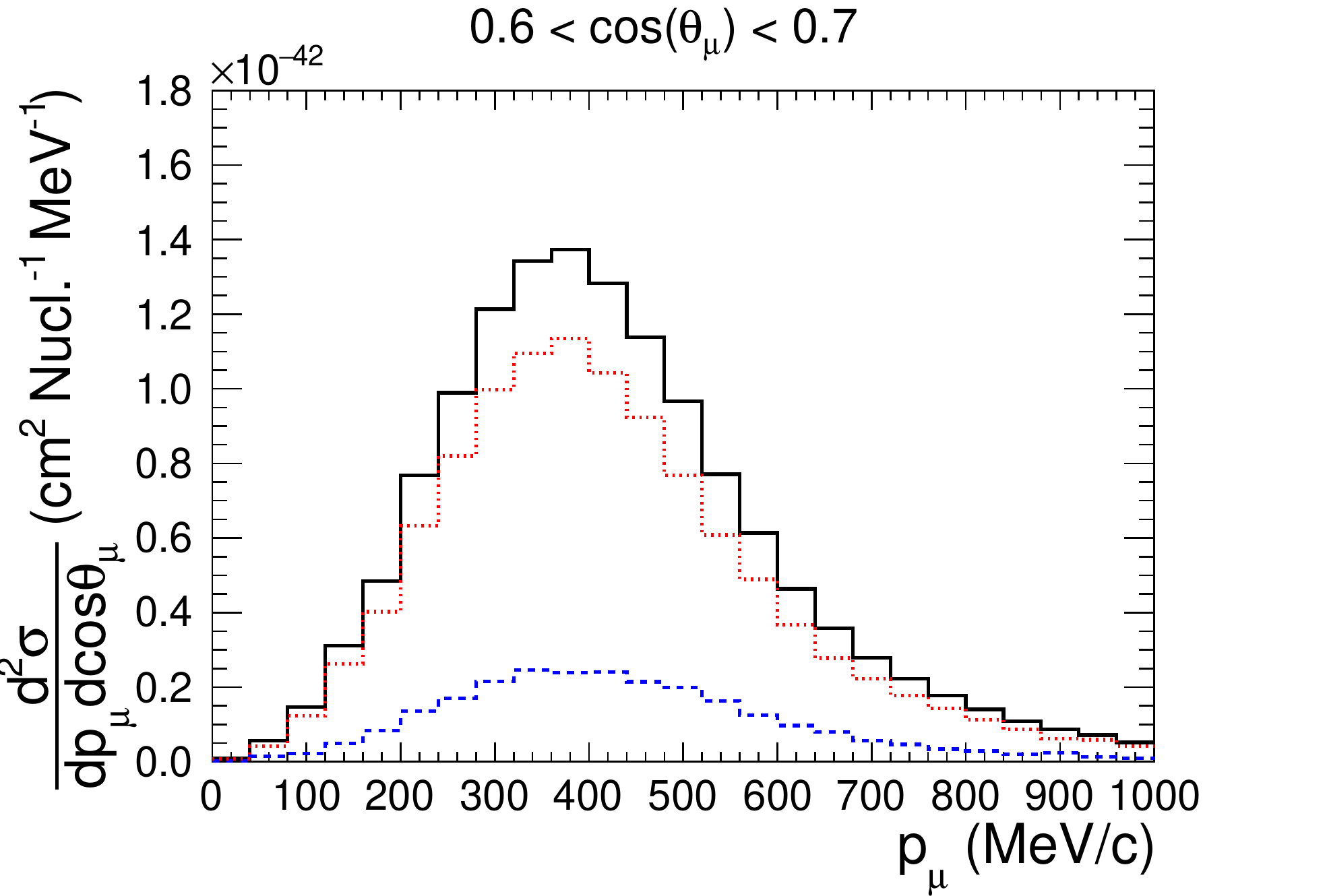}
\includegraphics[width=0.32\linewidth]{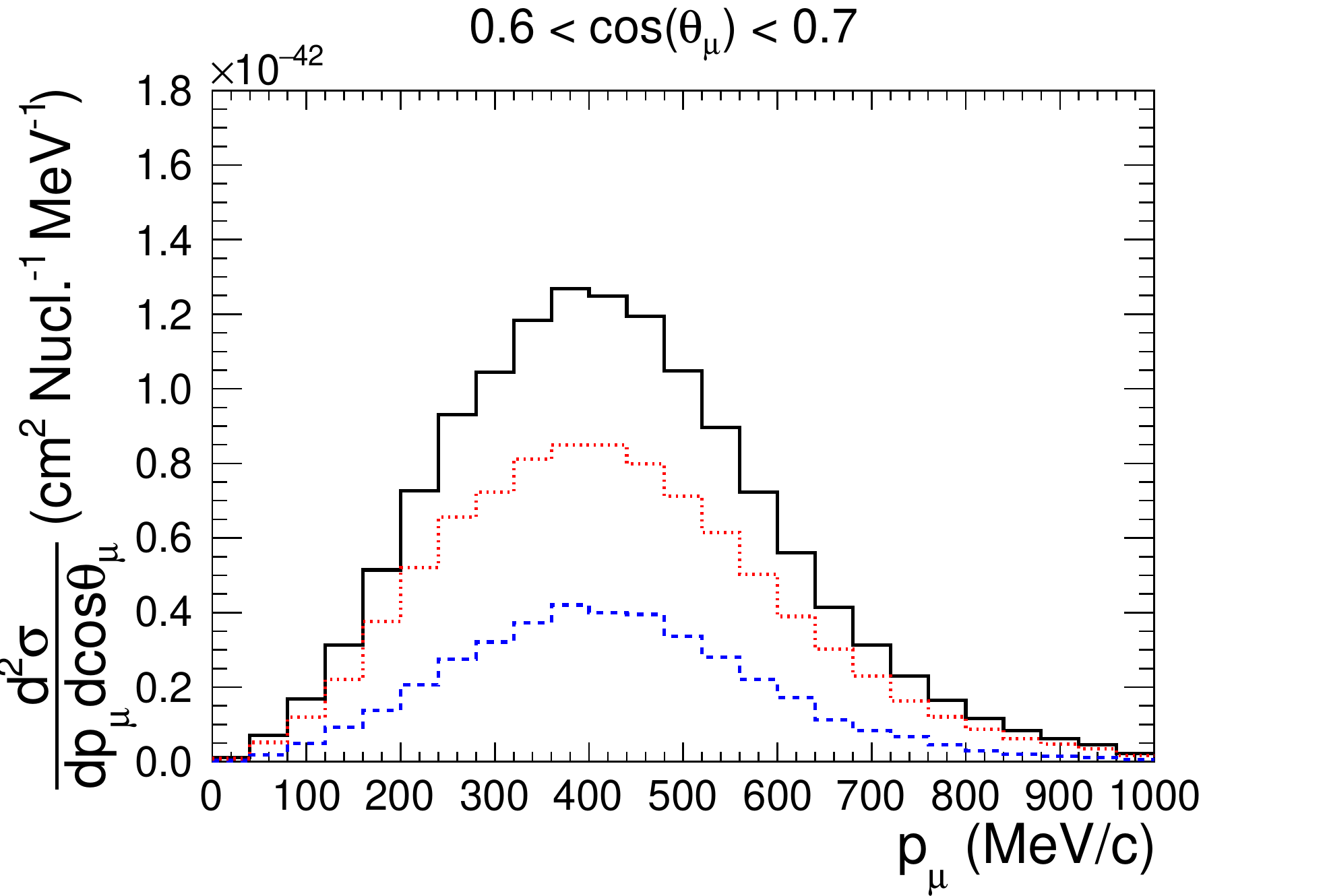}

\includegraphics[width=0.32\linewidth]{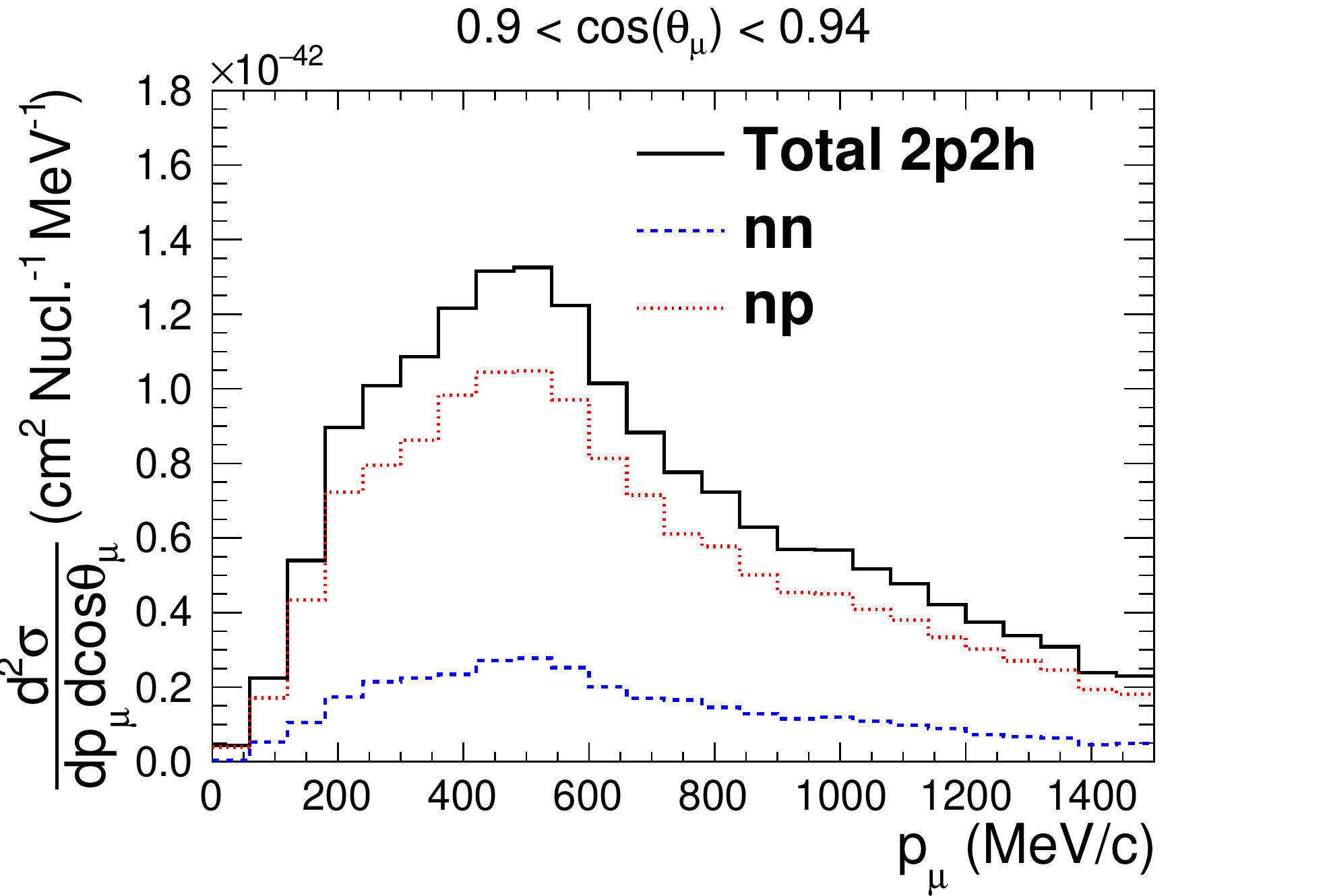}
\includegraphics[width=0.32\linewidth]{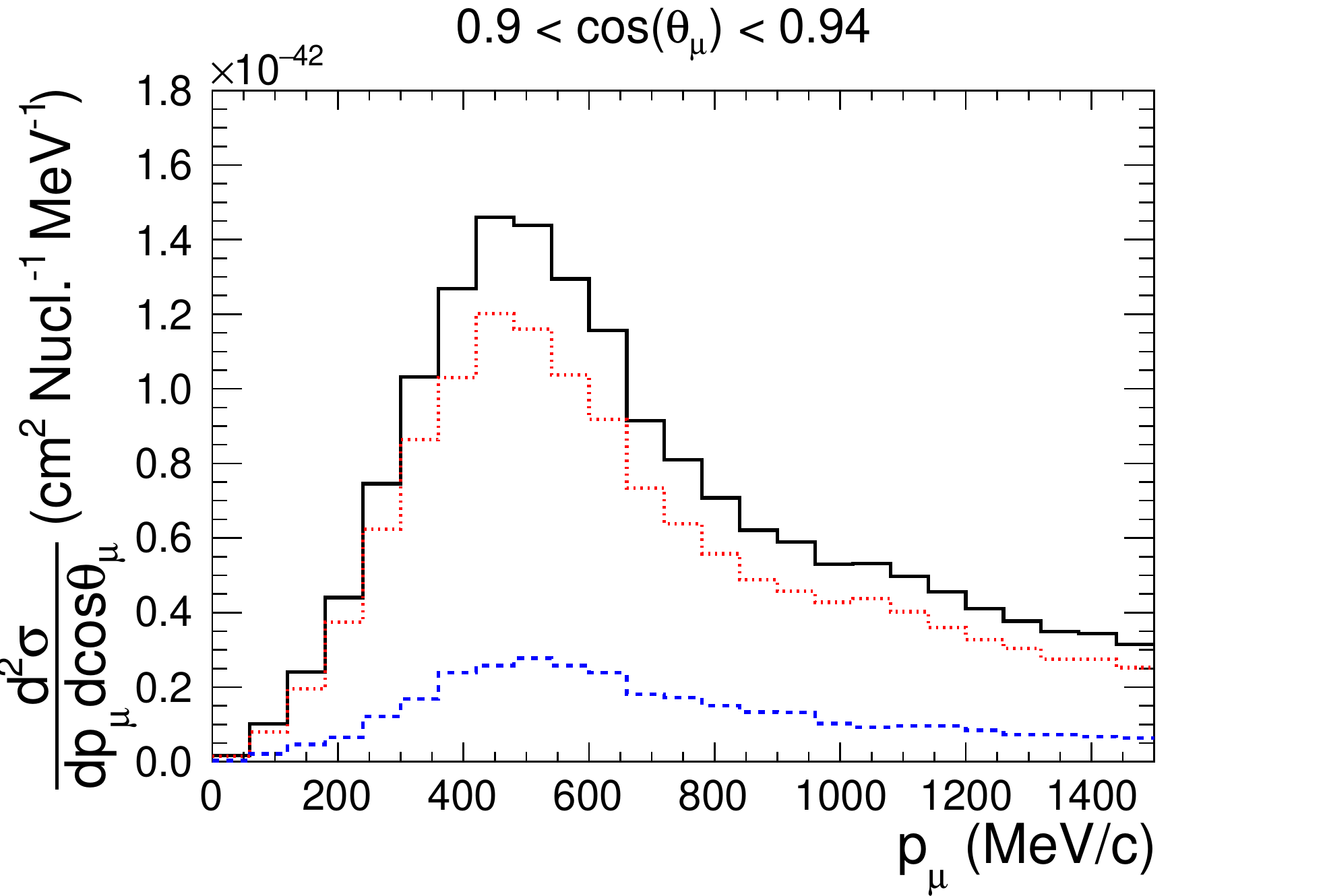}
\includegraphics[width=0.32\linewidth]{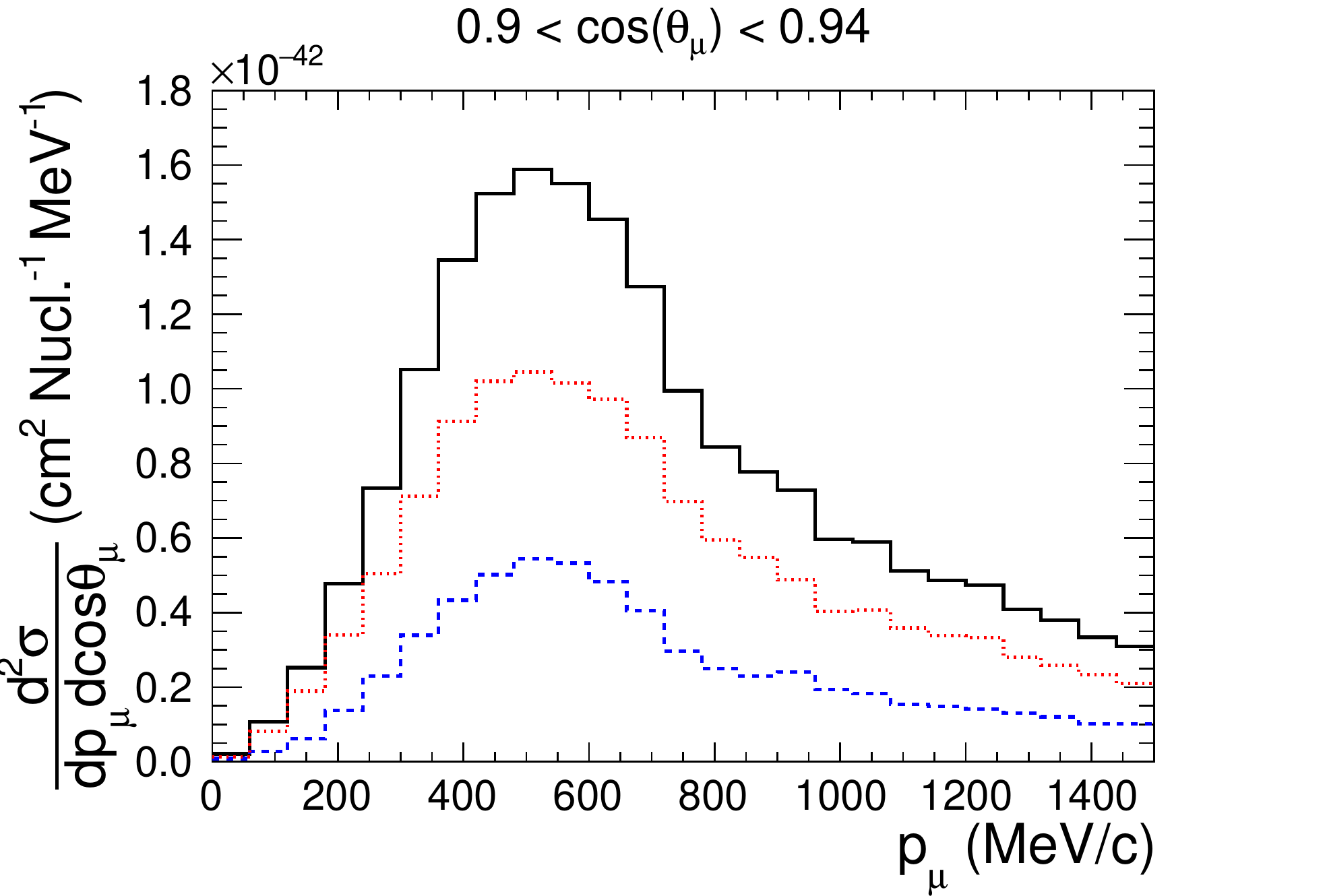}
\end{center}
\caption{A comparison of 2p2h double-differential cross sections in muon momentum for two different angular slices for muon-neutrino interactions on Carbon for different models, split by the contribution from the different initial state correlated pairs: neutron-neutron ($nn$) or neutron-proton ($np$). The left plots are from the GENIE ``Empirical'' model, the centre are from the implemented SuSAv2 prediction and the right plots are from the GENIE implementation of the Valencia model.}
\label{fig:2p2hpairs}
\end{figure*}

More of the fundamental differences between the models are made evident when comparing the T2K flux-integrated cross-section as a function of $q_3$,$q_0$ as in Fig.~\ref{fig:2p2hq0q3}. Two components are clearly visible in the Valencia model: one at relatively high $q_3$,$q_0$, in the region of $\Delta$ resonance, which is related with $\Delta$ excitation diagrams (also called $\Delta$ pion-less decay) and a second component at lower $q_3$,$q_0$, in the Quasi-Elastic kinematic region. The SuSAv2 2p2h model instead predicts a single wide region of cross-section enhancement in the 'dip' region between $\Delta$ and Quasi-Elastic kinematics.  Fig.~\ref{fig:2p2hpairs} shows that these starkly different model predictions are observable in experimentally accessible flux-averaged differential cross sections as a function of muon kinematics. The largest differences are visible at larger scattering angles and lower muon momentum. However, despite their notable size, such differences would be difficult to observe in any CC0$\pi$ or inclusive measurement because of the large uncertainty on the 1p1h component which dominates the cross-section. More exclusive measurements, including information of the proton(s) in the final state have been performed in T2K~\cite{Abe:2018pwo} and Minerva~\cite{Lu:2018stk} in order to enhance the sensitivity to 2p2h and will be discussed in Sec.~\ref{results-sinc}.

Although the microscopic 2p2h models available in GENIE are based on a predominantly inclusive calculation, they remain able to predict the relative contributions of neutron-neutron ($nn$) and neutron-proton ($np$) initial state nucleon pairs, which are shown in Fig.~\ref{fig:2p2hpairs}. While the variations in the total 2p2h prediction is fairly small, it is very interesting to note the large differences observed between the SuSAv2 and Valencia models when considering the relative contribution of $nn$ and $np$ pairs. These differences largely stem from the omission of the direct-exchange interference terms in the Valencia model, which are fully included in the SuSAv2 model. The effect of neglecting the direct-exchange interference of the MEC matrix elements in the 2p2h channel has been shown in previous works~\cite{Simo:2016ab,RuizSimo:2016ikw,Simo:2016ikv}, to result in a negligible effect for $np$ initial states but a reduction of a factor of $\sim$2 in $nn$ initial states (and so $np$ emission), thereby largely affecting the $nn/np$ ratio. This can be observed in Fig.~\ref{fig:2p2hpairs} when comparing the SuSAv2 model, which fully accounts for these interference terms, with the Valencia one, in which they are absent. Since protons typically deposit much more energy than neutrons, this observation suggests that following a neutrino 2p2h CC interaction the SuSAv2 model would produce final states that leave a substantially larger observable calorimetric energy deposit than would be predicted using the Valencia model. The opposite would occur for antineutrinos. These effects are especially relevant for neutrino oscillation experiments which use a calorimetric method of neutrino energy reconstruction, which may see a substantial alteration to neutrino energy reconstruction performance when switching models. Since the $pn$ and $nn$ pairs have notably different hadron tensors, the different relative contributions also lead to different inclusive kinematic predictions. This difference in initial state pair predictions may also act as a signature to allow model differentiation, in particular through semi-inclusive measurements of proton multiplicity which will be discussed in section~\ref{results-sinc}. Complementary future measurements of neutrons, such as those which can be performed in scintillator detectors as shown in~\cite{Elkins:2019vmy,Abe:2019whr}, may also prove to be a powerful probe of 2p2h. 

Further comparisons of the 2p2h (and 1p1h) predictions (including comparisons with T2K data) can be found in Appendix~\ref{appendix-nieves}.



\section{`Semi-semi-inclusive' results with SuSAv2-RMF}\label{results-ssinc}

Although the available 2p2h models differ substantially, inclusive measurements struggle to distinguish them due to the aforementioned dominant 1p1h contribution and the lack of a region of lepton kinematics particularly enhanced in 2p2h. This is discussed in more detail in Appendix~\ref{appendix-inc}. However, more exclusive measurements which include information about the final state nucleons, such as those which have recently been performed by T2K~\cite{Abe:2018pwo} and Minerva~\cite{Lu:2018stk}, have been demonstrated to have a much more acute sensitivity to the different nuclear effects involved in neutrino-nucleus interactions. Unfortunately a comparison of these measurements directly to microscopic models requires semi-inclusive predictions which the majority of models are not able to make, as they simplify their calculations by integrating over outgoing nucleon kinematics. An exception to this is the RMF model, used to construct the SuSAv2 predictions, which is capable of `semi-semi-inclusive' predictions for neutrino reactions: it is able to calculate outgoing nucleon momenta but not angles\footnote{The RMF model has proven its validity to address full semi-inclusive predictions for electron scattering~\cite{PhysRevC.64.024614} and work is underway to extend it to neutrino reactions.}. As described in Sec.~\ref{implementation}, the simulations used by experiments circumvent this limitation by factorising the leptonic and hadronic components of the interaction. Among other approximations, this approach relies strongly on a semi-classical description of FSI and the distribution of initial state nucleon kinematics seen by the probe being independent of its energy and momentum transfer.

The implementation of the SuSAv2 1p1h model in GENIE provides a first opportunity to test this factorisation approach. The RMF model is first used to predict an inclusive double-differential T2K flux-integrated cross section in muon kinematics and then another semi-inclusive cross section where the final state proton is below 500 MeV/c (a topology that was measured by T2K by analysing simultaneously events in which protons were and were not observed). The same exercise is then repeated using the SuSAv2 GENIE implementation where the inclusive prediction should match the original SuSAv2 model (which is identical to RMF over a large portion of the kinematic phase space) almost exactly for low to mid angle muons, and minor differences are expected at very forward angles due to different integration and interpolation methods. The semi-inclusive SuSAv2-GENIE prediction comes from the factorisation method described in Sec.~\ref{implementation}. To help understand the different elements of the factorisation approximation, the GENIE semi-inclusive prediction is made with/without FSI and with both a kinematic dependent binding energy (as described in Sec.~\ref{implementation}) and with a fixed value of 25 MeV (around the value often used, e.g.~\cite{Abe:2015awa}). A comparison of the inclusive and semi-inclusive results from RMF and the GENIE SuSAv2 implementation (alongside the inclusive predictions from the SuSAv2 model) is shown in Fig.~\ref{fig:ssIncToIncComp} in a few bins corresponding to T2K measurements (although the data is not overlaid as the 2p2h and pion absorption components are not evaluated here). The full comparison is available in Appendix~\ref{appendix-fact}.


In Fig.~\ref{fig:ssIncToIncComp}, we can observe a very good agreement between the original SuSAv2 inclusive results and its implementation in GENIE, only minor differences can be observed at very forward angles due to different interpolation and integration methods. When comparing SuSAv2 with the RMF model we observe very similar results at intermediate angles (0.6-0.9) while noticing a decrease of the RMF predictions at very forward angles and backward ones, where a small shift to low muon momentum values, i.e. large energy transfer, can also be observed. These discrepancies are both related to the implementation of RMF effects in SuSAv2: those at backward angles are from an implemented data-motivated transition to weaker FSI than in RMF at larger energy transfers, whilst those at forward angles stem from low energy transfer scaling violations in the full RMF creating difficulties in encapsulating it completely into the super-scaling formalism used. The comparison between SuSAv2 and RMF is discussed further in Appendix~\ref{appendix-fact}.

Beyond the inclusive comparison, the `semi-semi-inclusive' predictions within the kinematic region where SuSAv2 is a good description of RMF allows us to study the validity of the factorisation approach used in event generators. Here it can be seen that the implementation with both the kinematic-dependent binding energy and with FSI is closest to reproducing the RMF microscopic model prediction, but still appears to peak at too low muon momentum and also fails to describe the higher momentum region. It can also be seen that variations to the hadronic component of the interaction cause substantial alterations to the predictions, highlighting the role of these nonphysical freedoms available within the factorisation approach. Further work will focus on more stringent tests through the implementation of the RMF spectral function into event generators and by exploring the predictions in a wider region of hadronic kinematic phase-space (ideally using a fully semi-inclusive version of RMF).

\begin{figure*}[!hp]
	\begin{center}
		\includegraphics[width=0.34\linewidth, angle=0]{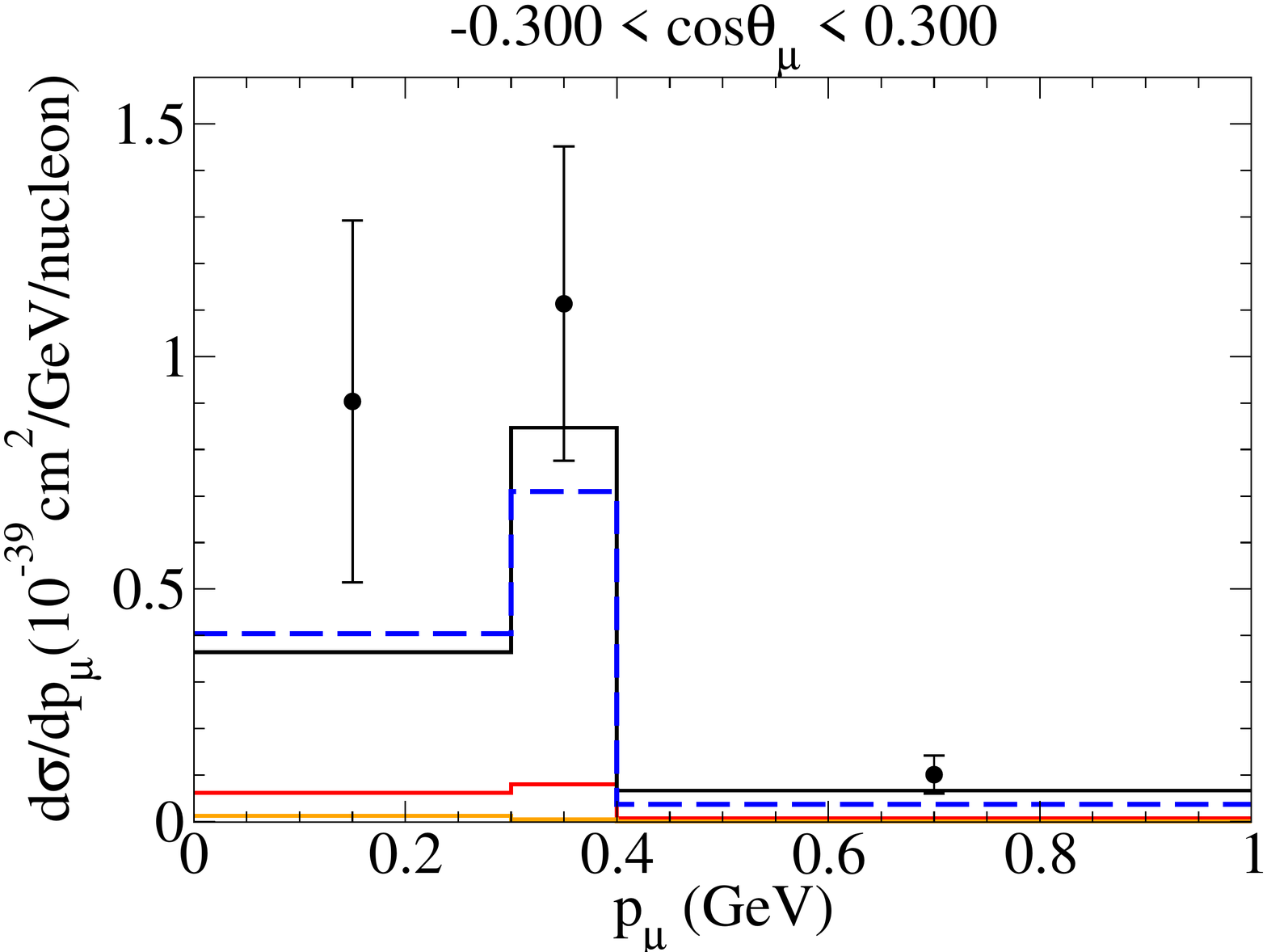}\hspace*{-0.295cm}
		\includegraphics[width=0.34\linewidth, angle=0]{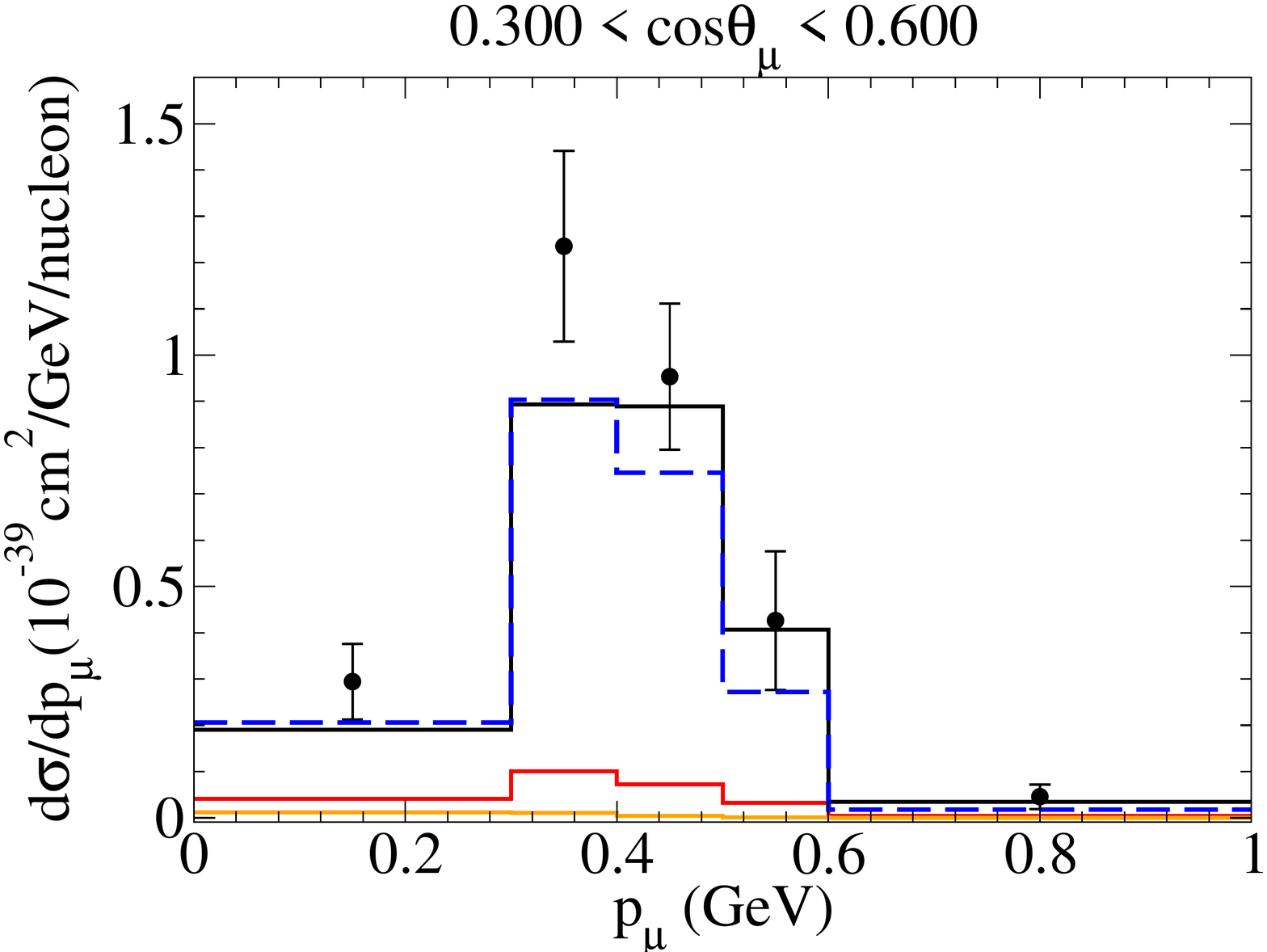}\hspace*{-0.295cm}
		\includegraphics[width=0.34\linewidth, angle=0]{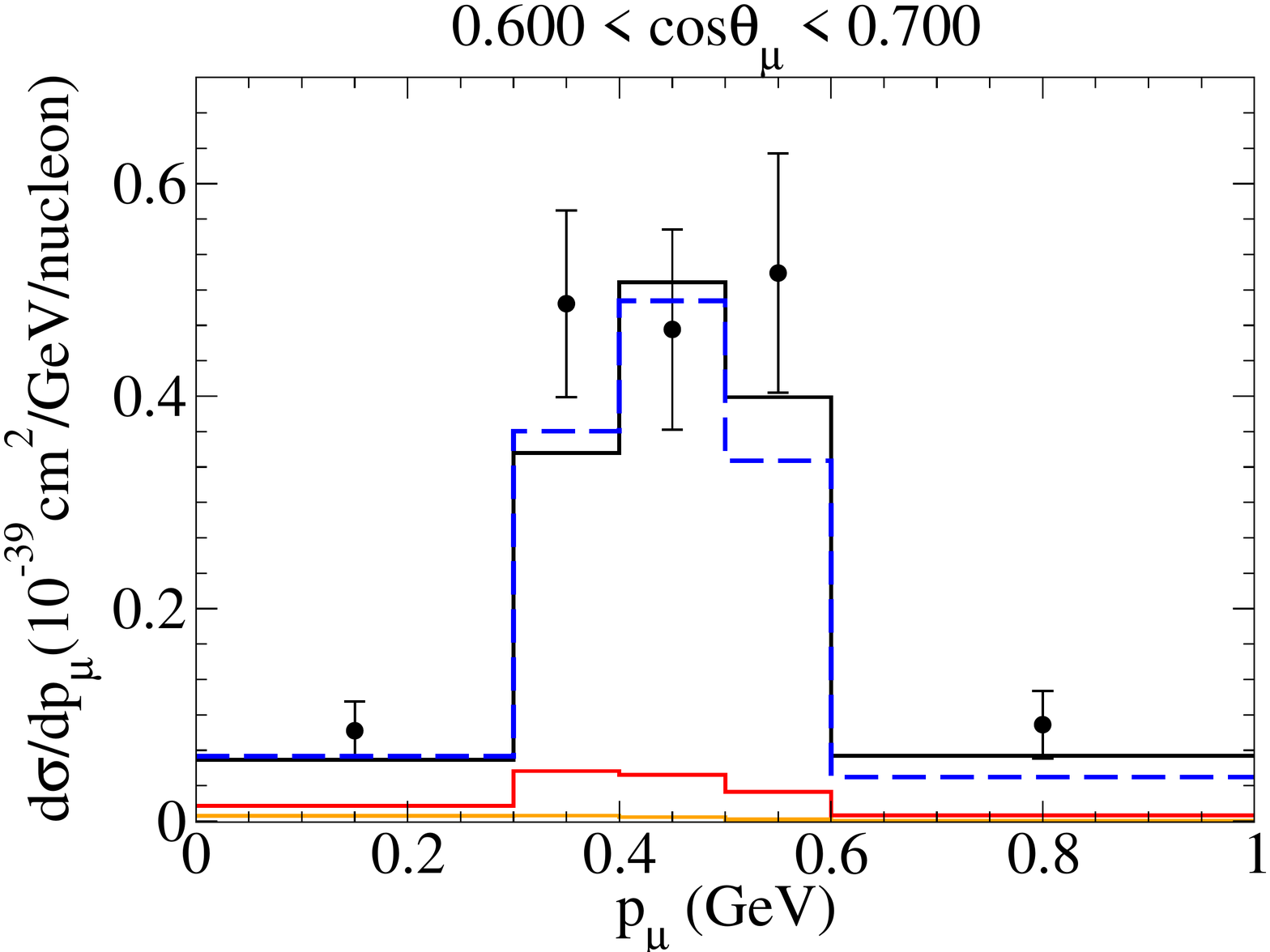}
		\includegraphics[width=0.34\linewidth, angle=0]{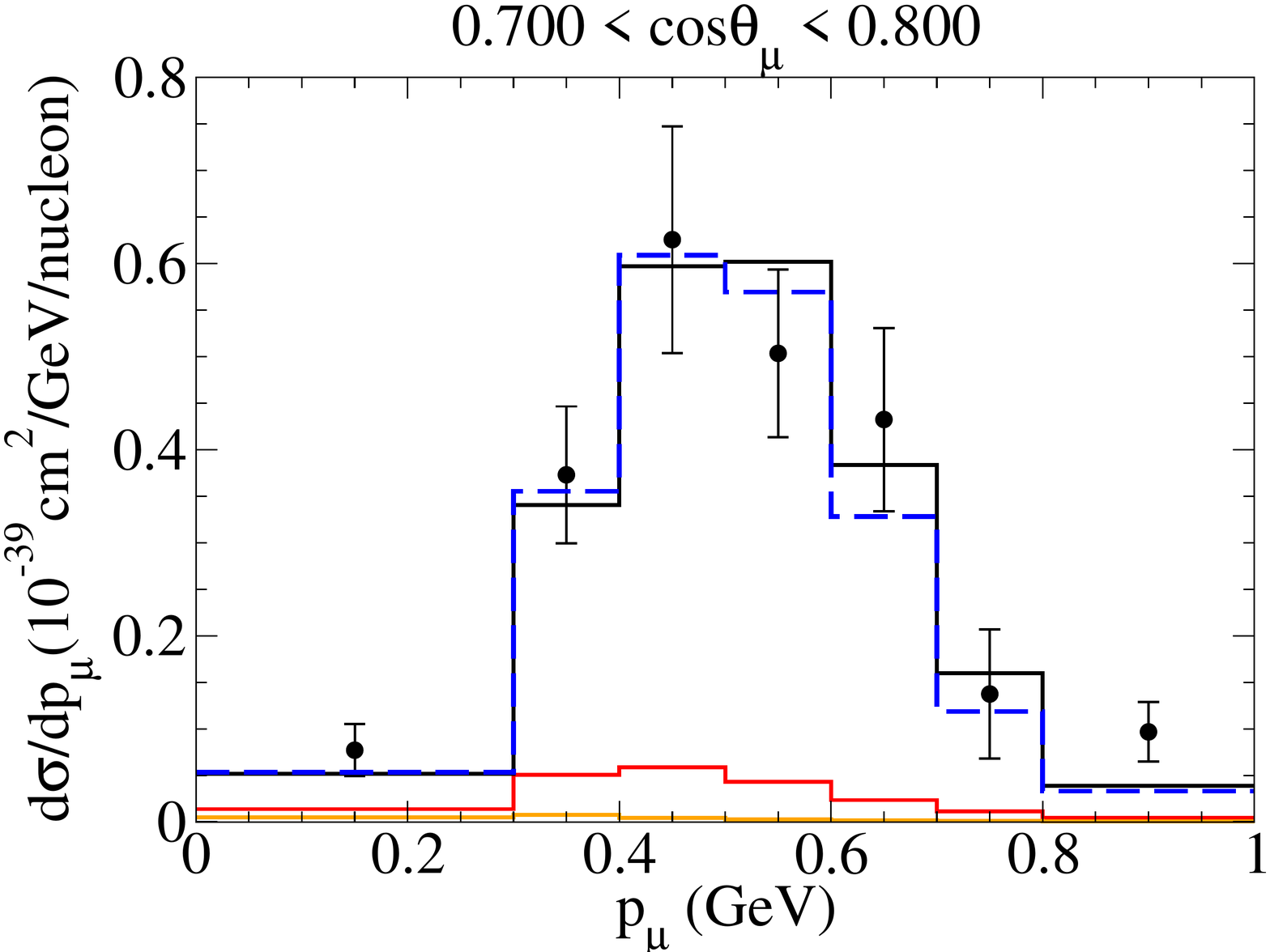}\hspace*{-0.295cm}
		\includegraphics[width=0.34\linewidth, angle=0]{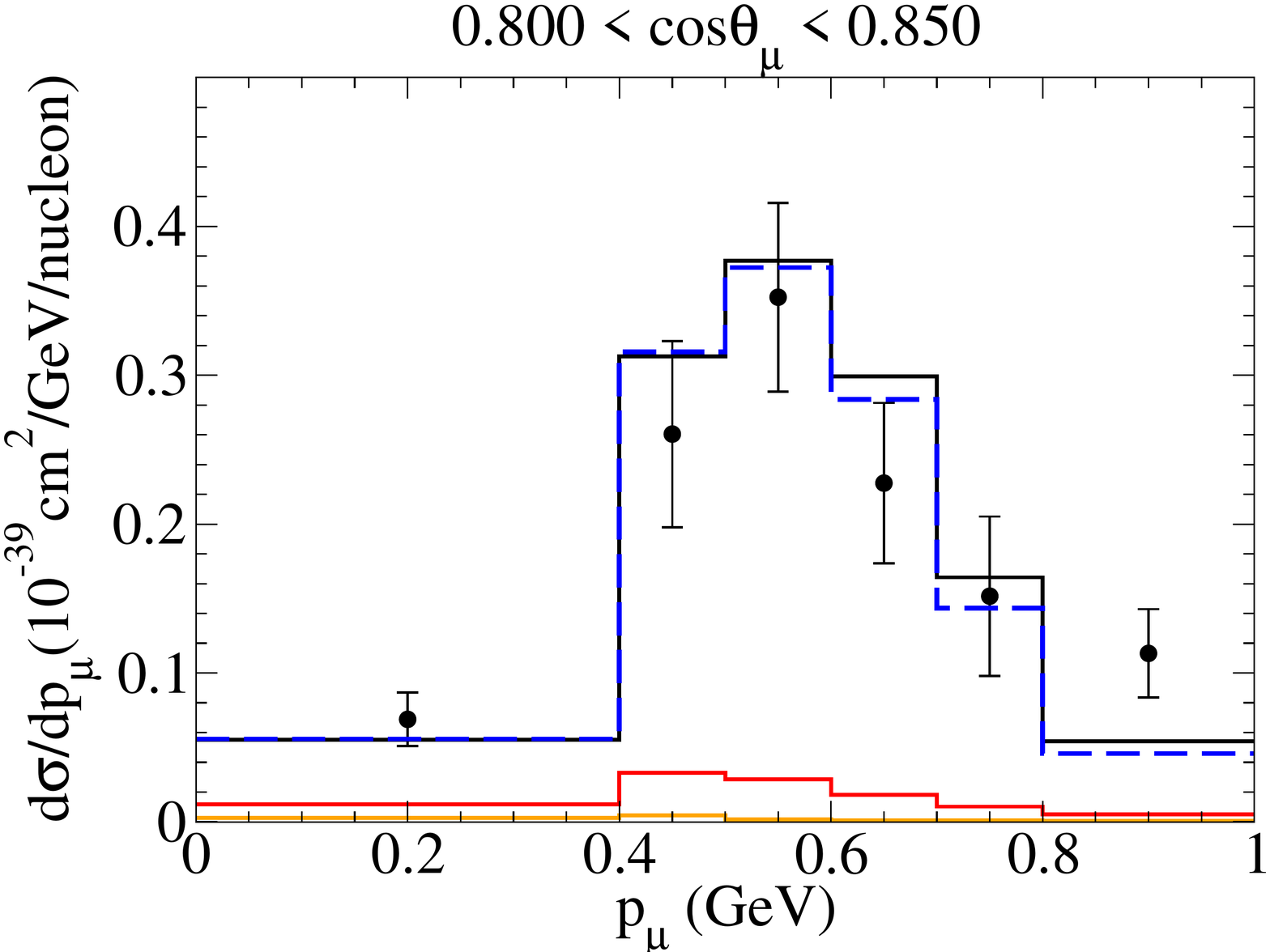}\hspace*{-0.295cm}
		\includegraphics[width=0.34\linewidth, angle=0]{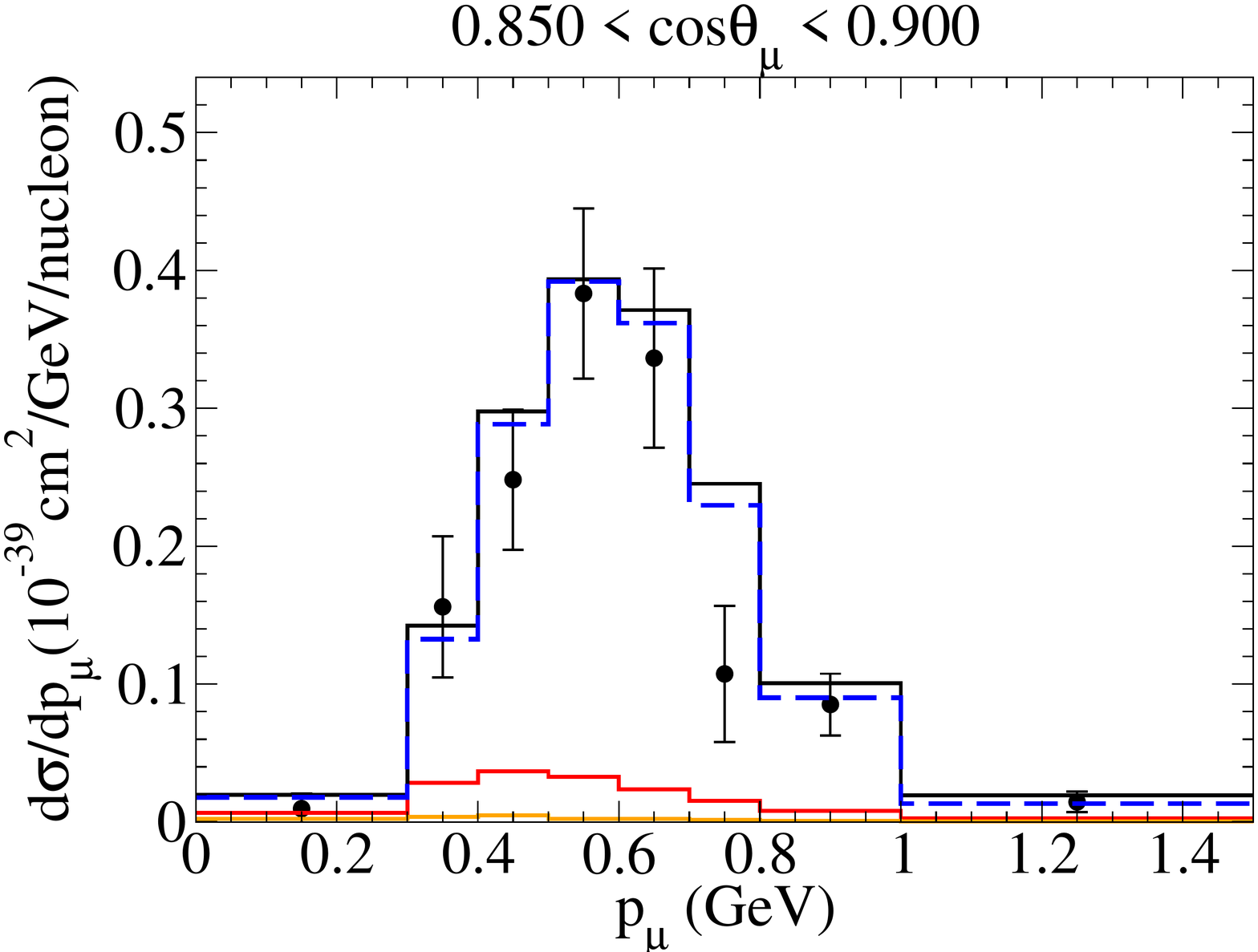}
		\includegraphics[width=0.34\linewidth, angle=0]{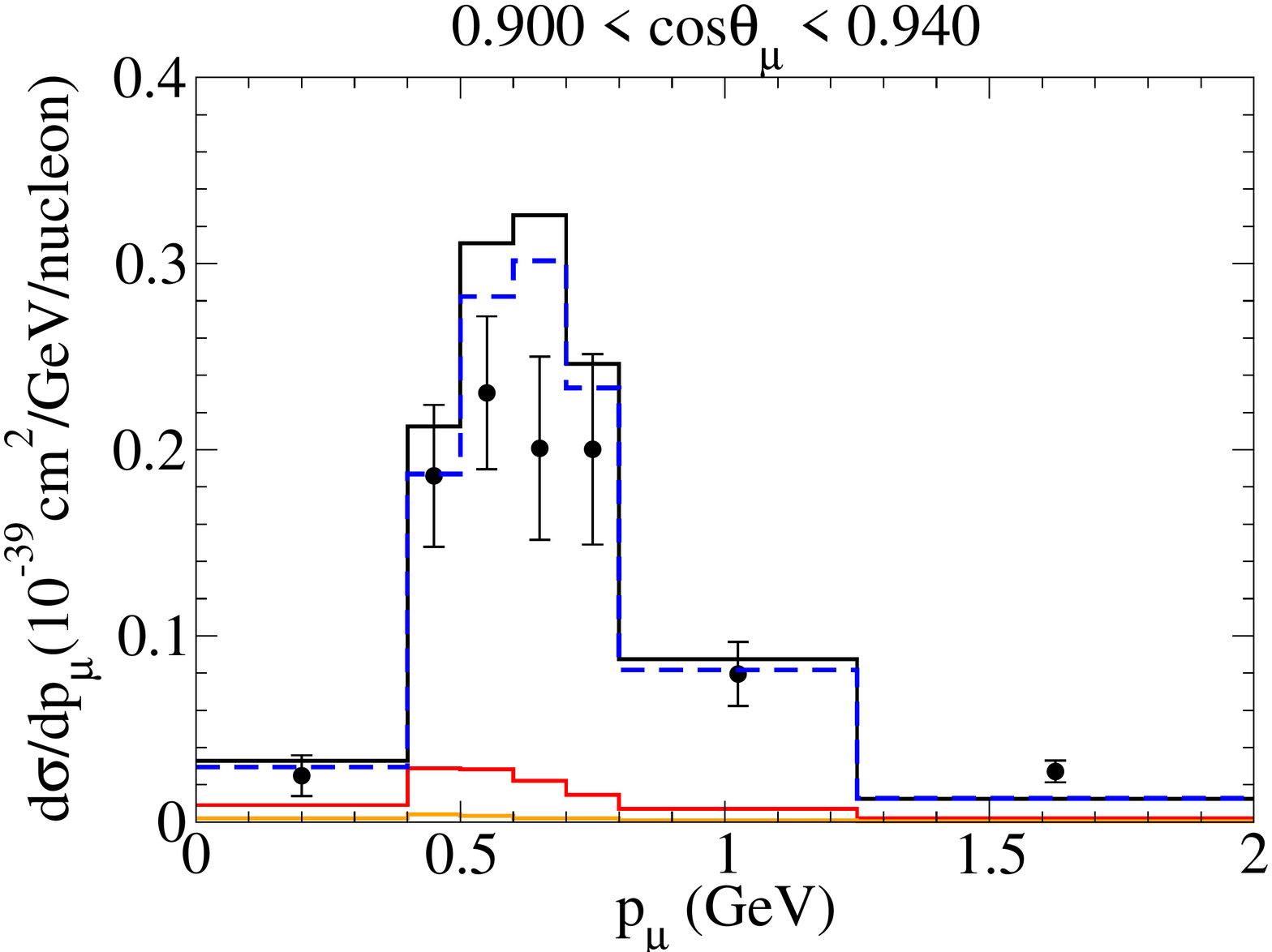}\hspace*{-0.295cm}
    	\includegraphics[width=0.34\linewidth, angle=0]{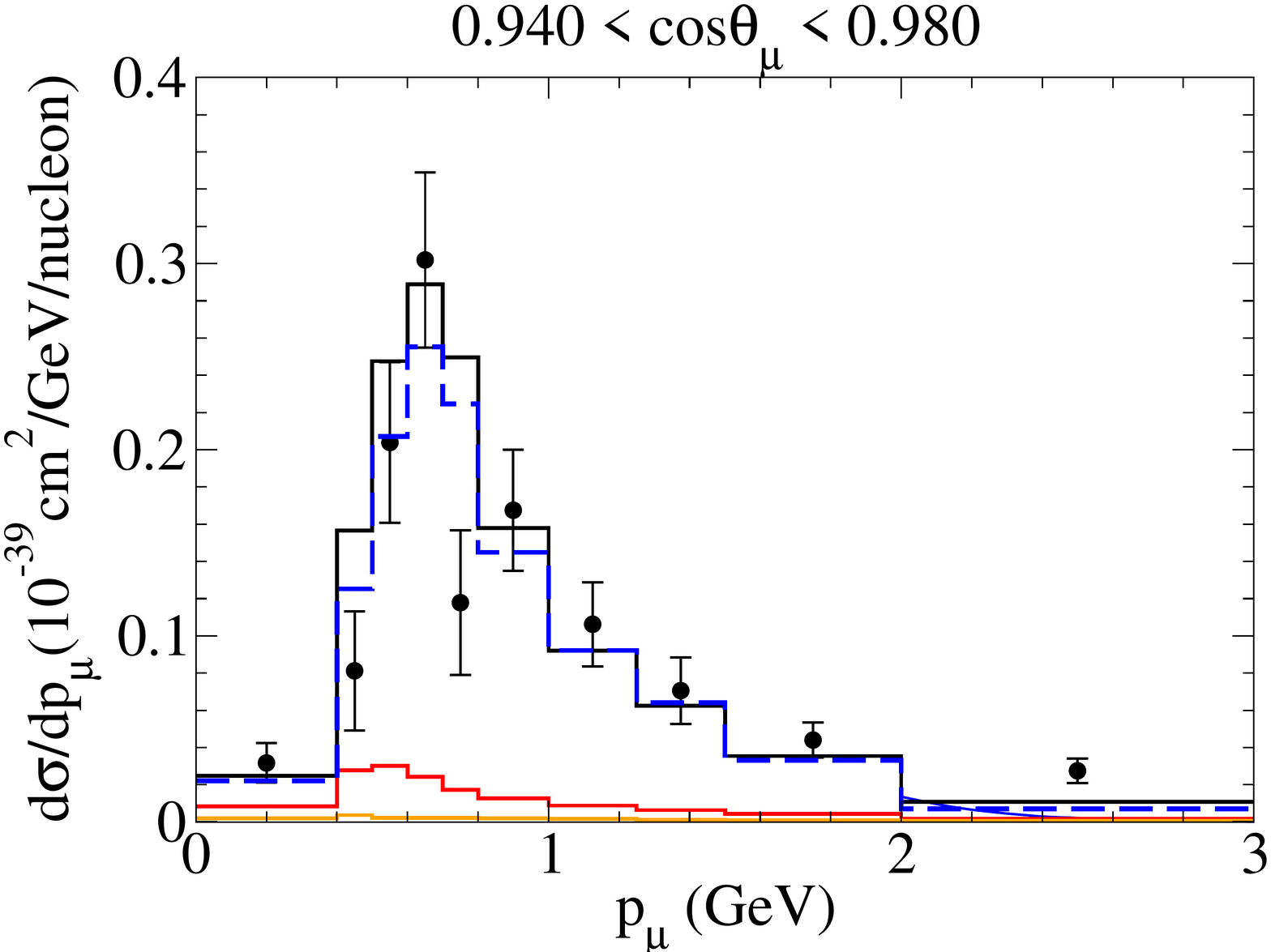}\hspace*{-0.295cm}
    	\includegraphics[width=0.34\linewidth, angle=0]{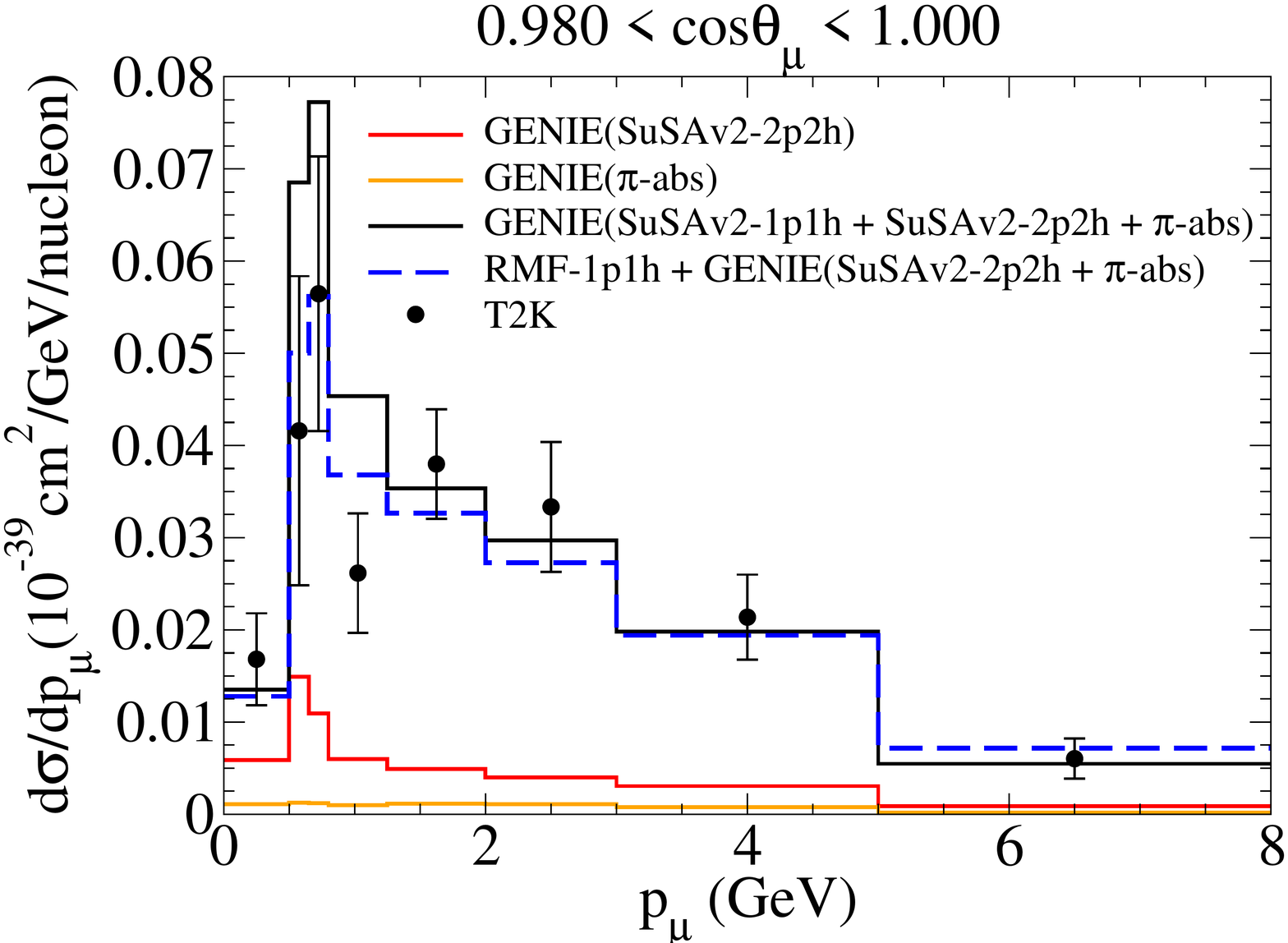}
	\end{center}
	\caption{Comparisons of data and model predictions for differential CC0$\pi$ muon-neutrino cross sections on Carbon in the T2K neutrino beam as a function of the muon kinematics when there are no protons with momenta above 500 MeV (making the model predictions `semi-semi-inclusive'). Two 1p1h predictions are shown (one from RMF, the other from SuSAv2 implemented in GENIE), in addition to the SuSAv2 2p2h and pion absorption contributions from GENIE. The (unstacked) contribution from 2p2h and pion-absorption are shown, as are the total contributions when using each of the two 1p1h models. Goodness of fit are calculated to be $\chi^2_{RMF}=171.87$ (59 bins) and $\chi^2_{SuSA}=168.92$ (60 bins), where the latter includes a single extra bin from -1.0 to -0.3 $\cos{\theta}$ (not shown). The data points are taken from~\cite{Abe:2018pwo}.}
	\label{fig:ssincT2KComp}
\end{figure*}

The T2K semi-inclusive CC0$\pi$ measurement of interactions with protons less than 500 MeV~\cite{Abe:2018pwo} provides an opportunity to compare the RMF `semi-semi-inclusive' model predictions to data, which is shown in Fig.~\ref{fig:ssincT2KComp} alongside the SuSA-GENIE predictions using the factorisation approach. In order to make this comparison the RMF predictions are added to the SuSAv2 2p2h and pion-absorption predictions from GENIE (that is to say, the SuSAv2 1p1h is replaced by an RMF prediction). In this manuscript latter always stems from GENIE's implementation of the Berger-Sehgal pion-production model~\cite{Berger:2007rq}, from which the predictions are fed through GENIE's `hN' FSI model to account for pion absorption. A comparison with the T2K measurement of proton multiplicity above 500 MeV is also shown in Tab.~\ref{tab:protonMult}.

In general, Fig.~\ref{fig:ssincT2KComp} demonstrates a fair agreement of both RMF+GENIE (SuSAv2-2p2h+$\pi$-abs) and GENIE (SuSAv2-1p1h+SuSAv2-2p2h+$\pi$-abs). It is also clear that the models predict a total dominance of 1p1h when looking at CC0$\pi$ interactions where the proton has a momentum below 500 MeV/c, thereby allowing an evaluation of the 1p1h contribution in almost isolation from 2p2h and pion absorption components. SuSAv2-GENIE's overestimation of the data at very forward angles can be ascribed to the aforementioned low energy transfer scaling violations absent in the SuSAv2-model but present in RMF, thereby explaining the better agreement achieved with the latter. Conversely, the larger results from SuSAv2 1p1h at very backward angles compared to RMF are related to the previously discussed FSI treatment alterations. Although the $\chi^2$ statistics obtained from RMF and SuSAv2 are very similar, it is clear that RMF performs better within the most forward angular bins (where additional RMF effects are most important). The fairly large $\chi^2$ for RMF likely stems from an imperfect treatment of the higher angle bins (where, as discussed, it is known that FSI effects may be too strong); a shape discrepancy in the 0.9-0.94 $\cos{\theta}$ bin and an underestimation of the data in the final muon momentum bins (which is shared by many models~\cite{Abe:2018pwo}). It can also be seen in Tab.~\ref{tab:protonMult} that, like many models, RMF and SuSAv2 in GENIE predict the inclusive cross-section well but then predict too few low momentum protons and too many at high momentum. Stronger nucleon FSI (lower nucleon transparency) may help alleviate this discrepancy.


\begin{center} 
\begin{table}[htbp!]
\begin{tabular}{ |l|c|c|c|c| } 
 \hline
  & $\sigma_{0p>500MeV}$ & $\sigma_{Np>500MeV}$  \\
 \hline
T2K Measurement &  $2.36\pm0.30$ & $1.97\pm0.25$  \\
\hline
RMF+2p2h+$\pi$abs. & 1.76 & 2.41  \\
GENIE-SuSAv2 (Full) &  1.91 & 2.49  \\
GENIE-Valencia (Full) &  1.71 & 2.34  \\
\hline
RMF-theory (1p1h-only)  &  1.50 & 1.64  \\
GENIE-SuSAv2 (1p1h-only) &  1.65 & 1.72  \\
GENIE-Valencia (1p1h-only) &  1.43 & 1.76  \\
\hline
\end{tabular}
\caption{1p1h and full CC0$\pi$ predictions of the multiplicity of protons with momentum above 500 MeV/c alongside the T2K measurement. }
 \label{tab:protonMult}
\end{table}
\end{center}

\section{Comparison of SuSAv2 implementation with semi-inclusive measurements}\label{results-sinc}

\begin{figure*}[!hp]
\begin{center}
\includegraphics[width=0.99\linewidth]{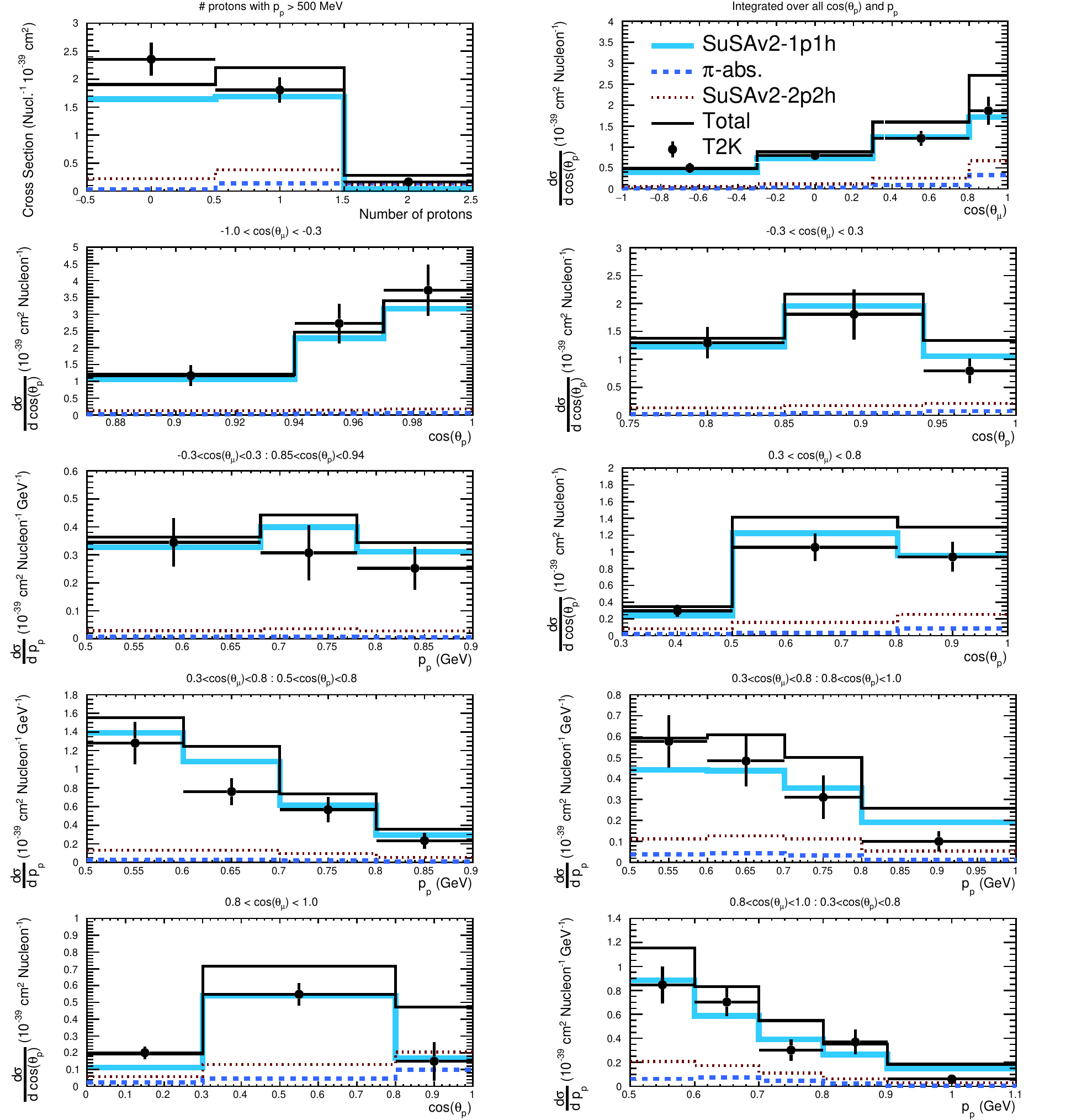}
\end{center}
\caption{The differential CC0$\pi$ muon-neutrino cross sections on Carbon in the T2K neutrino beam as a function of the muon kinematics when there is at least one proton with momentum above 500 MeV in the final state as measured by T2K and calculated using the SuSAv2 1p1h and 2p2h models implemented in GENIE alongside the pion absorption contribution. The (unstacked) contribution from each interaction mode 2p2h are each shown. The goodness of fit is calculated to be $\chi^2_{1p}$=100.72 (33 bins). The combined goodness of fit for the SuSAv2 prediction here and in Fig.~\ref{fig:ssincT2KComp} is: $\chi^2_{combined}$=490.98 (93 bins). The data points are taken from~\cite{Abe:2018pwo}.}

\label{fig:sincT2KComp}
\end{figure*}

\begin{figure*}[!hp]
\begin{center}
\includegraphics[width=0.49\linewidth]{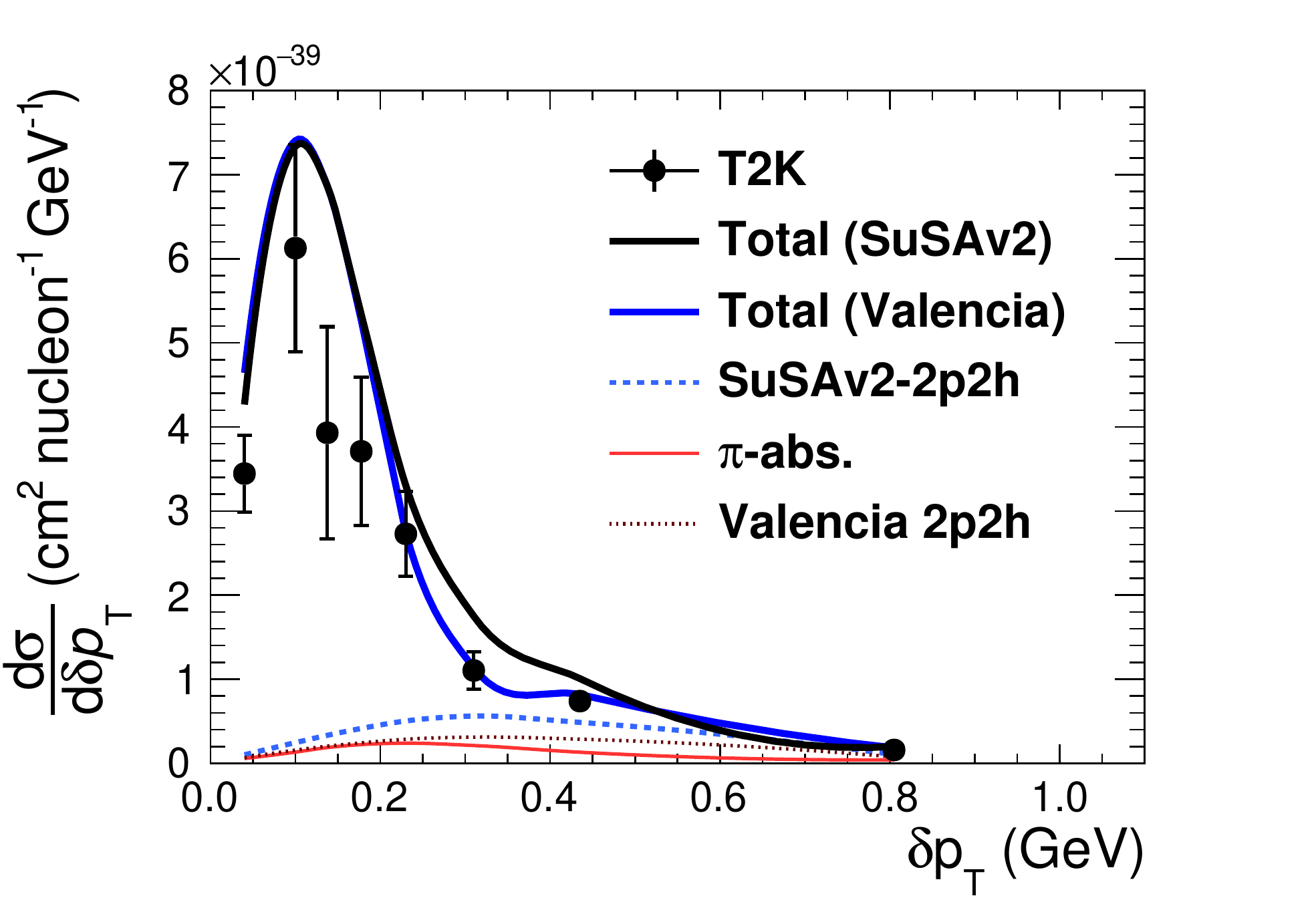}
\includegraphics[width=0.49\linewidth]{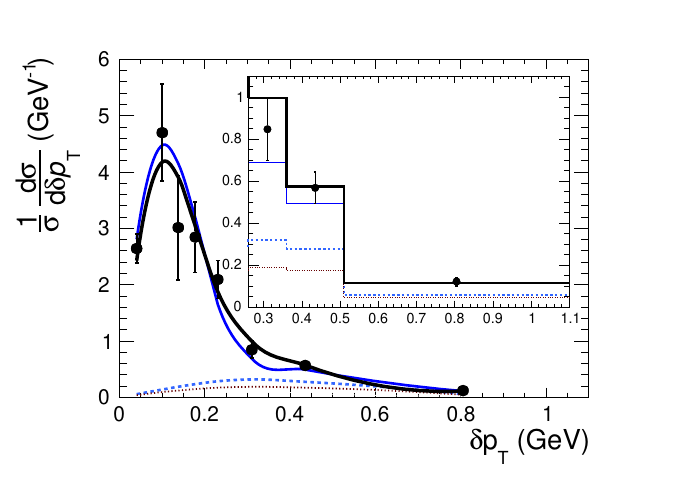}

    
\includegraphics[width=0.49\linewidth]{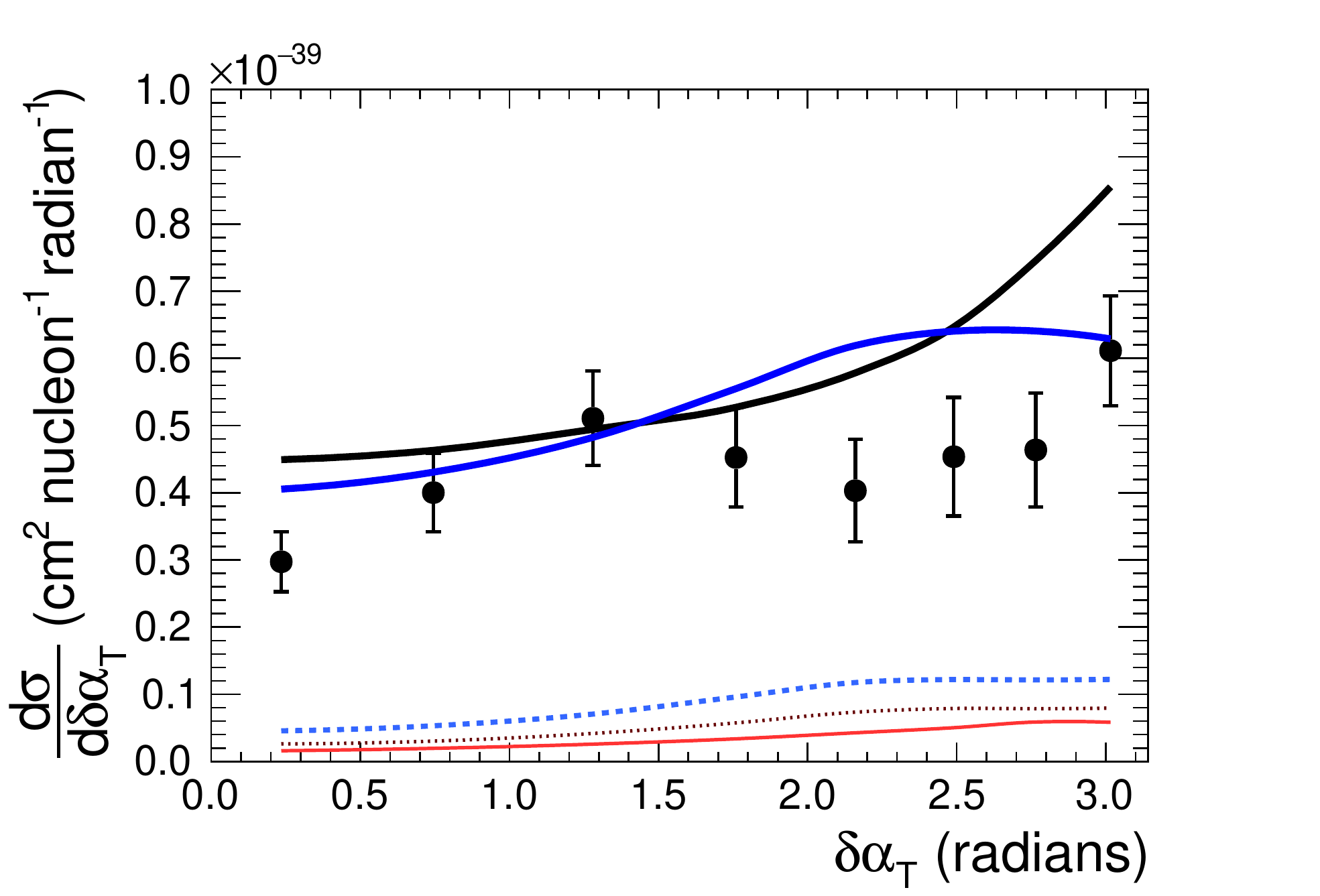}
\includegraphics[width=0.49\linewidth]{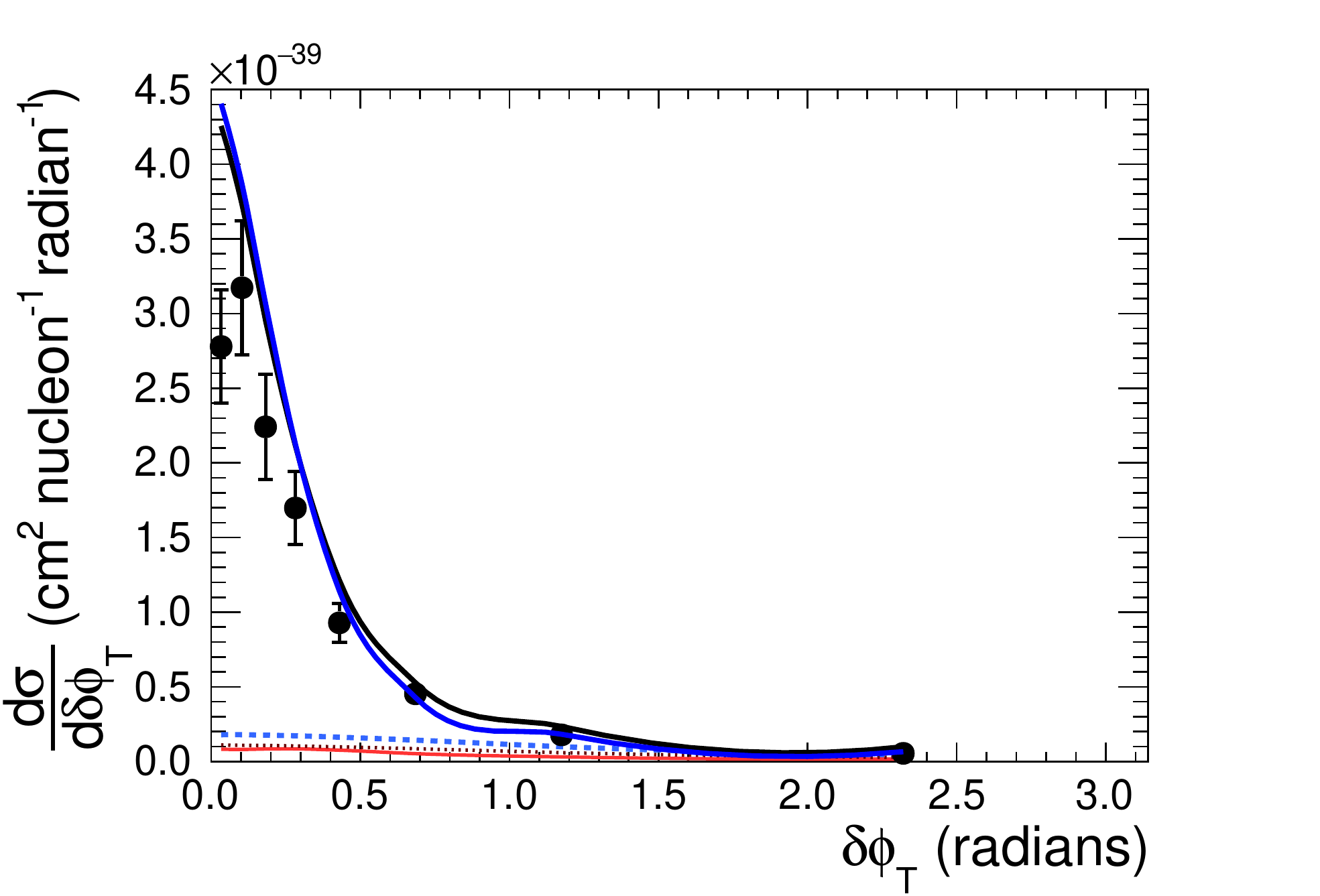}

\end{center}
\caption{The regularised T2K measurement of CC0$\pi$ muon-neutrino cross sections on Carbon at T2K kinematics as a function of the Single Transverse Variables~\cite{Lu:2015tcr} compared to predictions from the GENIE-implemented SuSAv2 and Valencia 1p1h+2p2h models, each of which is added to GENIE's pion absorption prediction. The (unstacked) contribution from 2p2h and pion-absorption are shown, as are the total contributions when using each of the two 1p1h models. For $\delta p_T$ a shape only comparison is also shown. Goodness of fit are calculated as follows. For $\delta p_T$: $\chi^{2}_{SuSA}=20.5$, $\chi^{2}_{Valencia}=27.1$. For $\delta \alpha_T$: $\chi^{2}_{SuSA}=45.3$, $\chi^{2}_{Valencia}=31.4$. For $\delta \phi_T$: $\chi^{2}_{SuSA}=40.1$, $\chi^{2}_{Valencia}=36.8$. The data points are taken from~\cite{Abe:2018pwo}.}
\label{fig:stvT2KComp}
\end{figure*}

Although the `semi-semi-inclusive' comparisons with microscopic RMF predictions provides a powerful test of both the factorisation approach and the model itself, it remains difficult to draw clear conclusions regarding the size and shape of the 2p2h contribution due to the aforementioned dominance of the 1p1h component when analysing CC0$\pi$ interactions with low momentum protons. Instead, the 2p2h contribution can be explored further using full semi-inclusive measurements which can be analysed using GENIE and the factorisation approach. To do this we compare the 2p2h SuSAv2 prediction with measurements of proton and muon kinematics in Fig.~\ref{fig:sincT2KComp}, and in Fig.~\ref{fig:stvT2KComp}, as a function of the momentum imbalances between the outgoing muon and highest momentum proton in the plane transverse to the incoming neutrino (see~\cite{Lu:2015tcr} for more details of how these imbalances are defined).

It is immediately clear that 2p2h plays a more important role when considering higher momentum protons. In general the data-simulation agreement is fair, but it is clear from Fig.~\ref{fig:sincT2KComp} that the model slightly over-predicts the number of protons above 500 GeV, particularly at more forward muon angles (suggesting the discrepancy is for more high energy neutrinos, since the interaction's energy transfer must already be enough to produce the proton) where, interestingly, the 1p1h prediction alone tends to be in good agreement with the data. This should be considered in conjunction with the slight under-prediction of the number of protons below 500 MeV, shown in detail in Fig.~\ref{fig:ssincT2KComp}. Overall this might suggest slightly too large 2p2h strength and/or too little FSI, but within the confines of the factorisation approach it is difficult to be certain. 

Similar conclusions can be drawn by analysing the the T2K measurement of transverse kinematic imbalance~\cite{Abe:2018pwo, Lu:2015tcr}, which better isolates the 2p2h contribution. This is shown best through the consideration of the the transverse momentum imbalance, $\delta p_T$, between the outgoing muon and highest momentum proton in CC0$\pi$ interactions. In $\delta p_T$ the 1p1h contribution is not expected to contribute strongly beyond the maximum initial state nucleon momentum (the Fermi surface, $\sim$230~MeV/c for Carbon) and so the high-$\delta p_T$ tail is expected to be dominated by 2p2h and pion absorption, as indeed is predicted in the top panels of Fig.~\ref{fig:stvT2KComp}. The over-prediction in the $\delta p_T$ tail in the top left panel could therefore be seen to suggest that the 2p2h may be too strong, but this cannot account for the simultaneous over prediction in the bulk. As has been discussed in detail in Ref.~\cite{Dolan:2018zye}, this overall over prediction could potentially be alleviated by stronger nucleon FSI, which may also bring the total SuSAv2 prediction into agreement in the tail. It is also possible that this could simply be a product of the approximations described in Sec.~\ref{implementation}. It is, however, interesting to note that the SuSAv2 model is able to almost perfectly describe the shape of $\delta p_T$.

By definition, the other transverse kinematic imbalance predictions share the normalisation discrepancy and it appears that they also show a general agreement in the shape. It can be seen that $\delta \phi_T$, the angle between the outgoing muon and highest momentum proton transverse momentum vector, shows a similar trend to $\delta p_T$ but of particular note is the difference in the SuSAv2 and Valencia model predictions in $\delta \alpha_T$. As first discussed in Ref.~\cite{Lu:2015tcr}, for 1p1h interactions $\delta \alpha_T$ can be interpreted as characterising the deceleration of the outgoing nucleon as it moves through the nuclear potential and undergoes FSI, where larger $\delta \alpha_T$ implies stronger deceleration. The sharp rise in the SuSAv2-1p1h prediction of $\delta \alpha_T$ in comparison to the more gradual rise from the Valencia model may therefore be interpreted as the outgoing nucleon having a more severe deceleration from re-interactions inside the nucleus and this shape seems slightly preferred by the shape of the result (the last two bins have a weak positive correlation and the rise remains present in the unregularised results~\cite{Abe:2018pwo}). However, this is despite the two predictions having the the same FSI model applied. It was confirmed that the Valencia and SuSA 1p1h models share very similar $\delta \alpha_T$  predictions if nucleon FSI is disabled in GENIE and so this implies that the energy-momentum transfer predicted by SuSAv2-1p1h model tends to eject nucleons with kinematics which have a larger probability of rapid deceleration in the FSI cascade. This shows that $\delta \alpha_T$ can be sensitive to the inclusive interaction kinematics indirectly through FSI processes.

\section{Conclusions}\label{conc}

The SuSAv2 1p1h and 2p2h models have been implemented in the GENIE event generator and shown to produce results consistent with the inclusive model predictions. Both the 1p1h and 2p2h model make substantially different predictions than those currently implemented and so provide an important complementary addition. In particular, the 2p2h prediction differs substantially in the prediction of the relative number of neutrons and protons in the final state. Critically the SuSAv2 models have been well validated on electron scattering data, making this the first complete (1p1h+2p2h) implementation in GENIE to have been so.

Whilst the implemented models give reliable inclusive predictions, the semi-inclusive predictions are based on the widely used factorisation approach. This relies on: the impulse approximation; that FSI can be modelled using a semi-classical cascade; and the assumption that the initial state seen by an interaction is independent of its kinematics, although this last assumption is partially mitigated by a kinematic dependent removal energy in the implementation. The implementation of a model based on RMF, which is also capable of `semi-semi-inclusive' predictions, has allowed us to begin to address the validity of such approximations for 1p1h interactions by comparing the predictions for a CC0$\pi$ cross section with a constraint on the outgoing proton kinematics from the bare semi-inclusive model and GENIE. Here it was shown that the factorisation approximation was unable to recover the semi-inclusive model predictions, but further investigations with a full semi-inclusive version of the RMF model~\cite{PhysRevC.101.015503,PhysRevC.100.045501} will be shortly addressed to quantify the difference.

The current semi-inclusive RMF prediction is then combined with GENIE's pion absorption and newly-implemented SuSAv2 2p2h predictions to make an estimation of the measured T2K CC0$\pi$ cross section with a constraint on the outgoing proton kinematics which is free from factorisation approximations in the 1p1h. The agreement with the data is good, outside of the high-kinematical regions where RMF has known deficiencies, particularly compared to when using the factorisation approach. This demonstrates the importance of improving the treatment of hadronic kinematics in neutrino interaction event generators. It is also worth mentioning that the drawbacks of the RMF model at high kinematics will be solved in the future with an improved approach~\cite{PhysRevC.101.015503,PhysRevC.100.045501} that introduces an energy-dependent potential to keep the RMF strength and proper orthogonalization for slow nucleons while softening the potentials for increasing nucleon momenta.

Finally we compare the new model implementation, alongside existing GENIE models, to semi-inclusive T2K measurements sensitive to nuclear effects, including the measurement of transverse kinematic imbalance. Here we find generally fair agreement with the shape of the data but a notable normalisation difference.

The inability of event generators to reliably predict outgoing hadron kinematics represents a potentially serious issue for reaching the few-\% level understanding of neutrino nucleus interactions that will be required for the next generation of long-baseline oscillation experiments. An improved treatment will require increased availability of microscopic semi-inclusive neutrino interaction predictions and their implementation into event generators. 

\begin{acknowledgments}
We thank the GENIE collaboration for their helpful input regarding the model implementation. In particular we thank S. Gardiner for facilitating the simple update of our implementation into GENIEv3. We also acknowledge J. E. Amaro and I. Ruiz-Simo from the University of Granada for providing us detailed information and results about the 2p2h hadron tensors employed in this implementation. We also are grateful to many members of the T2K collaboration for interesting discussions pertinent to this work. This work was partially supported by the Spanish Ministerio de Economia y Competitividad and ERDF (European Regional Development Fund) under contracts FIS2017-88410-P, and by the Junta de Andalucia (grant No. FQM160). GDM acknowledges support from a Junta de Andalucia fellowship (FQM7632, Proyectos de Excelencia 2011), a P2IO-CNRS grant, and by the European Union's Horizon 2020 research and innovation programme under the Marie Sklodowska-Curie grant agreement No. 839481. We acknowledge the support of CEA, CNRS/IN2P3 and P2IO, France; and the MSCA-RISE project JENNIFER, funded by EU grant n.644294, for supporting the EU-Japan researchers mobility. 
\end{acknowledgments}

\appendix

\section{Comparison to T2K CC0$\pi$ inclusive analysis and implementation validations}\label{appendix-inc}

\begin{figure*}
	\begin{center}
		\includegraphics[width=0.36\linewidth, angle=0]{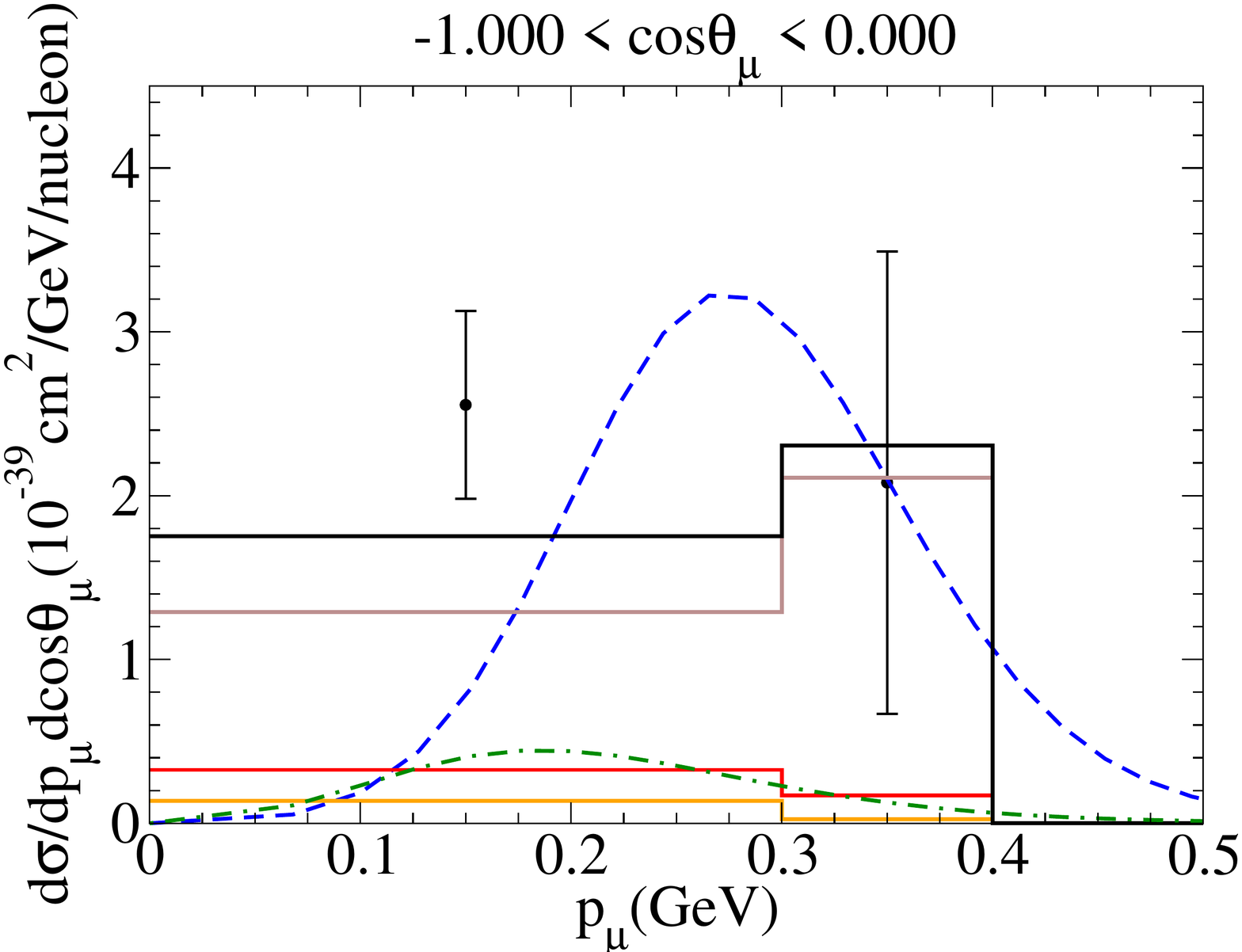}\hspace{-12mm}
		\includegraphics[width=0.36\linewidth, angle=0]{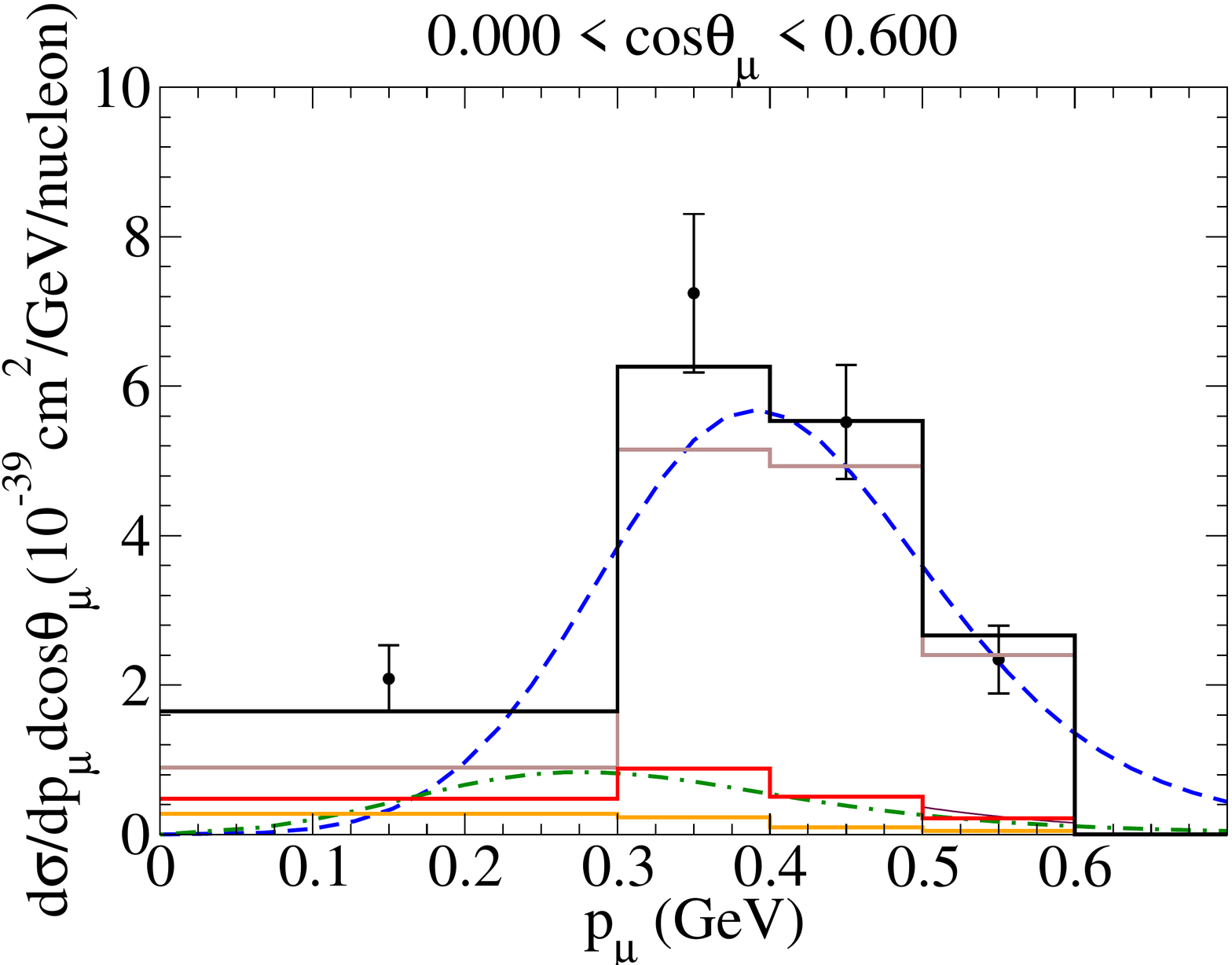}\hspace{-12mm}	\includegraphics[width=0.36\linewidth, angle=0]{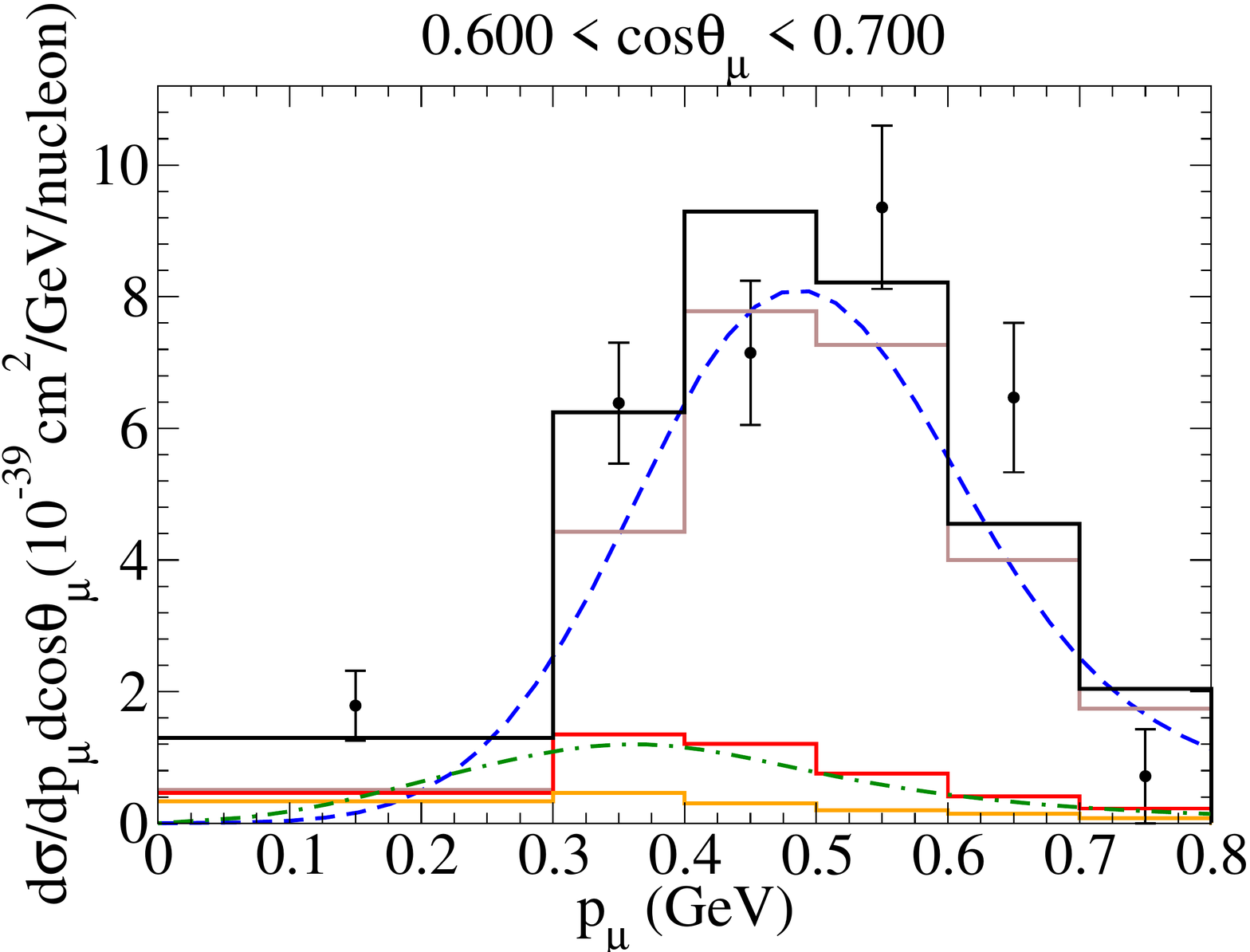}
		\includegraphics[width=0.36\linewidth, angle=0]{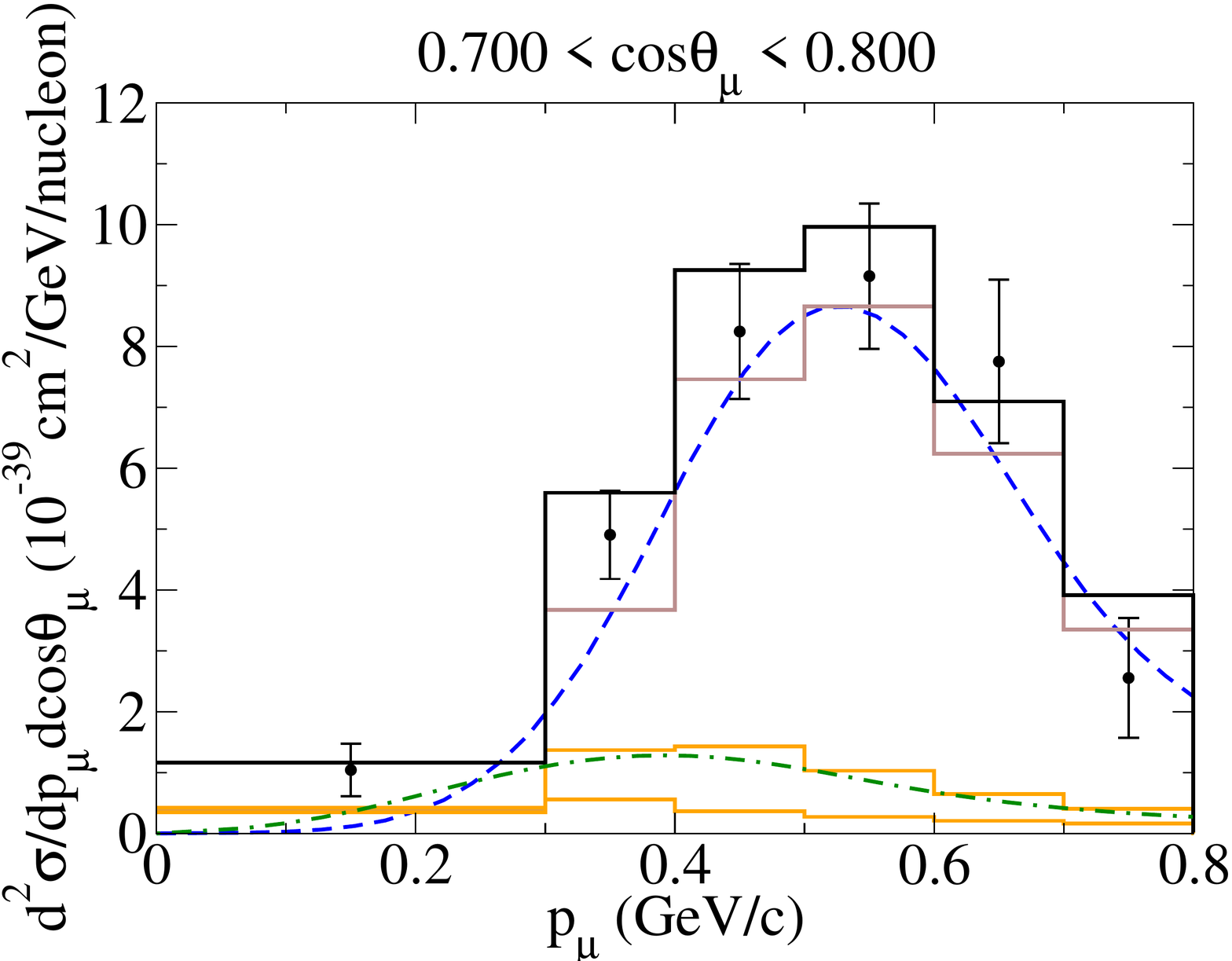}\hspace{-12mm}
		\includegraphics[width=0.36\linewidth, angle=0]{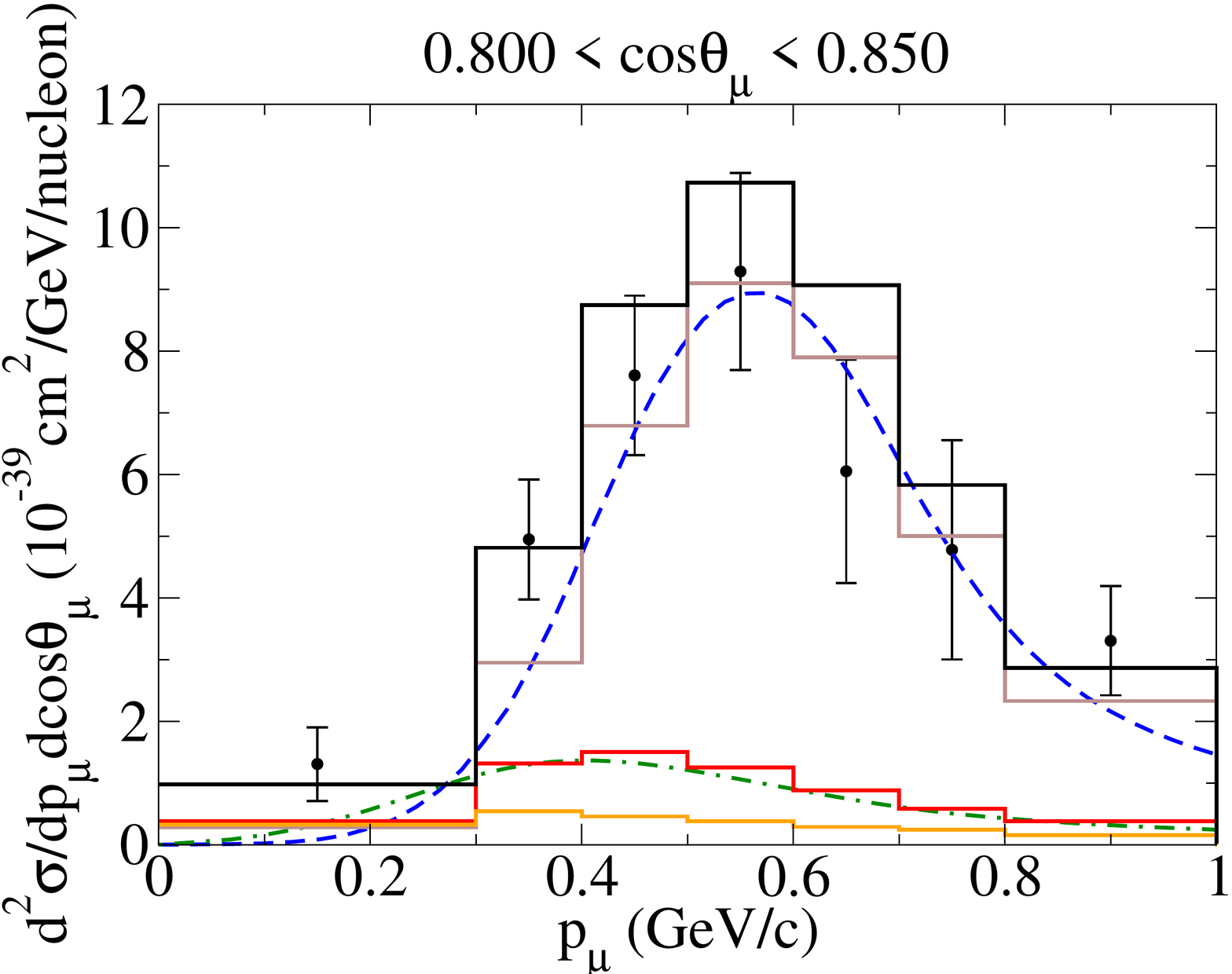}\hspace{-12mm}		\includegraphics[width=0.36\linewidth, angle=0]{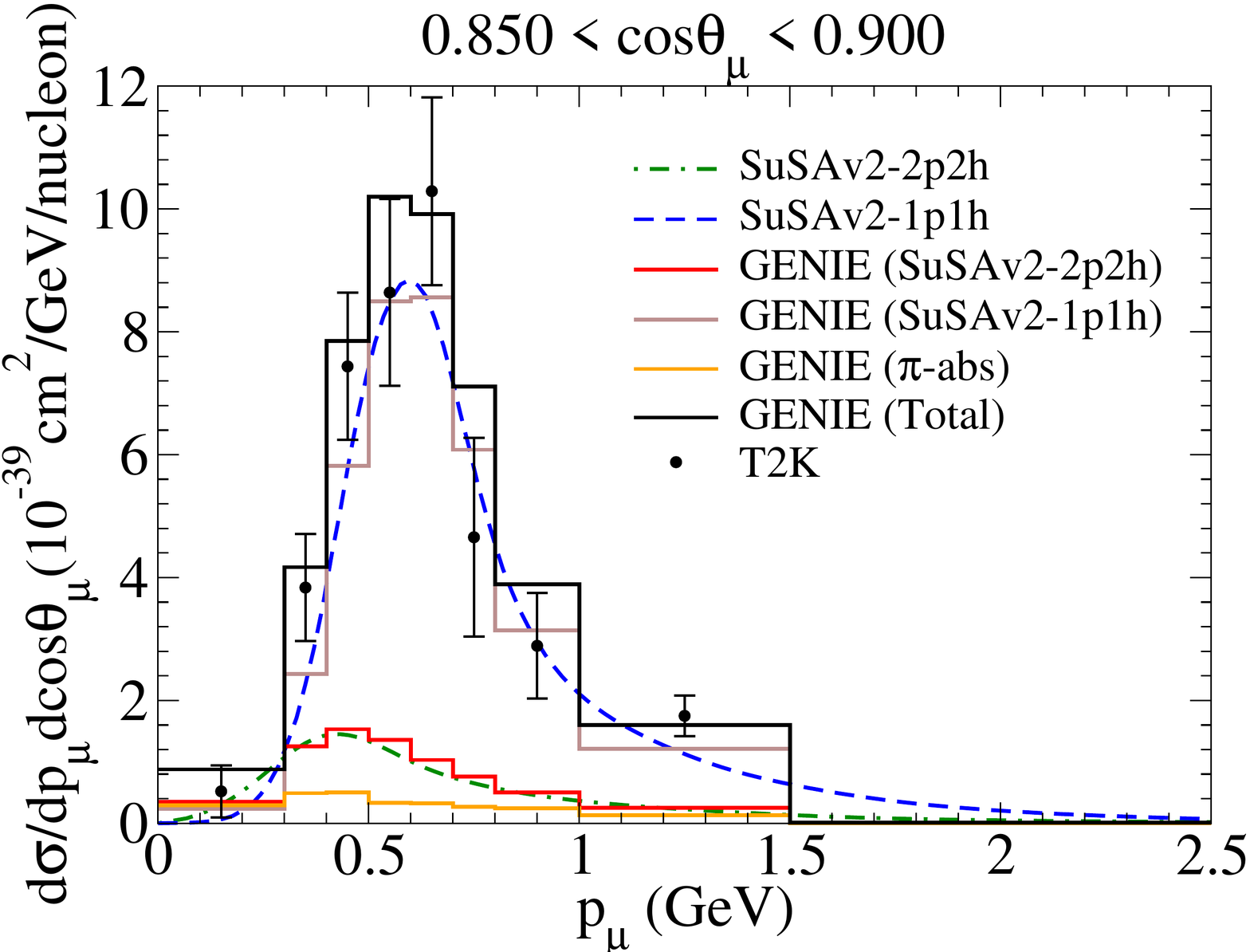}
		\includegraphics[width=0.36\linewidth, angle=0]{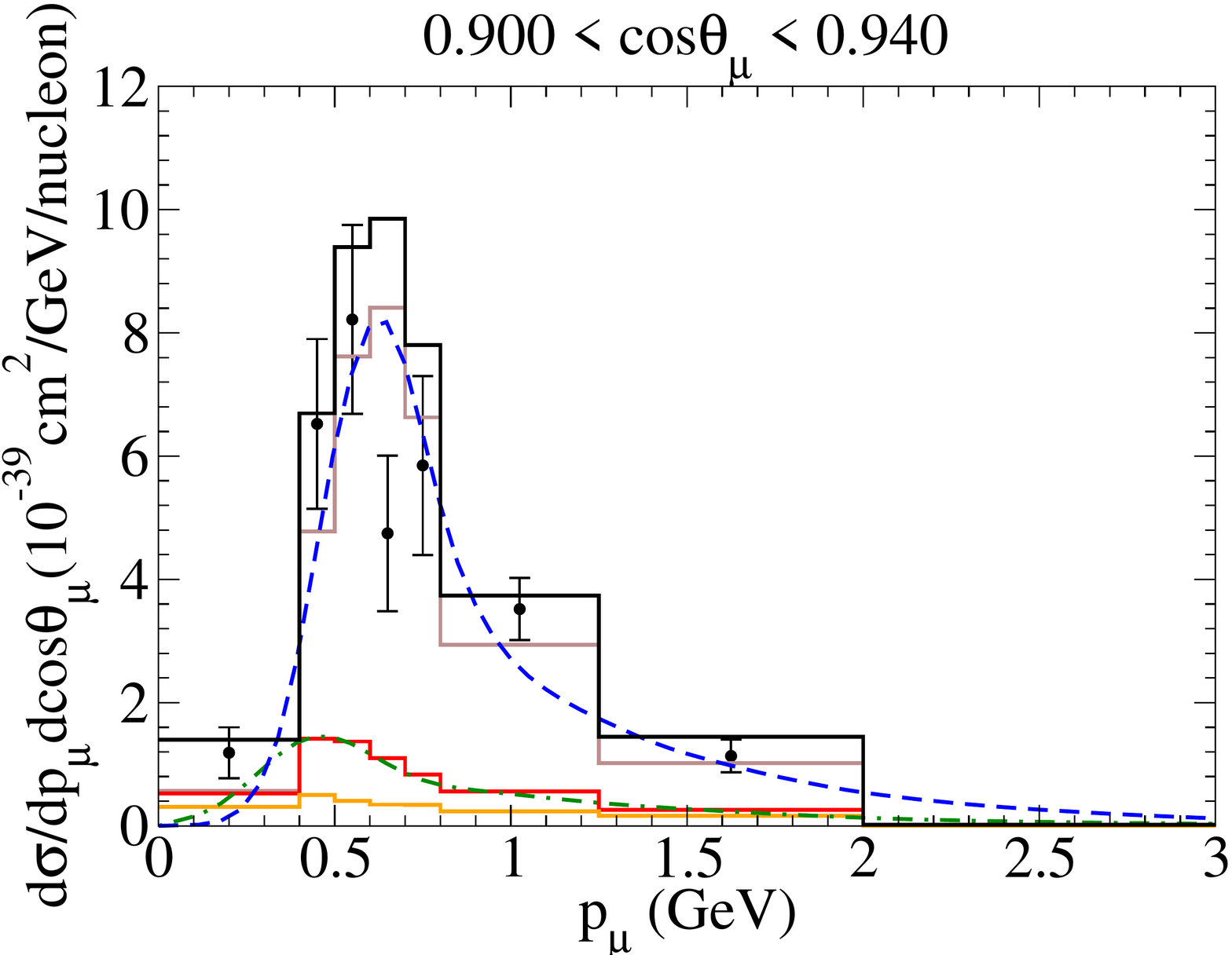}\hspace{-12mm}		\includegraphics[width=0.36\linewidth, angle=0]{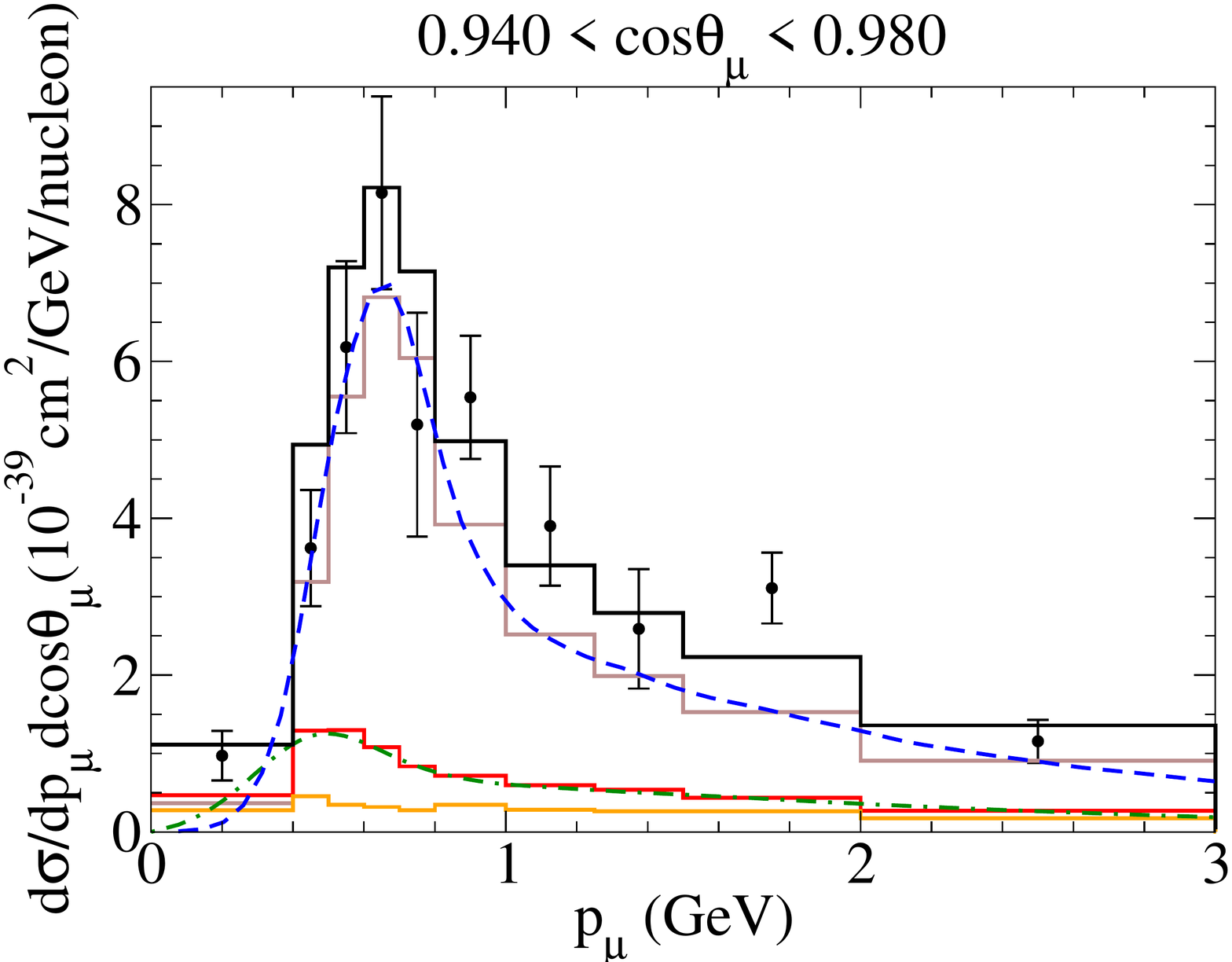}\hspace{-12mm}
		\includegraphics[width=0.36\linewidth, angle=0]{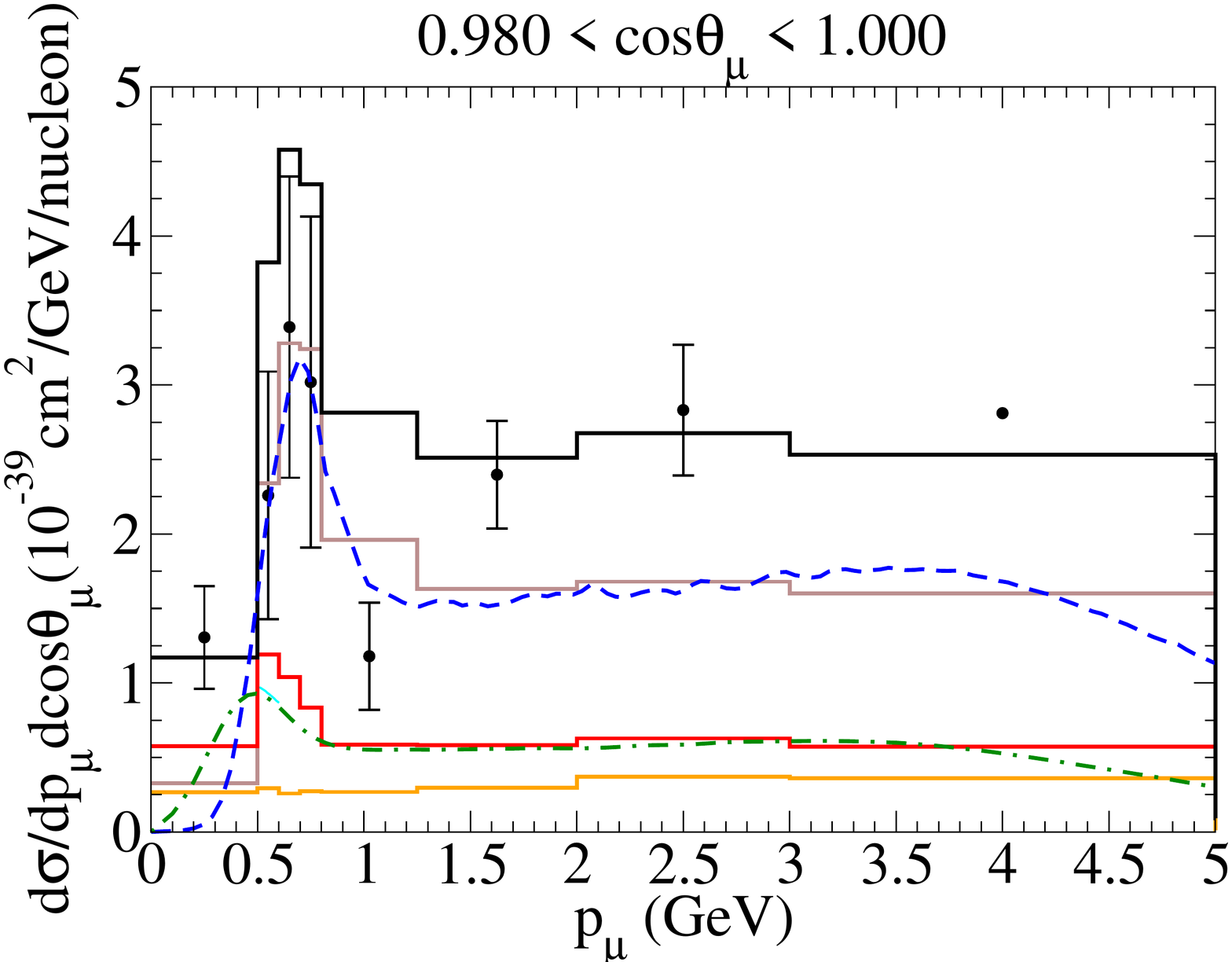}
	\end{center}
	\caption{Comparison of the T2K CC0$\pi$ measurement of the muon-neutrino cross section on Carbon with the SuSAv2 model (1p1h+2p2h) and a pion-absorption contribution as implemented in GENIE. The (unstacked) contribution from each interaction mode is shown separately, as well the total prediction. Comparison between 1p1h and 2p2h GENIE implementation (histograms) and the microscopic calculations (smooth curves) are also shown for model implementation validation. The goodness of fit is: $\chi^2=255.8$ (67 bins). The data points are taken from~\cite{Abe:2016tmq}.}
	\label{fig:incT2KSusa}
\end{figure*}

Fig.~\ref{fig:incT2KSusa} shows a comparison of the SuSAv2 1p1h and 2p2h calculation (in GENIE and directly from the model) on top of the GENIE absorption prediction to T2K CC0$\pi$ inclusive results~\cite{Abe:2016tmq} (i.e. there is no restriction on the outgoing protons), which are in good agreement with the data. As has been shown in Fig.~\ref{fig:ssIncToIncComp}, the slight discrepancies in the very forward going bins at intermediate momenta can be improved by using the full RMF. It can can also be seen that a contribution beyond the 1p1h seems essential at higher momentum and forward angles and that the SuSAv2 2p2h prediction appears to have the required strength. However, as discussed in Sec.~\ref{results-ssinc}, it is clear that it is difficult to draw more detailed conclusions regarding the 2p2h contribution as the 1p1h remains dominant. 

\begin{figure*}[!hp]
\begin{center}
\includegraphics[scale=0.425, angle=0]{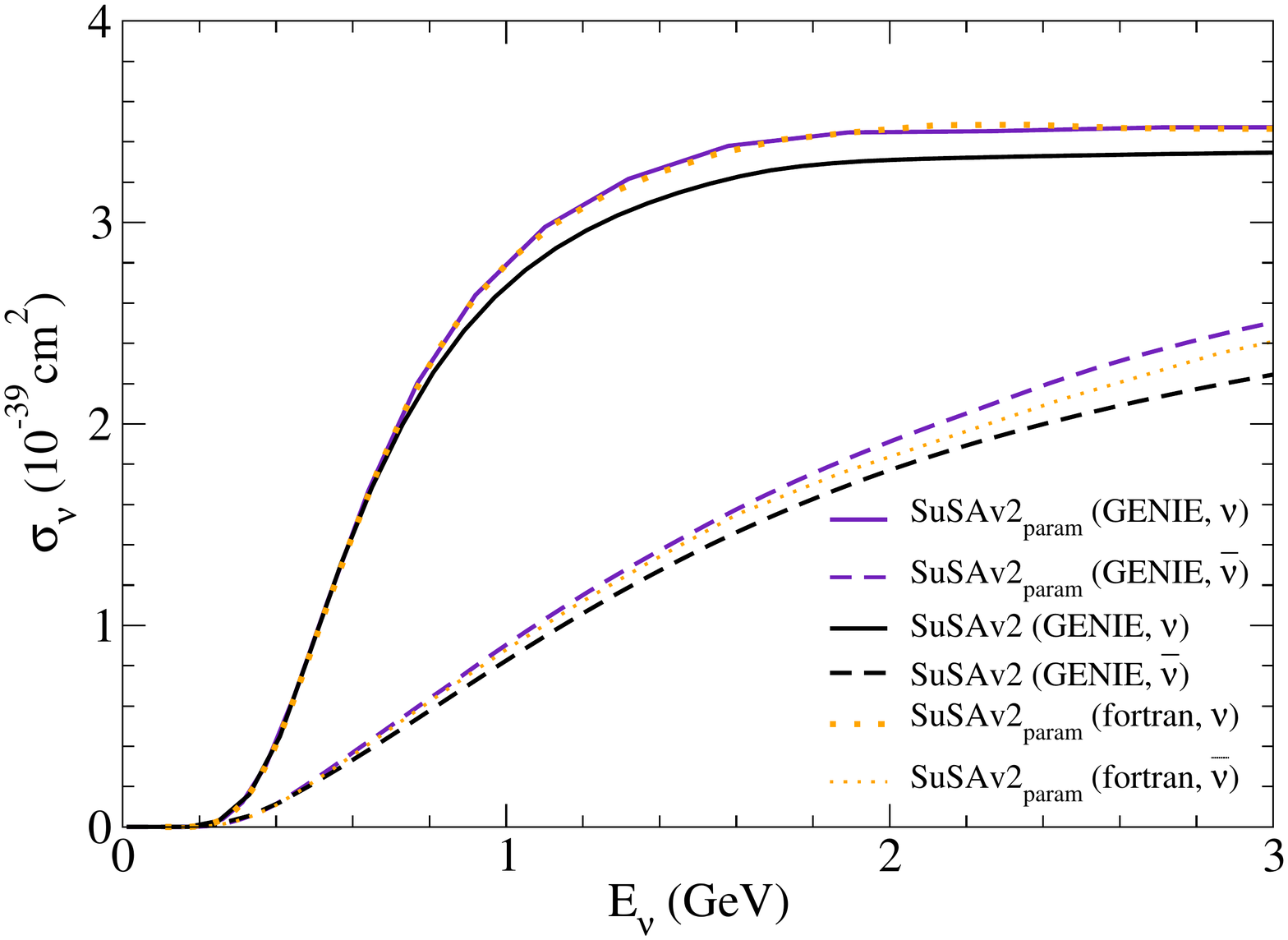}
\end{center}
\caption{Total 2p2h cross section for muon-neutrino and anti-neutrino interactions on Carbon for different implementations of SuSAv2. The $param$ label specifies that the calculation is made using the SuSAv2 parameterisation of the microscopic model (either in GENIE or using the original fortran code).}
\label{fig:2p2hEnu_appendixvalidation}
\end{figure*}

Importantly is can also be seen that there is in general very good agreement between the full SuSAv2 1p1h and 2p2h calculations and their implementations in GENIE. The remaining differences in the 1p1h channel stem from interpolation and integration method differences. Whilst these also affect the 2p2h case, the largest difference here stems from the SuSA group's use of a parametrisation of the microscopic model in order to speed up calculations, which is not necessary in the GENIE implementation. To validate that this is the primary source of the small differences observed, the SuSA 2p2h parametrisation was used to build a hadron tensor which was then implemented into GENIE. Fig~\ref{fig:2p2hEnu_appendixvalidation} shows the total cross-section predictions from the SuSA 2p2h model alongside the implementation in GENIE using the hadron tensor taken directly from the microscopic model or taken from the parametrisation. From this it can clearly be seen that GENIE is able to match the SuSA 2p2h when using the hadron tensor taken from their parametrisation and that small differences exist when using the hadron tensor from the full microscopic model. Some differences remain due to the aforementioned integration and interpolation methods (particularly for anti-neutrinos), but these are fairly small.

\section{Further comparisons to the GENIE-Valencia model predictions}\label{appendix-nieves}

\begin{figure*}[!hp]
\begin{center}
\includegraphics[width=0.32\linewidth]{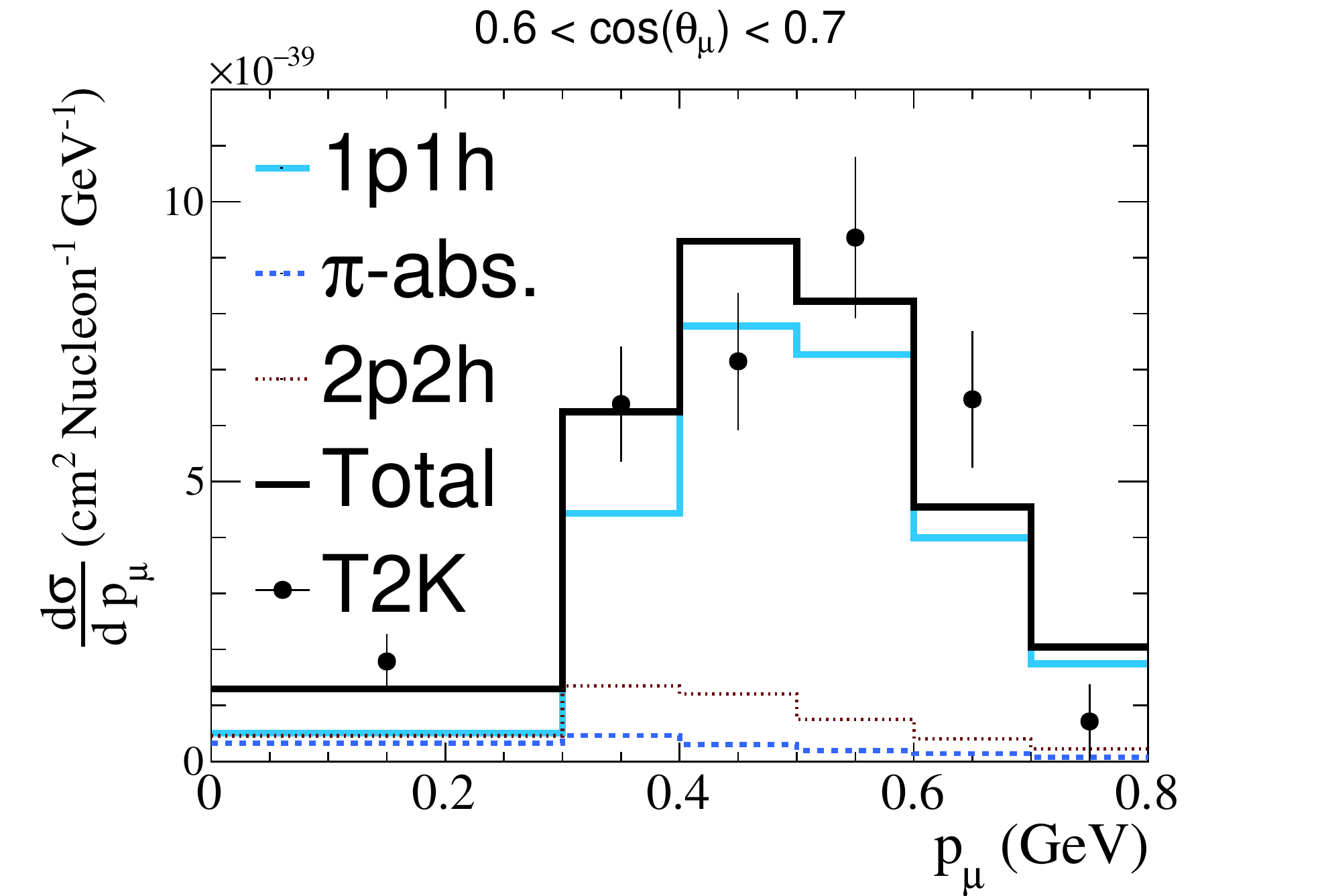}
\includegraphics[width=0.32\linewidth]{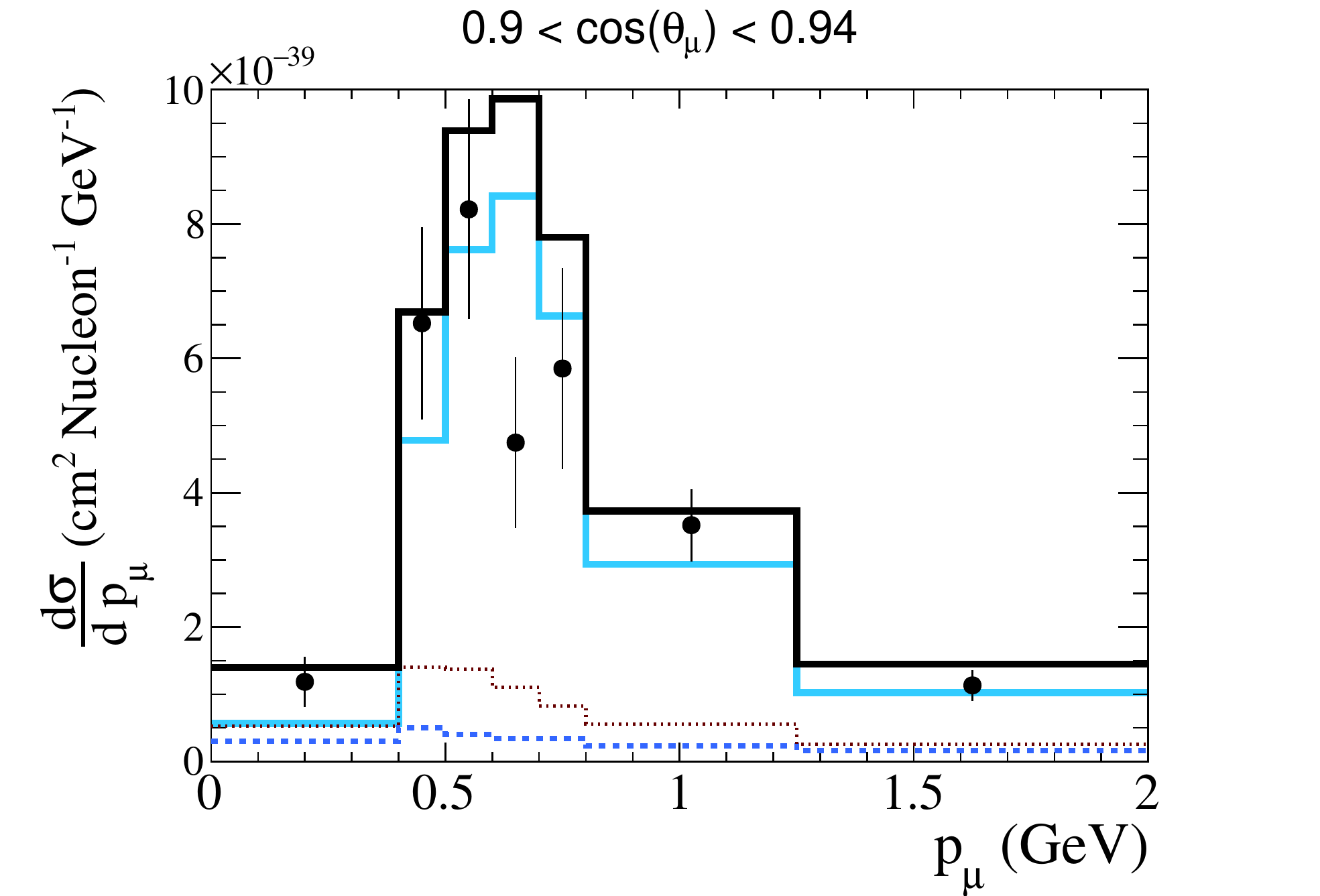}
\includegraphics[width=0.32\linewidth]{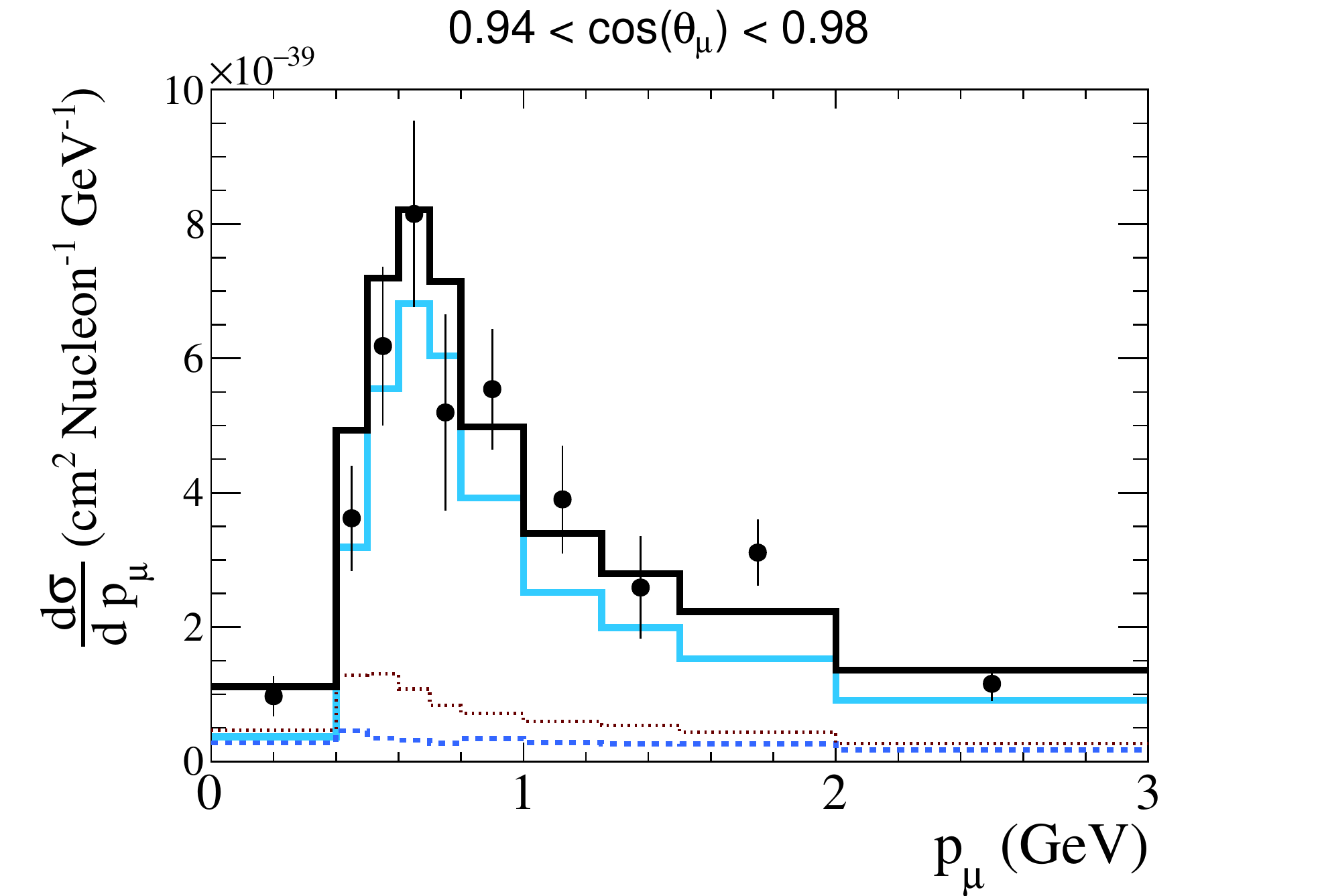}

\includegraphics[width=0.32\linewidth]{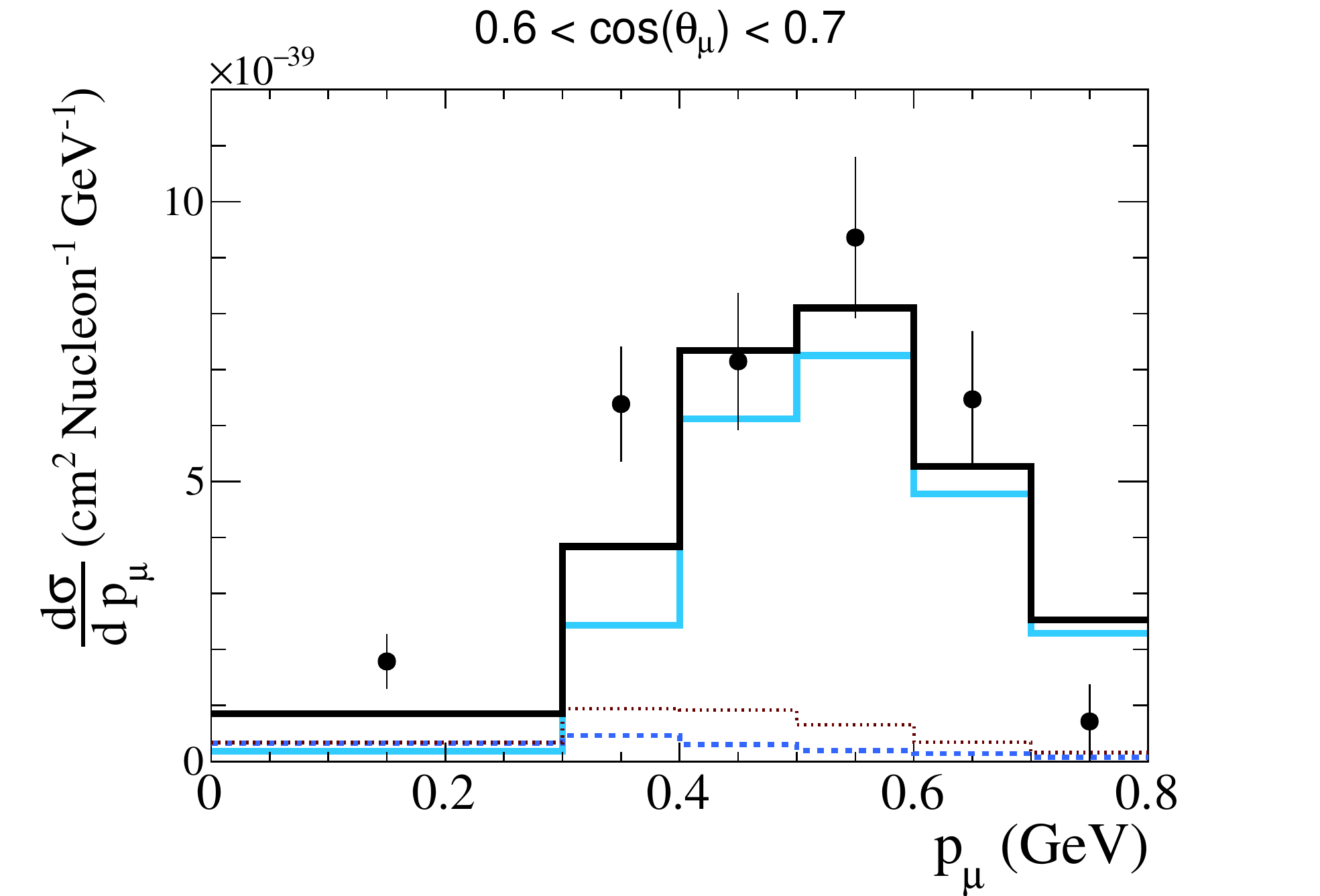}
\includegraphics[width=0.32\linewidth]{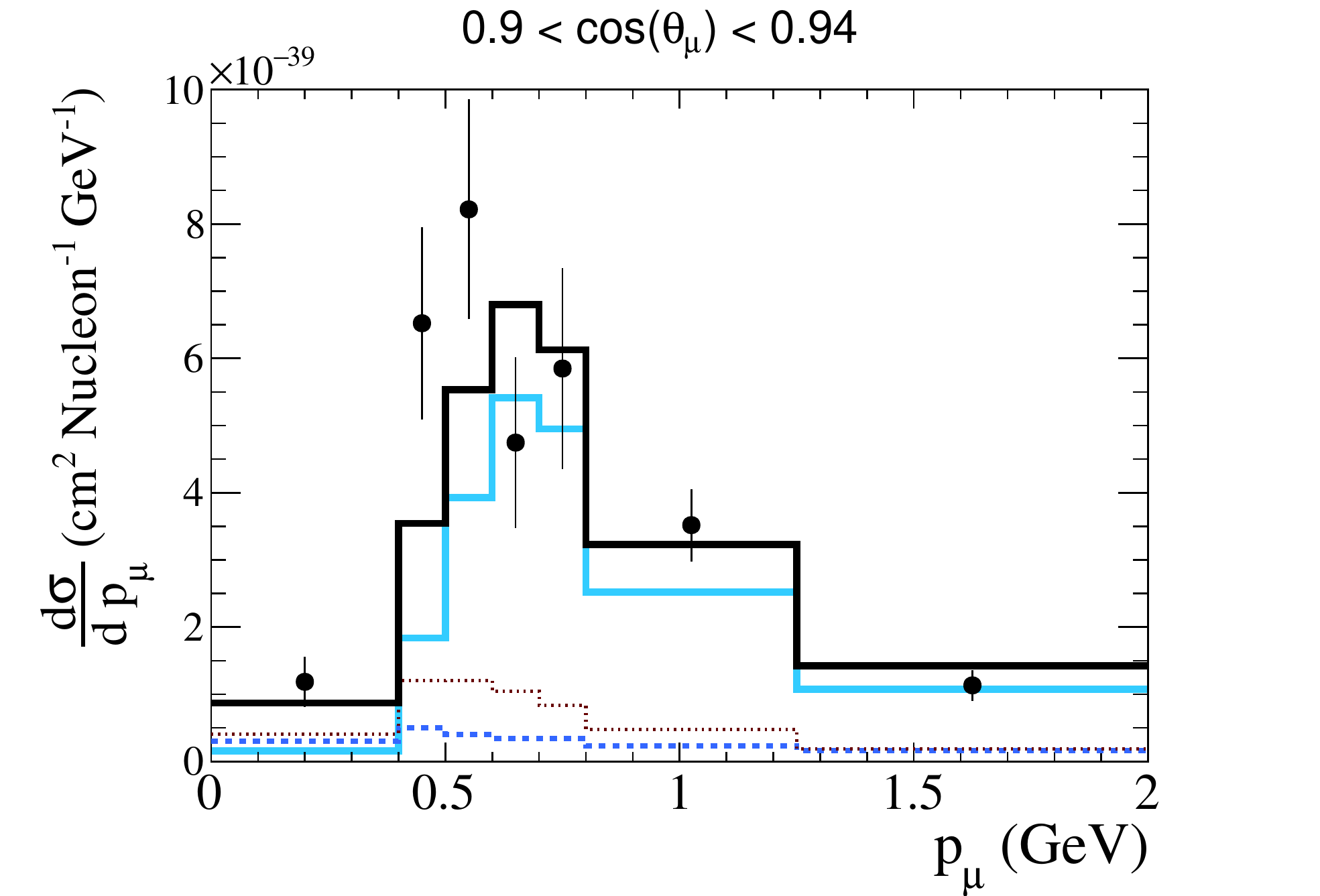}
\includegraphics[width=0.32\linewidth]{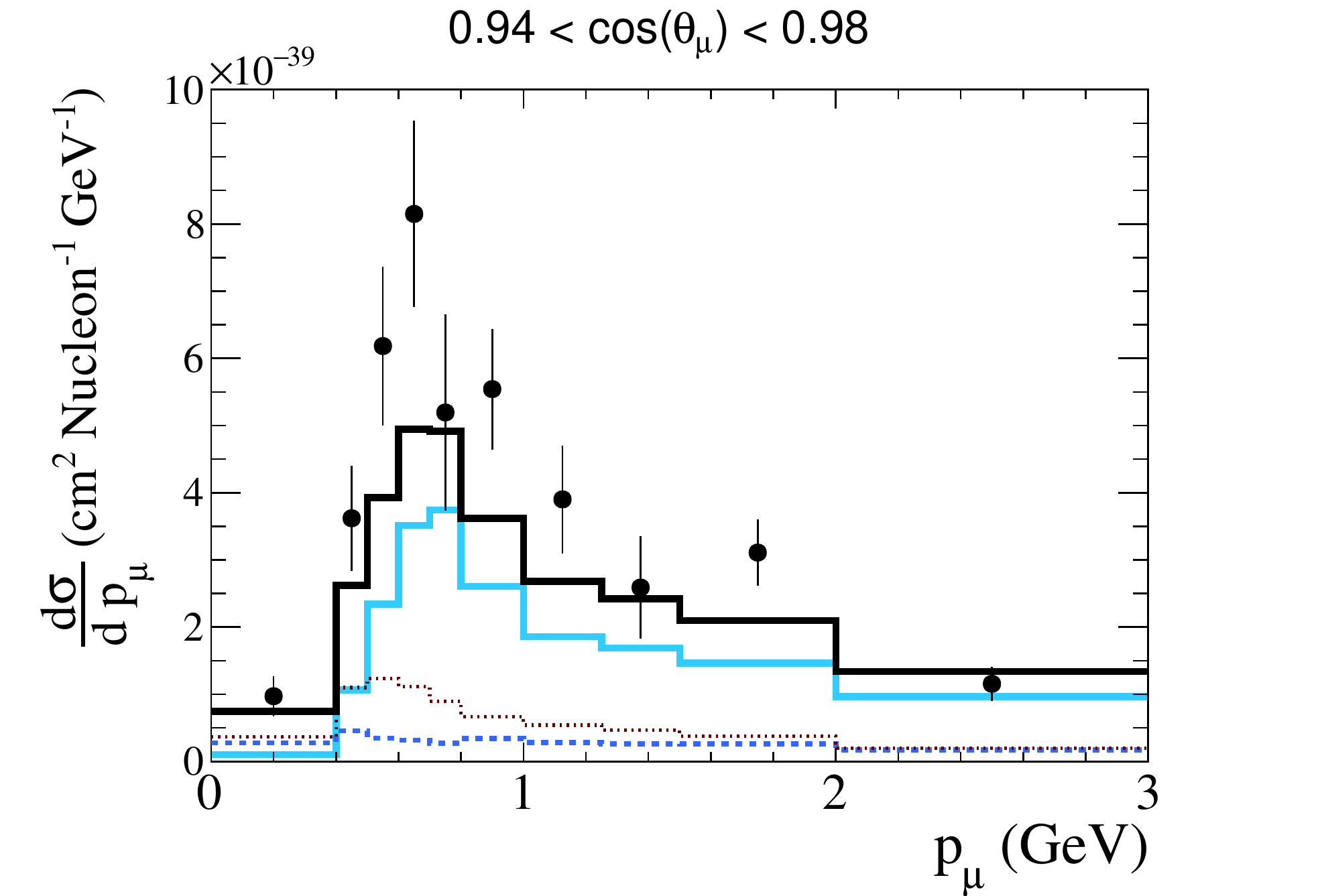}
\end{center}
\caption{Comparison of the T2K CC0$\pi$ measurement of muon-neutrino interactions on Carbon with the SuSAv2 and Valencia models (1p1h+2p2h) each with an additional pion-absorption contribution as implemented in GENIE.  The (unstacked) contribution from each interaction mode is shown separately, as well the total prediction. The top plots are the SuSAv2 predictions whilst the Valencia ones are below. The data points are taken from~\cite{Abe:2016tmq}.}
\label{fig:incT2KModelComp}

\end{figure*}

\begin{figure*}[!hp]
\begin{center}
\includegraphics[width=0.32\linewidth]{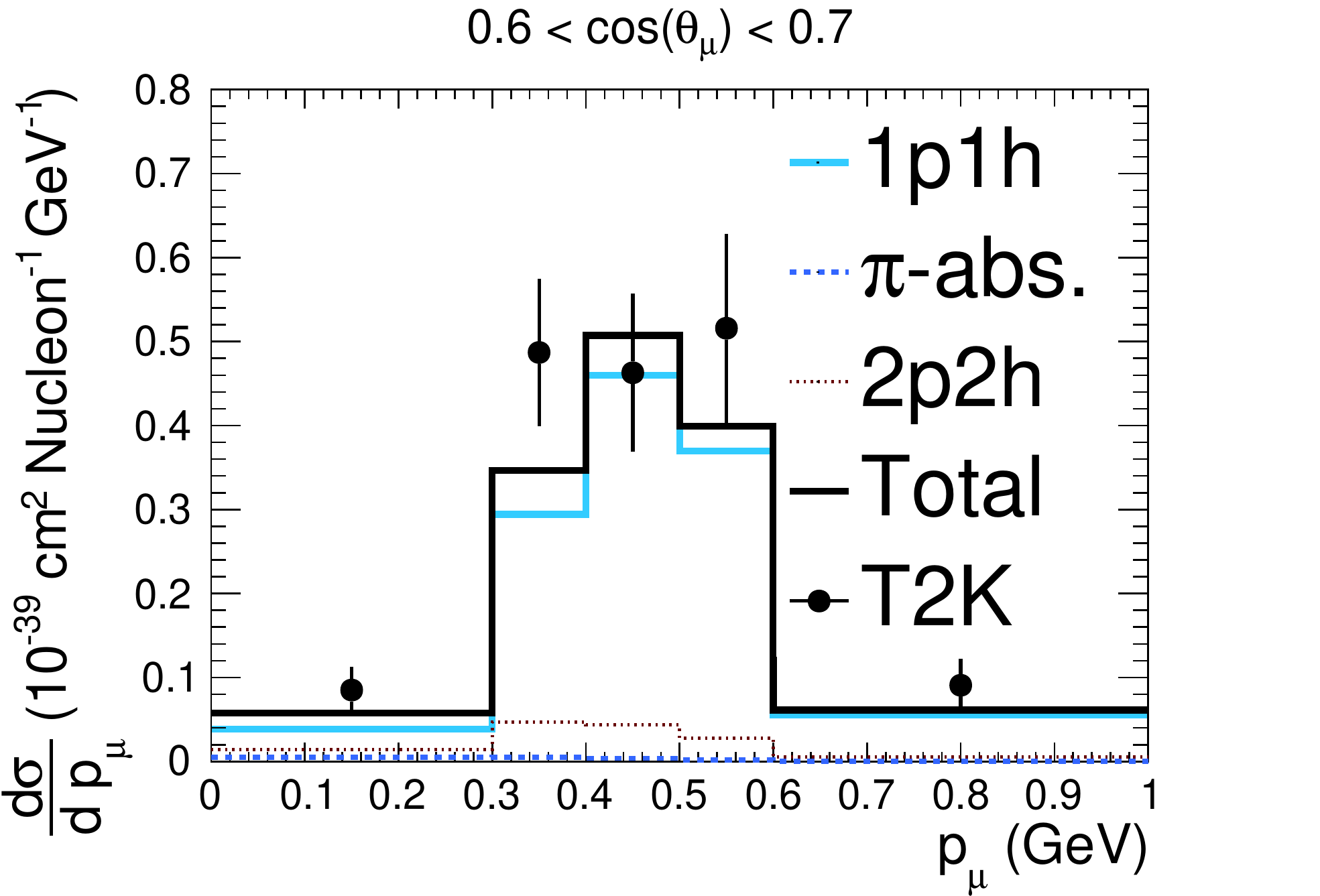}
\includegraphics[width=0.32\linewidth]{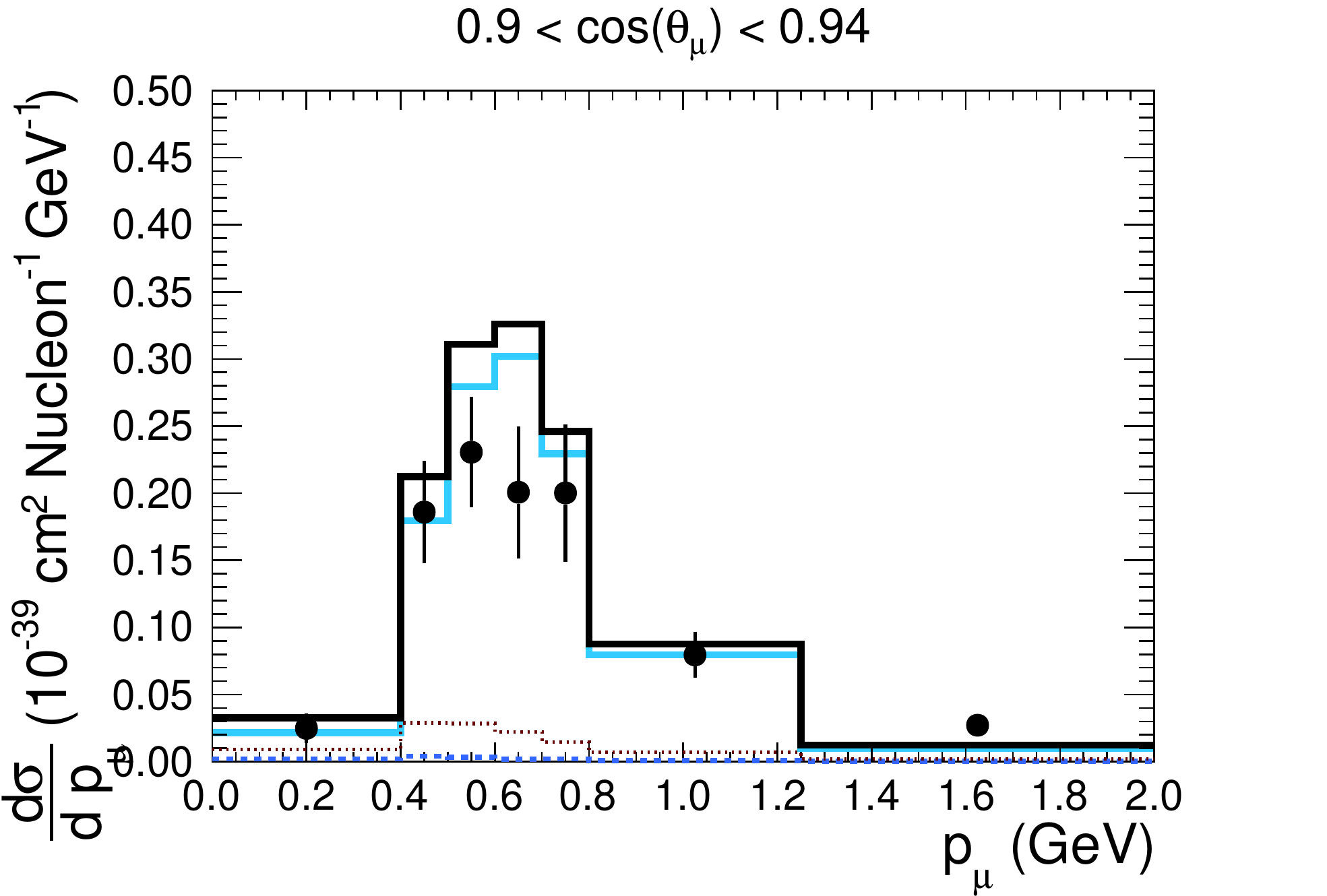}
\includegraphics[width=0.32\linewidth]{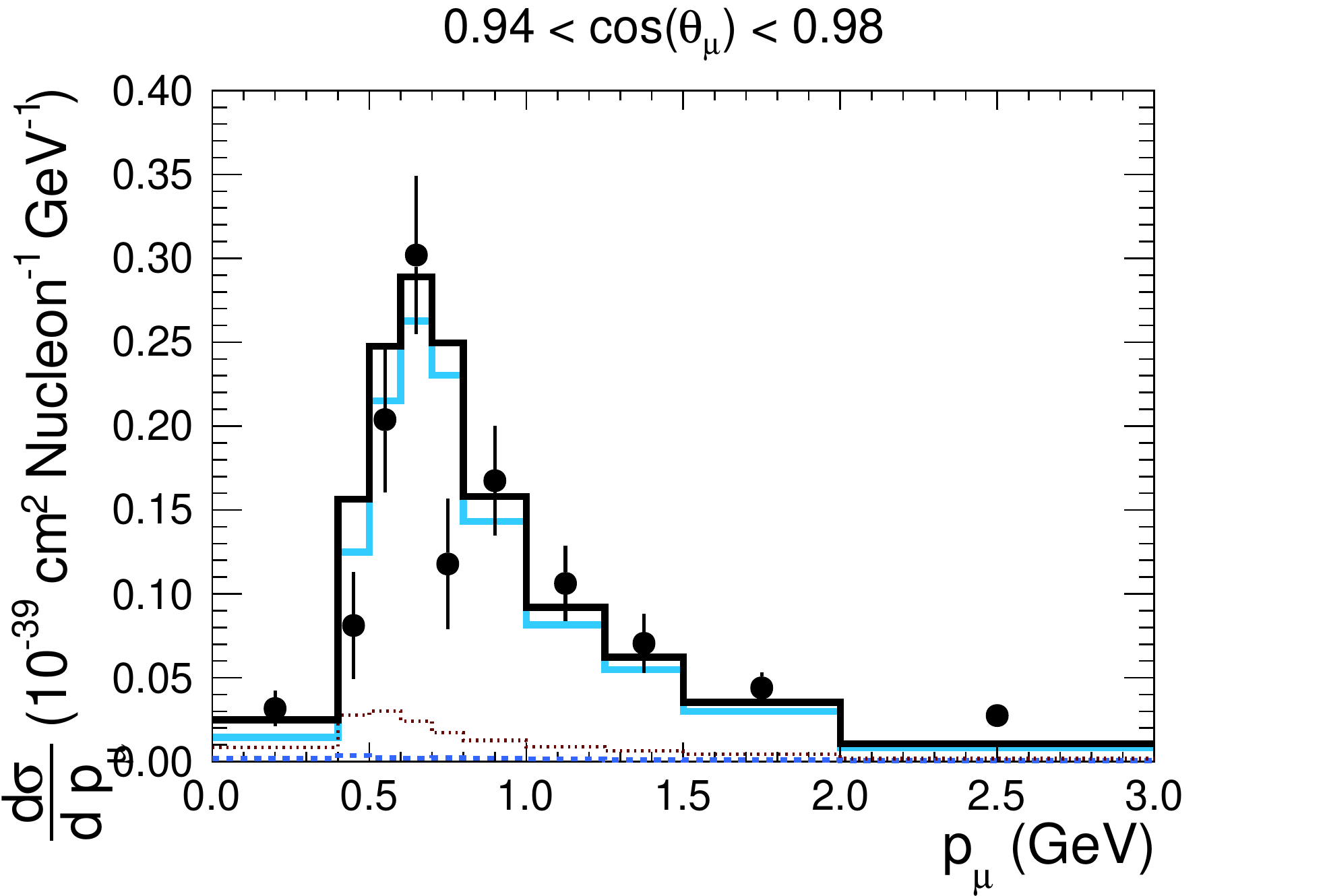}

\includegraphics[width=0.32\linewidth]{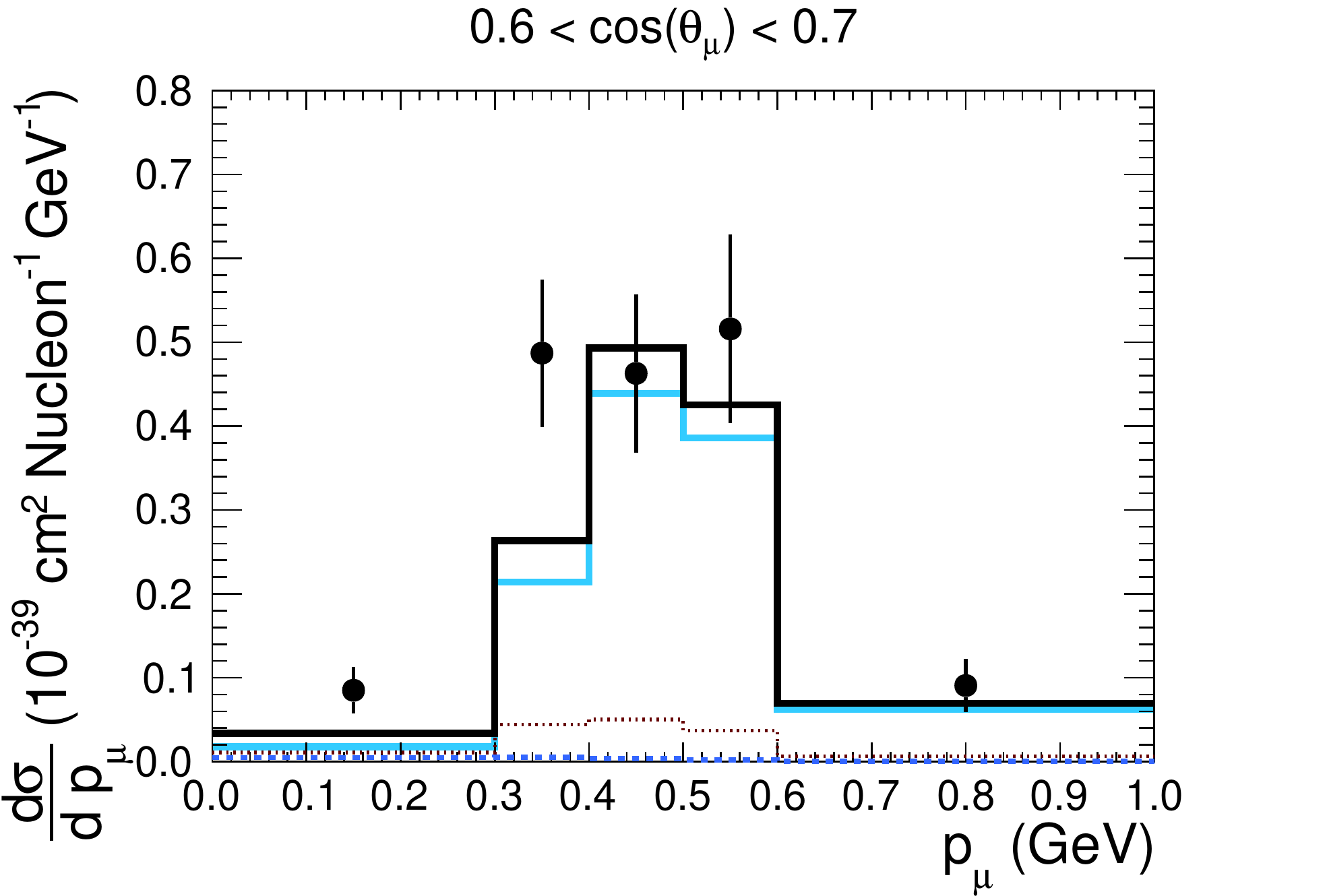}
\includegraphics[width=0.32\linewidth]{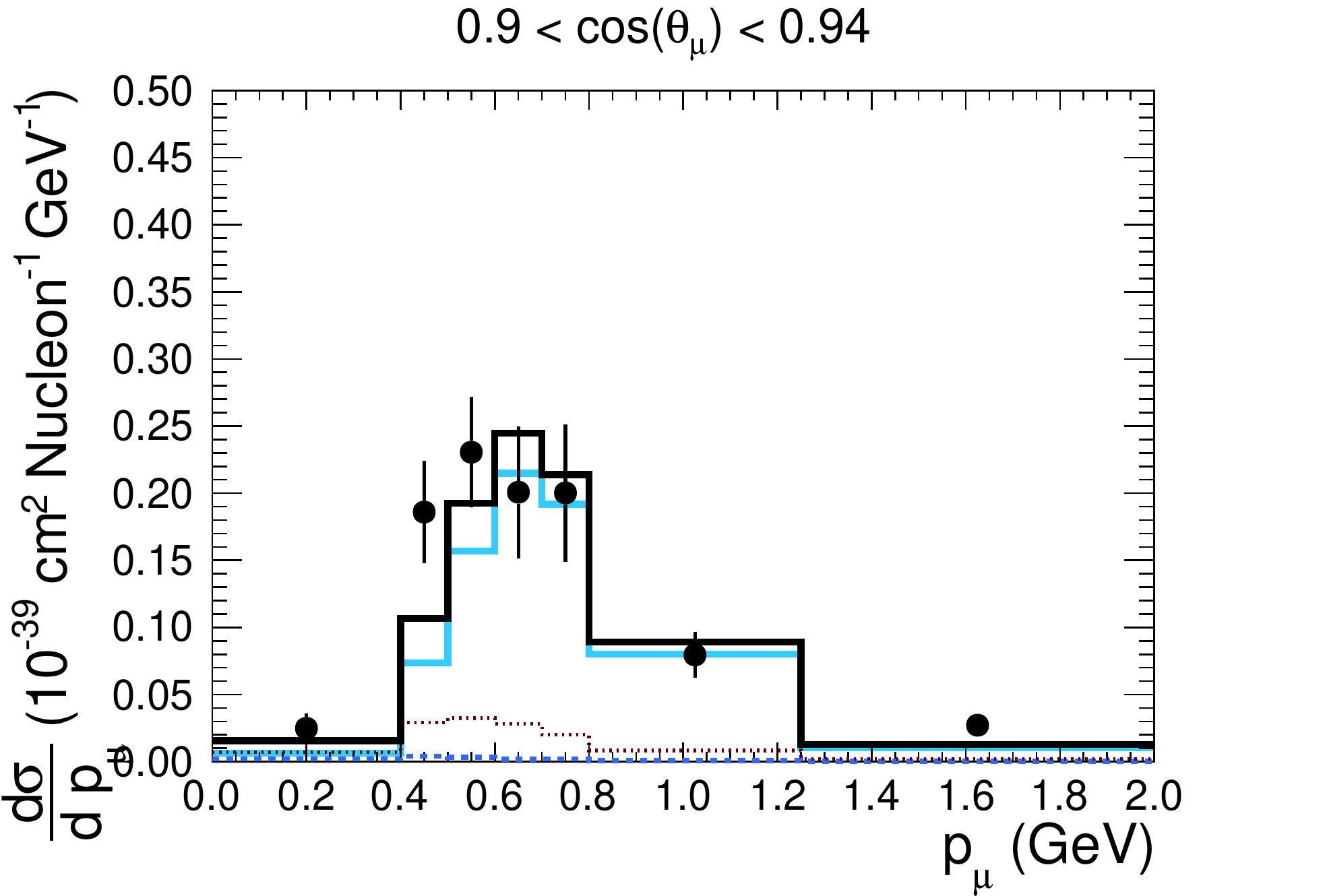}
\includegraphics[width=0.32\linewidth]{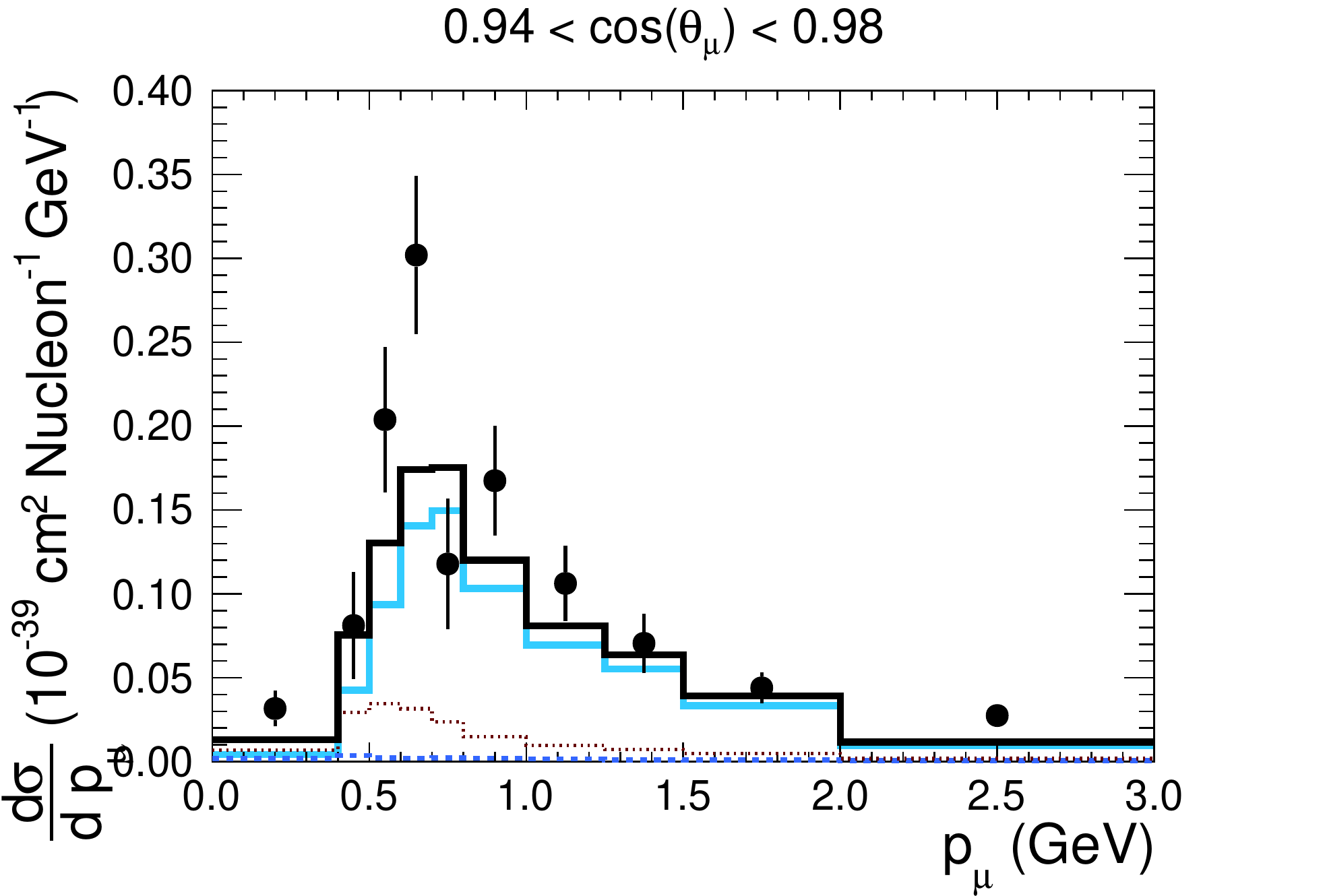}
\end{center}
\caption{Comparison of the T2K CC0$\pi$ measurement of muon-neutrino interactions on Carbon here there are no protons above 500 MeV with the SuSAv2 and Valencia models (1p1h+2p2h) each with an additional pion-absorption contribution as implemented in GENIE.  The (unstacked) contribution from each interaction mode is shown separately, as well the total prediction. The top plots are the SuSAv2 predictions whilst the Valencia ones are below. The data points are taken from~\cite{Abe:2018pwo}.}
\label{fig:ssincT2KModelComp}

\end{figure*}

Fig.~\ref{fig:incT2KModelComp} and~\ref{fig:ssincT2KModelComp} show a comparison of the SuSAv2 and Valencia model predictions (1p1h and 2p2h), as implemented in GENIE, on top of GENIE's pion absorption prediction for T2K inclusive and `semi-semi-inclusive' CC0$\pi$ results. This clearly shows that the implemented Valencia and SuSA models differ substantially, with only the SuSA model able to describe the very forward data and the Valencia model describing the mid-angle data a little better. The discrepancies between the model and data is consistent between the inclusive and semi-inclusive results, suggesting that they at least partially stem from the underlying inclusive cross section model.

\section{Full phase-space test of the factorisation approach}\label{appendix-fact}

\begin{figure*}\vspace{-0.35cm}
	\begin{center}\vspace{-0.35cm}
		\includegraphics[width=0.34\linewidth, angle=0]{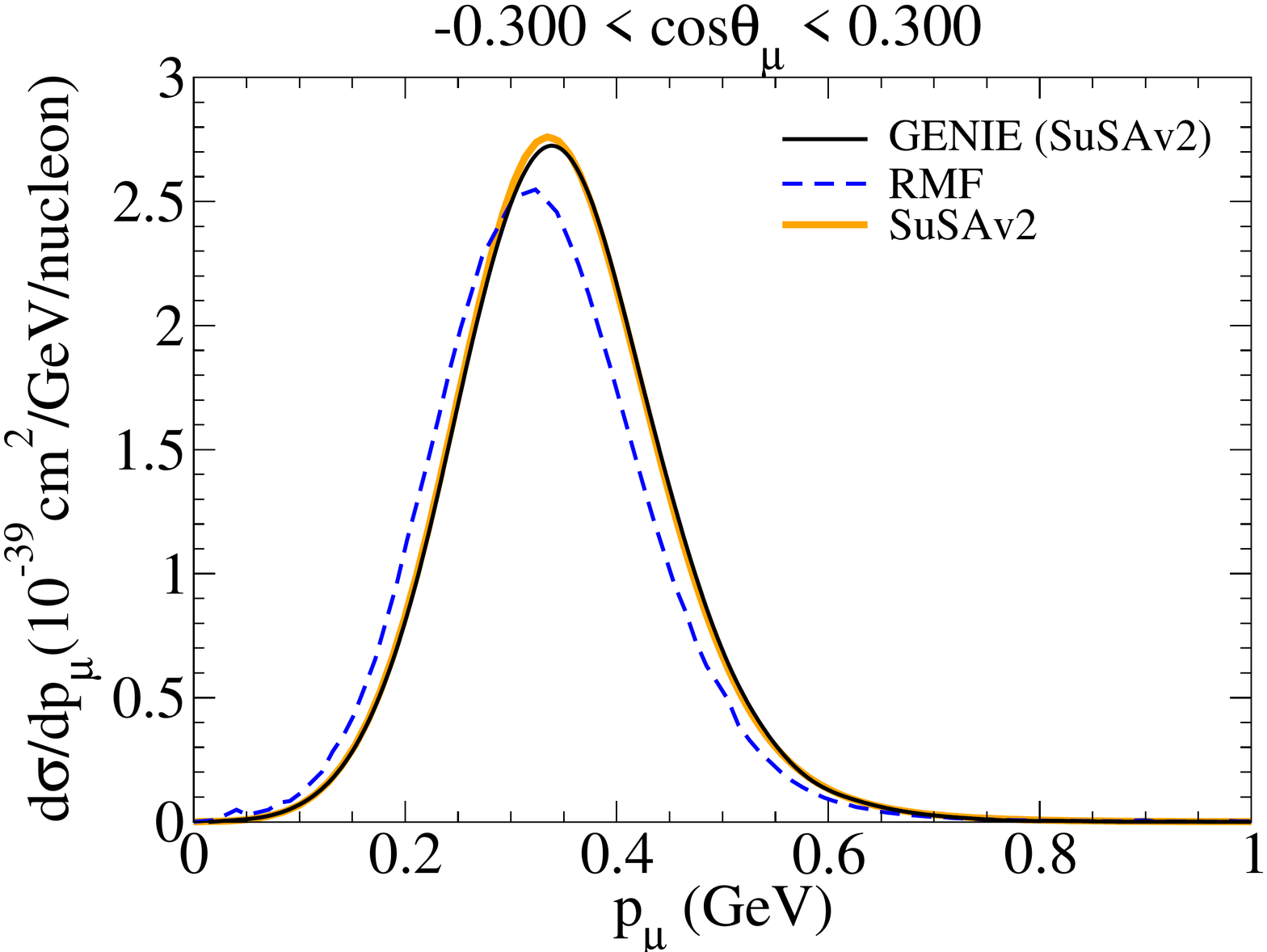}\hspace*{-0.295cm}
		\includegraphics[width=0.34\linewidth, angle=0]{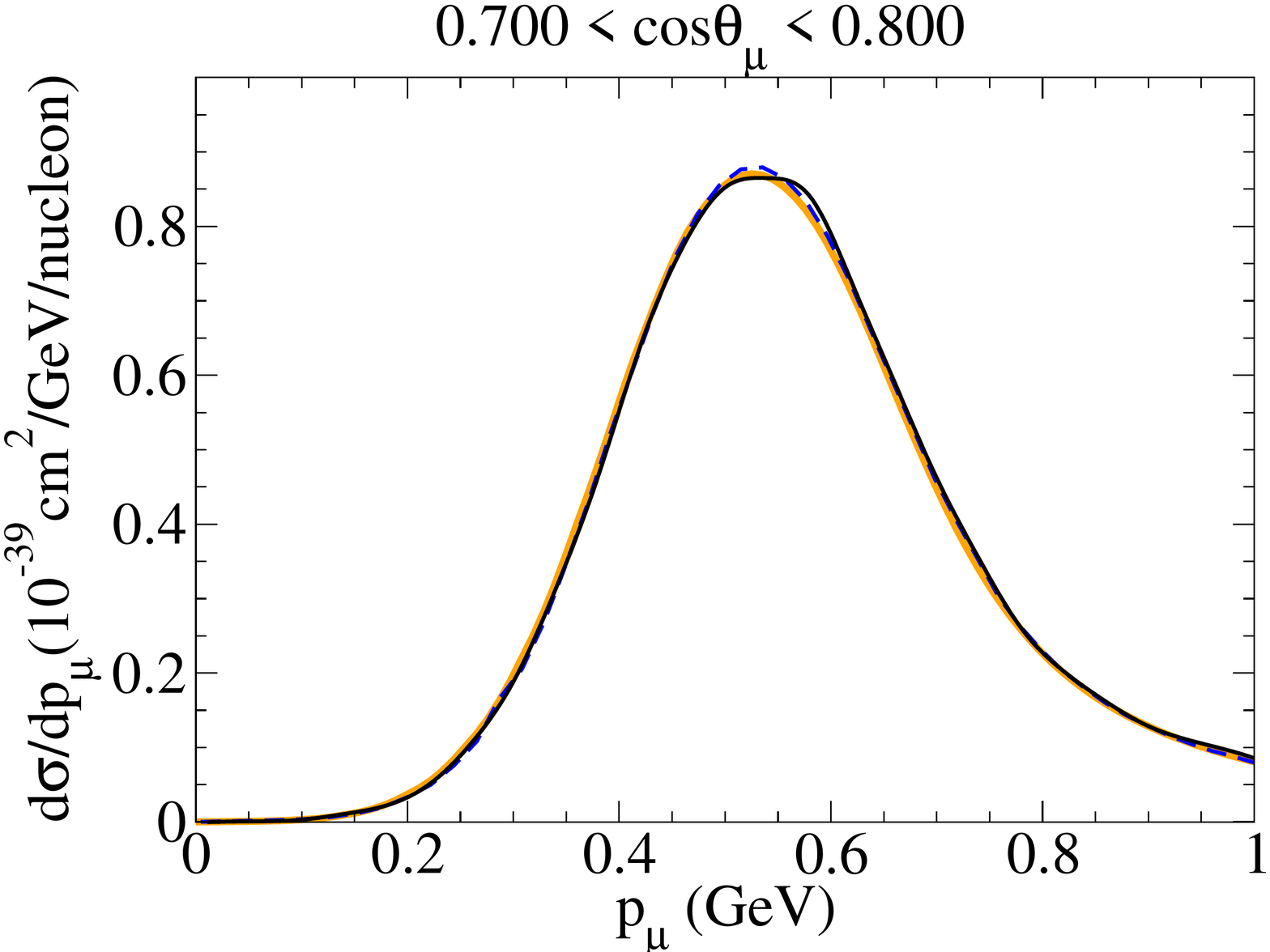}\hspace*{-0.295cm}
		\includegraphics[width=0.34\linewidth, angle=0]{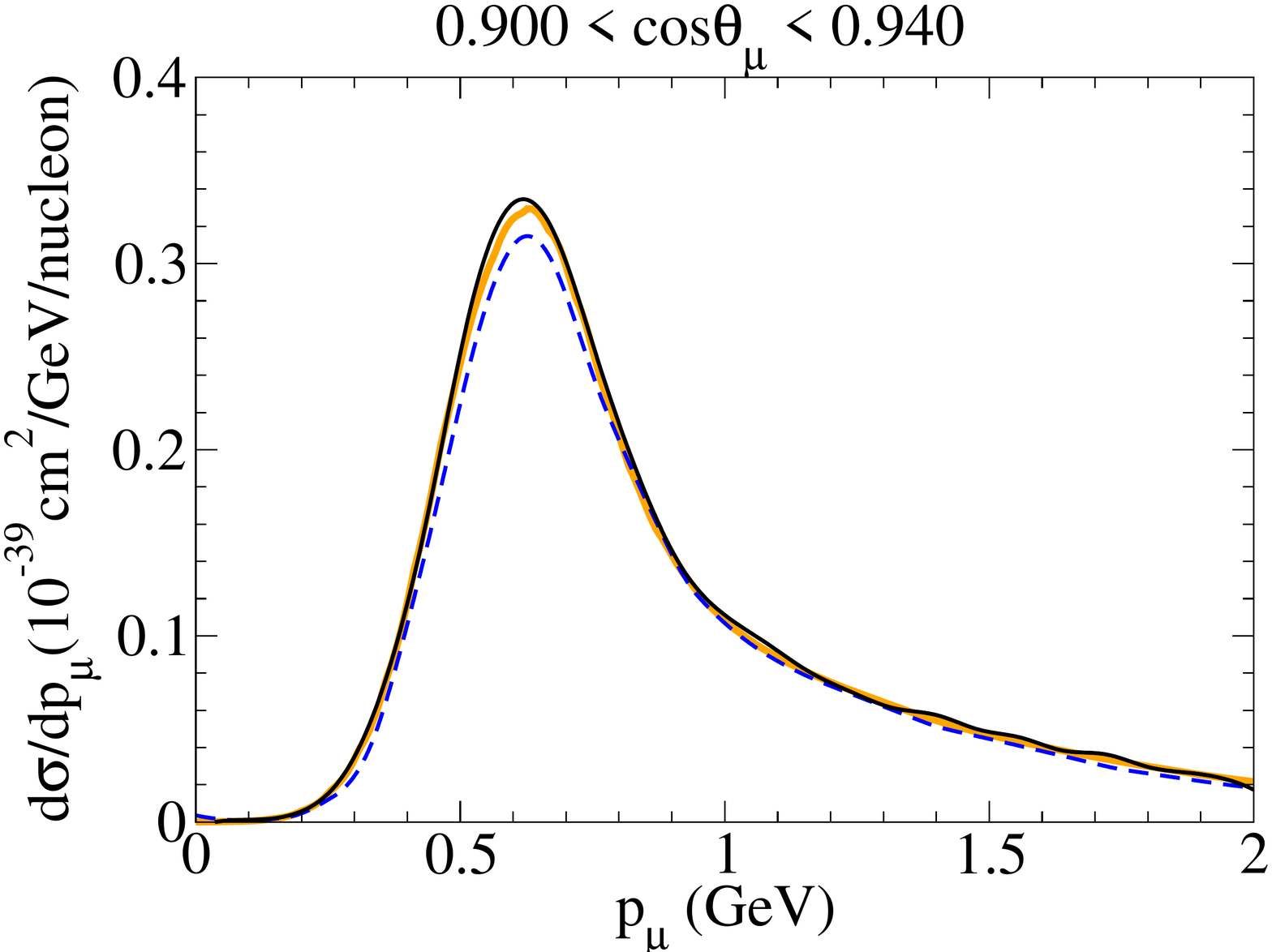}\\
				\includegraphics[width=0.34\linewidth, angle=0]{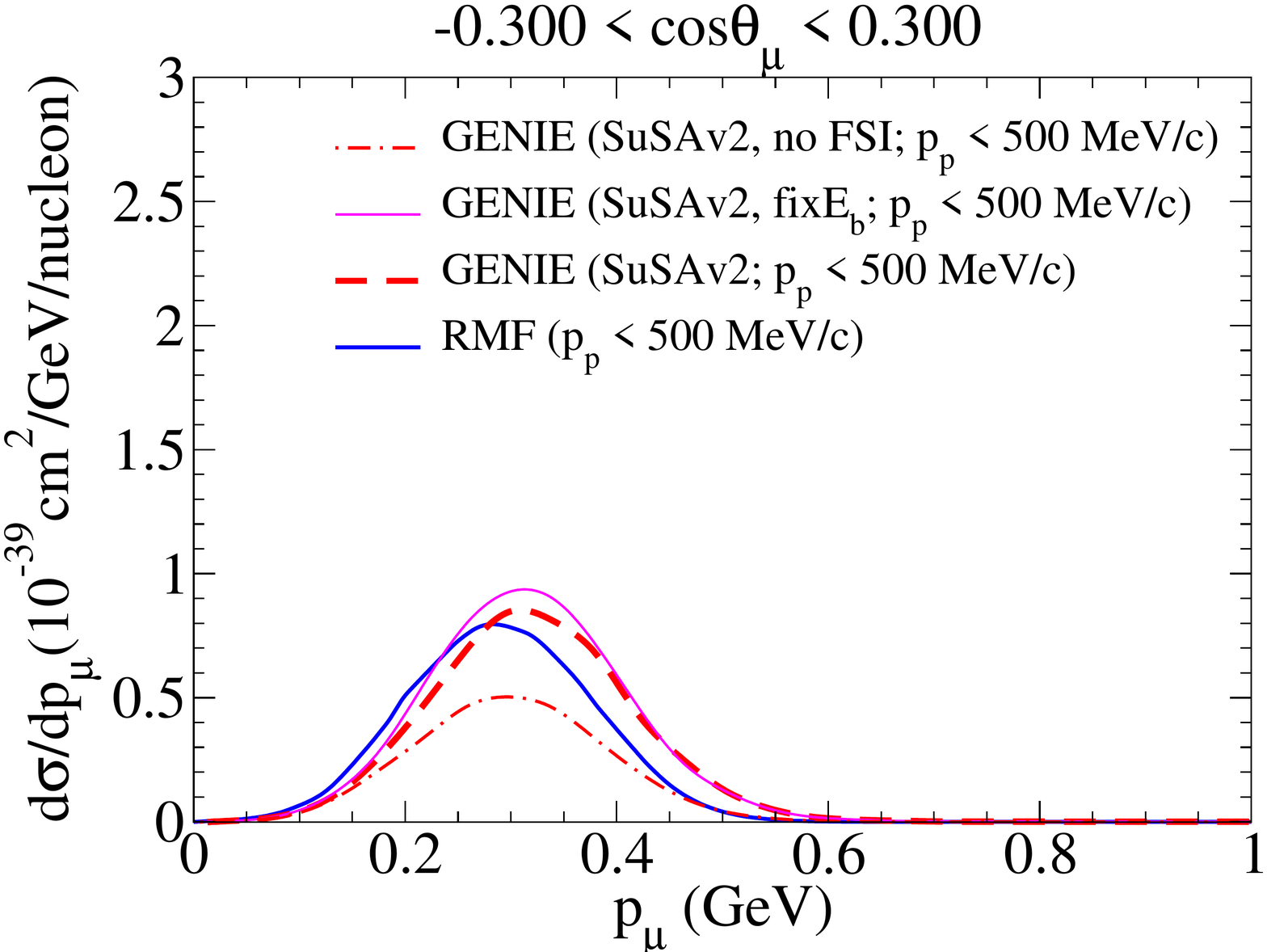}\hspace*{-0.295cm}
		\includegraphics[width=0.34\linewidth, angle=0]{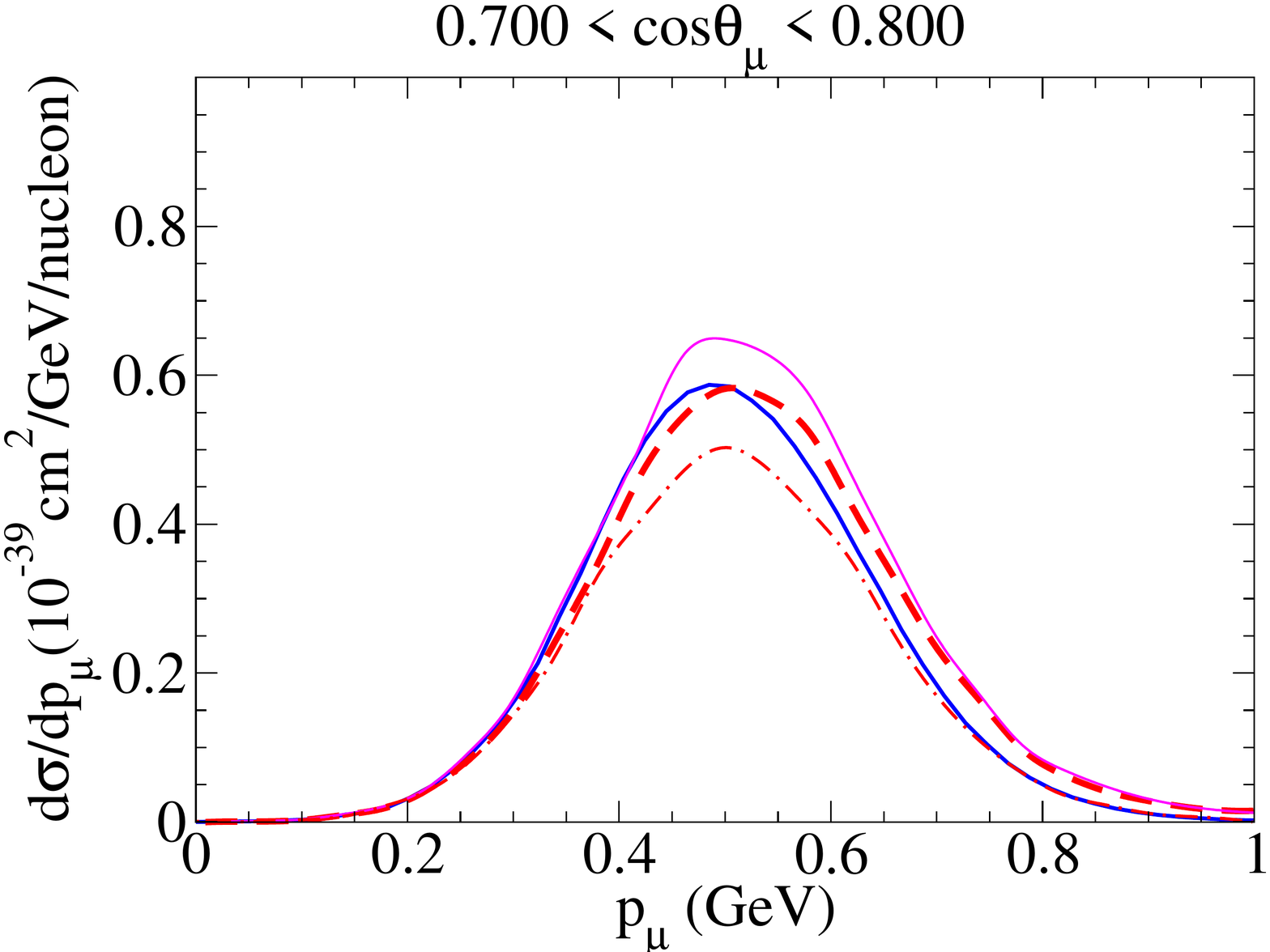}\hspace*{-0.295cm}
		\includegraphics[width=0.34\linewidth, angle=0]{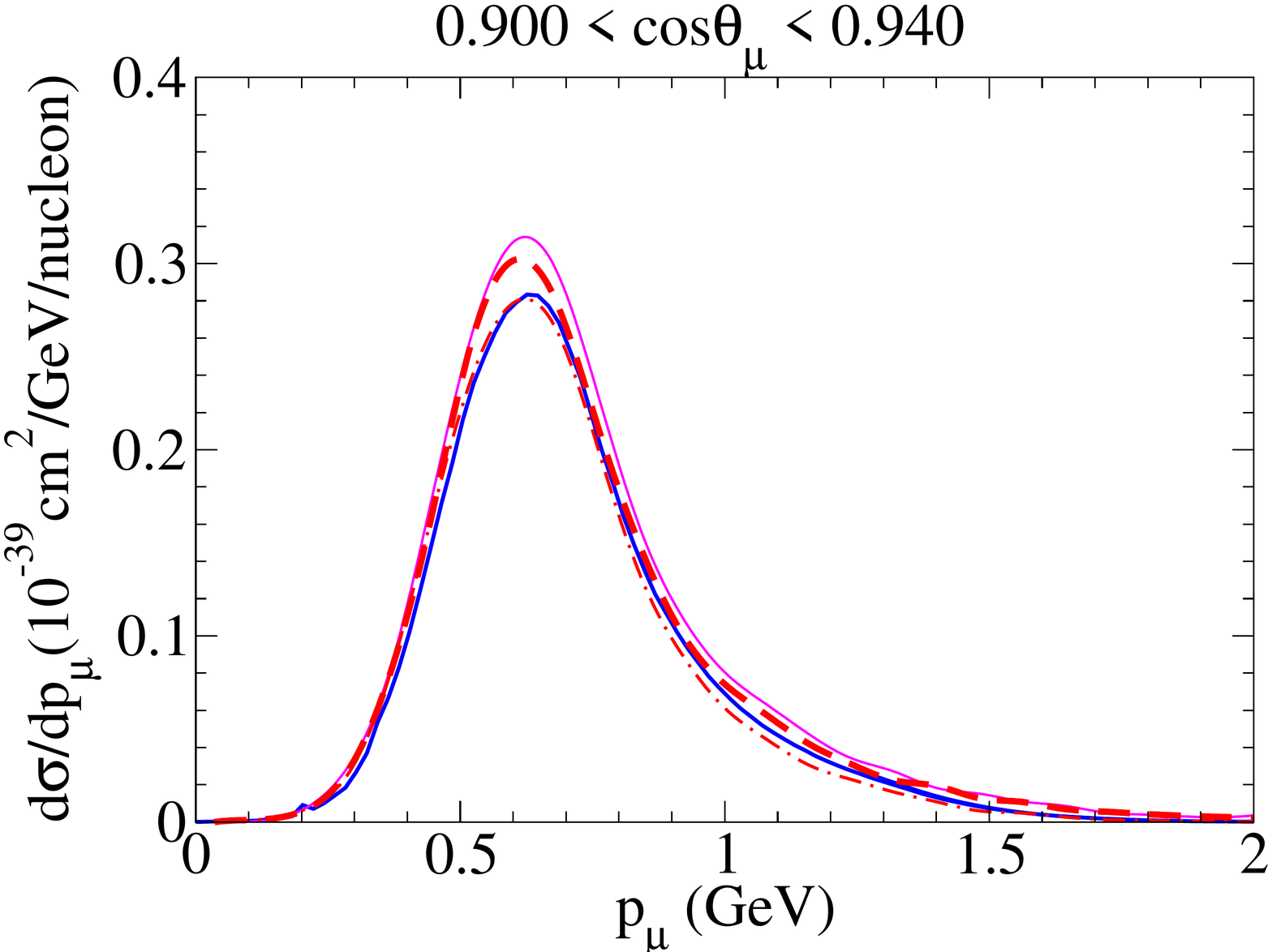}
	\end{center}
	\caption{Comparison of muon-neutrino single differential 1p1h cross sections on Carbon at T2K kinematics as a function of the muon kinematics as both an inclusive (top panels) and a `semi-semi-inclusive' cross section (bottom panels), the latter meaning that a restriction that there are no protons with momenta above 500 MeV is applied. In the inclusive case the RMF, SuSAv2 model and SuSAv2 GENIE implementation are compared. In the semi-semi-inclusive case the RMF prediction is compared to those of GENIE using the implemented SuSAv2 model and variations of the factorisation approach. These variations are split depending on whether an FSI cascade was applied and whether the nuclear removal energy is fixed or kinematic dependent (as described in Sec.~\ref{implementation}). The full set of angular slices are shown in Appendix~\ref{appendix-fact}.}
	\label{fig:ssIncToIncComp}
\end{figure*}

\begin{figure*}\vspace{-0.35cm}
	\begin{center}\vspace{-0.35cm}
		\includegraphics[width=0.34\linewidth, angle=0]{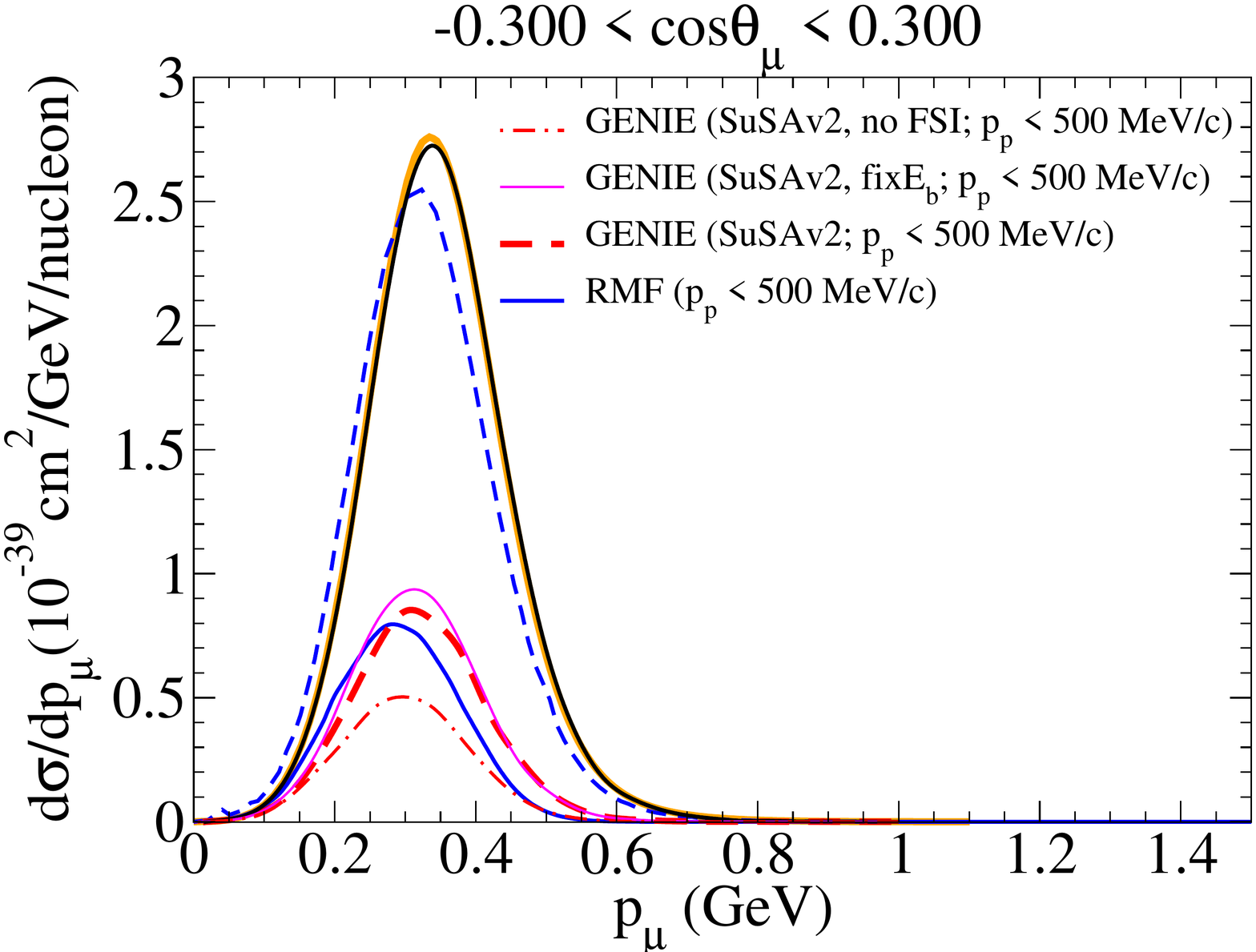}\hspace*{-0.295cm}
		\includegraphics[width=0.34\linewidth, angle=0]{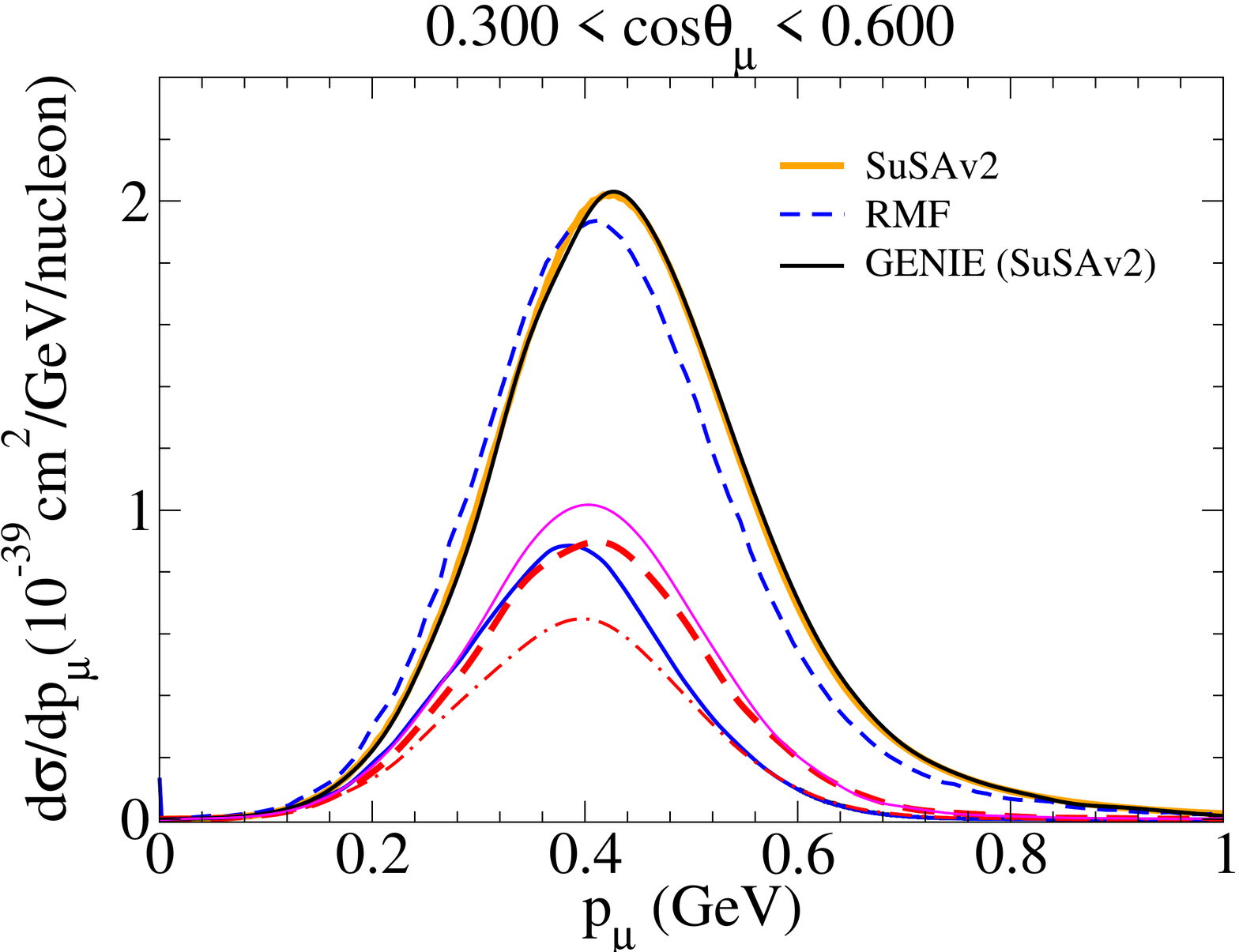}\hspace*{-0.495cm}
		\includegraphics[width=0.34\linewidth, angle=0]{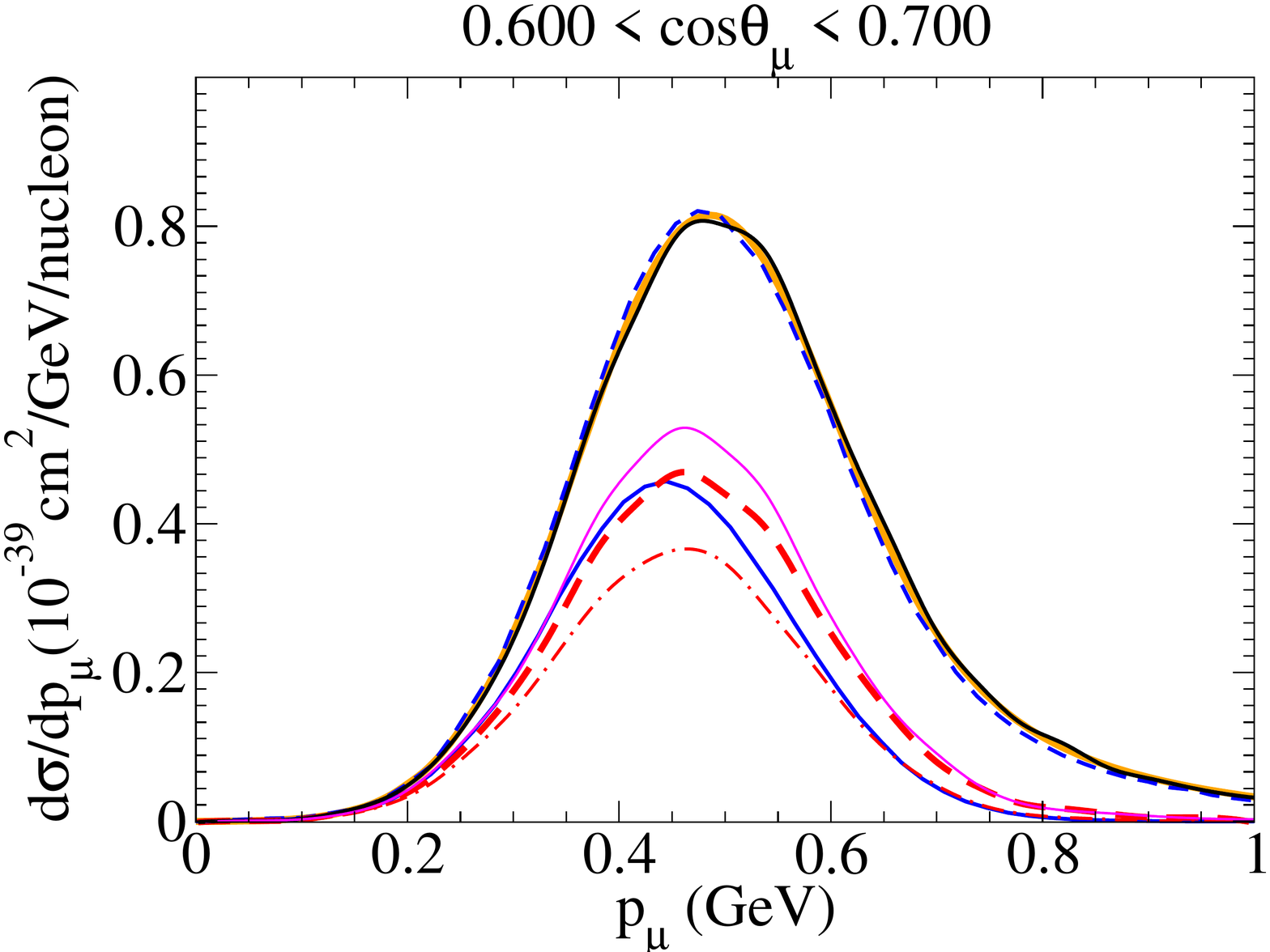}
		\includegraphics[width=0.34\linewidth, angle=0]{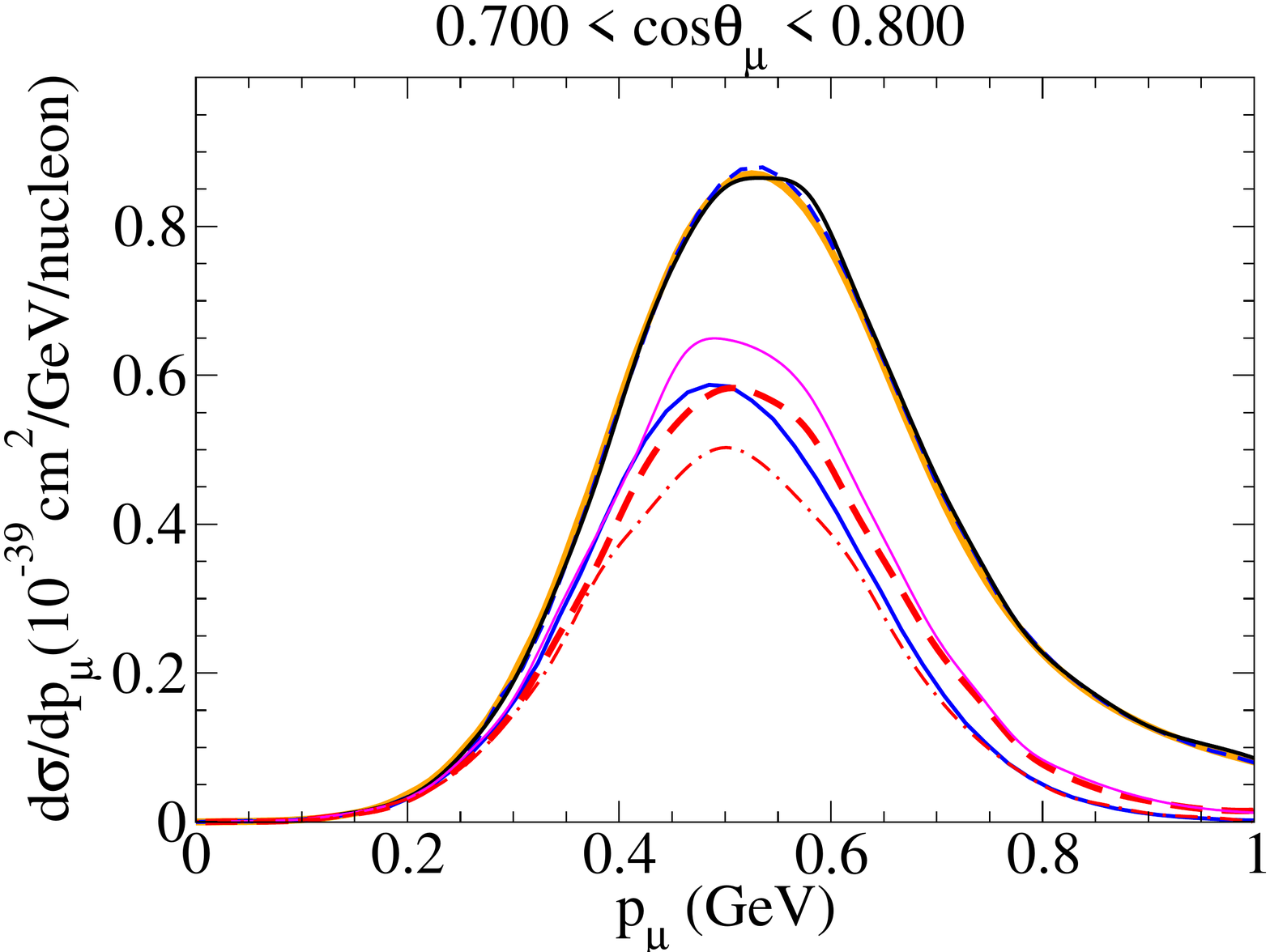}\hspace*{-0.295cm}
		\includegraphics[width=0.34\linewidth, angle=0]{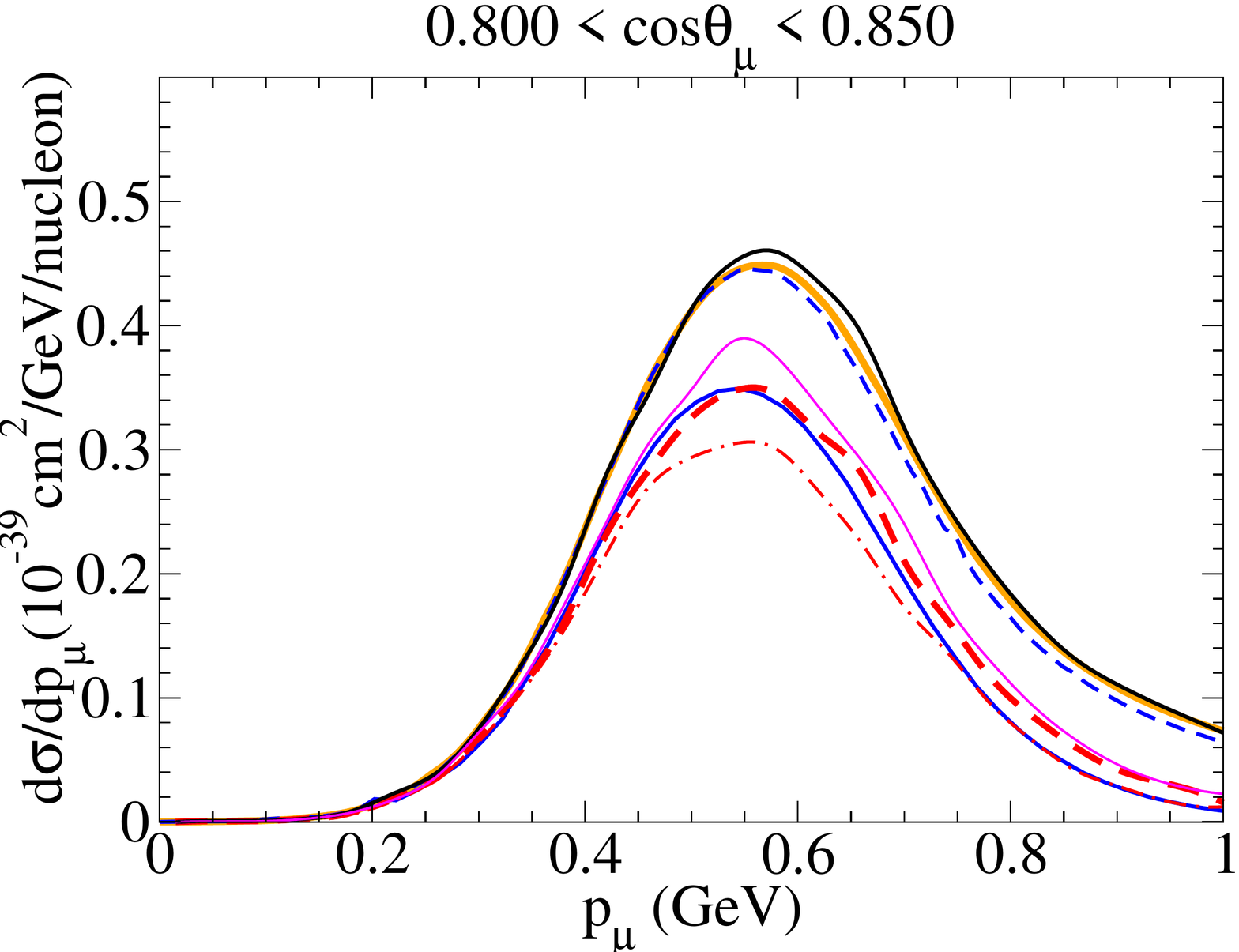}\hspace*{-0.495cm}
		\includegraphics[width=0.34\linewidth, angle=0]{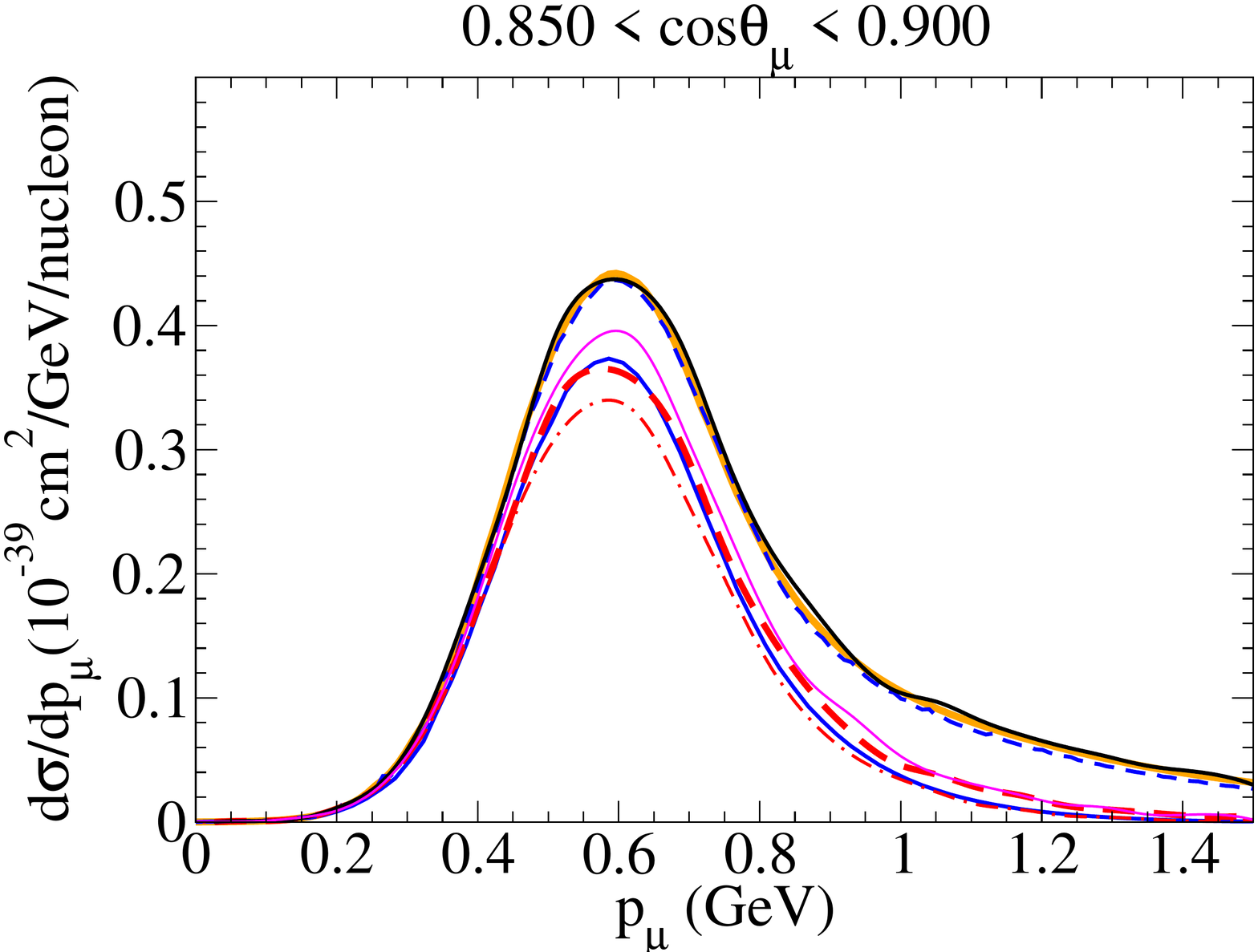}
		\includegraphics[width=0.34\linewidth, angle=0]{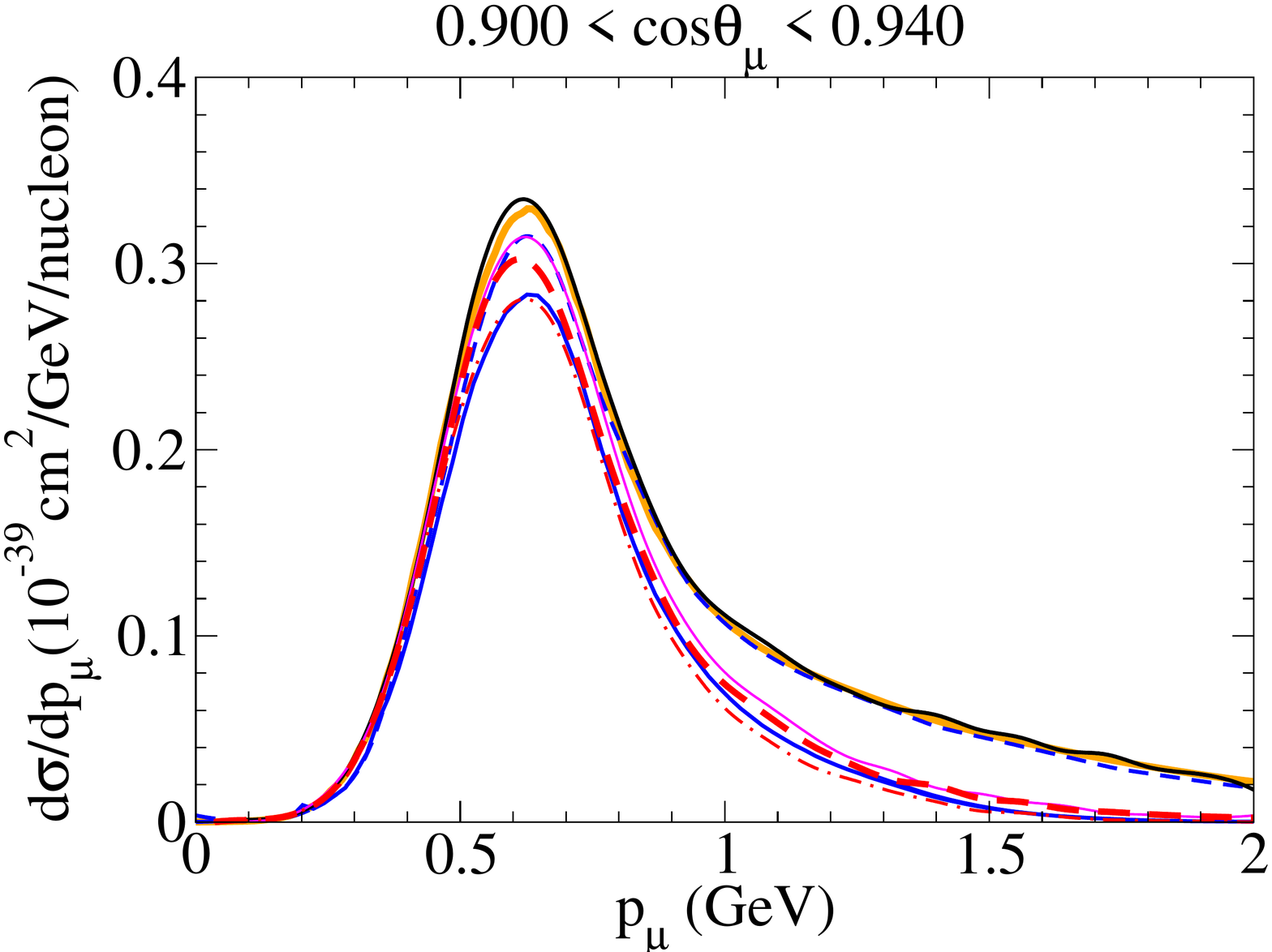}\hspace*{-0.295cm}
    	\includegraphics[width=0.34\linewidth, angle=0]{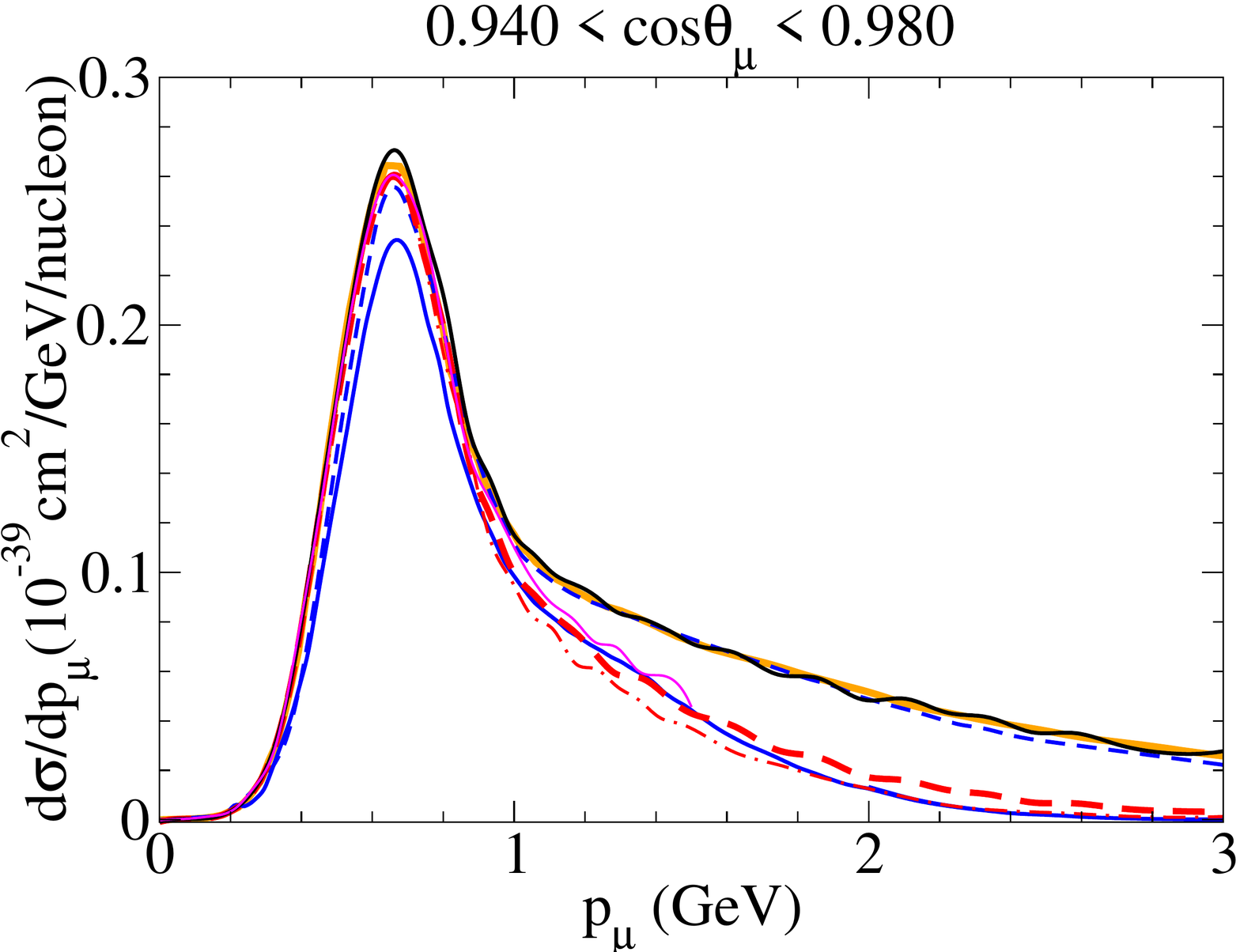}\hspace*{-0.495cm}
    	\includegraphics[width=0.34\linewidth, angle=0]{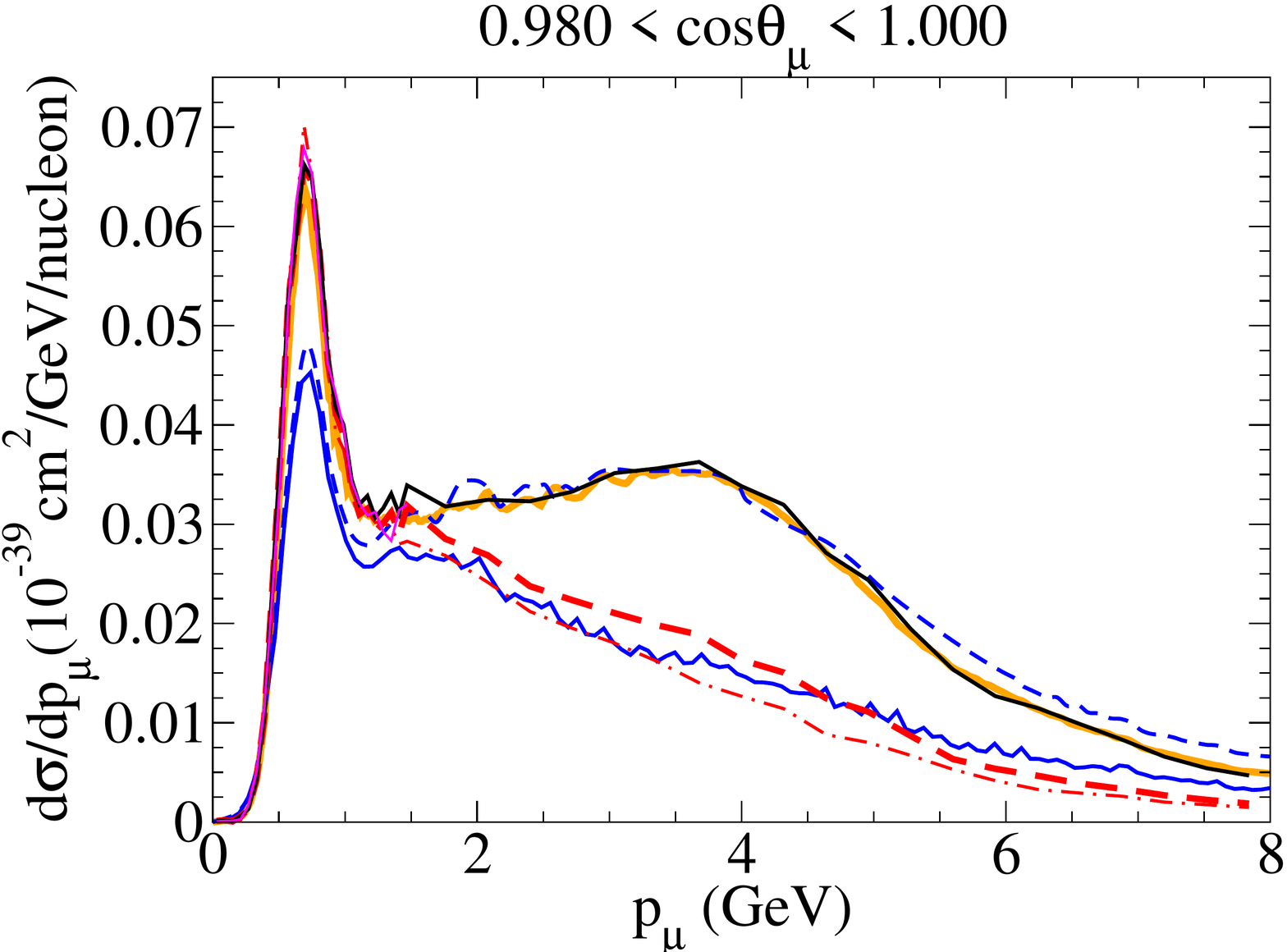}
    	\includegraphics[width=0.34\linewidth, angle=0]{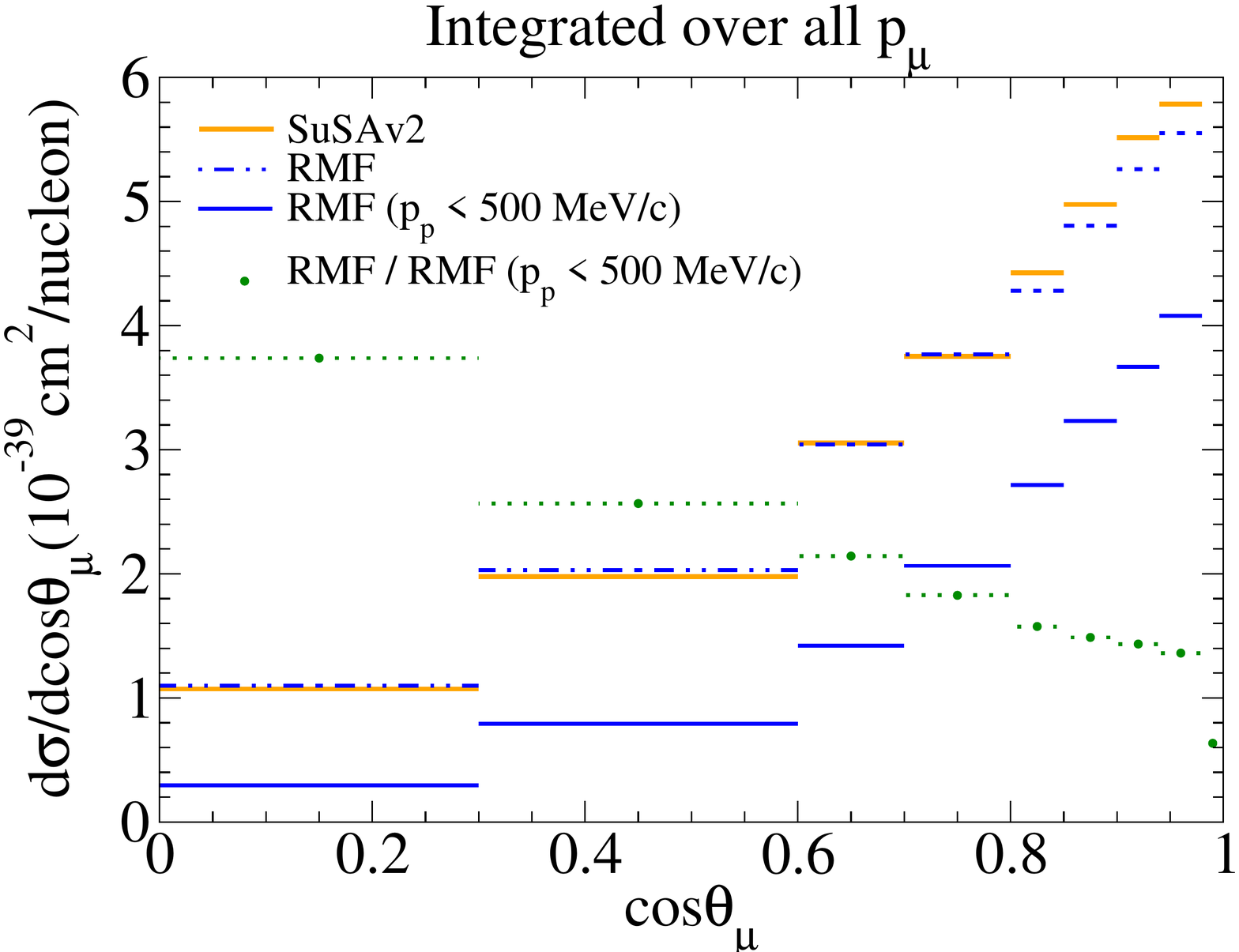}\vspace{-0.35cm}
	\end{center}
	\caption{An extended version of Fig.~\ref{fig:ssIncToIncComp}. A comparison of muon-neutrino single differential 1p1h cross sections on Carbon at T2K kinematics as a function of the muon kinematics as both an inclusive and a `semi-semi-inclusive' cross section, the latter applying a restriction that there are no protons with momenta above 500 MeV. In the inclusive case the RMF, SuSAv2 model and SuSAv2 GENIE implementation are compared. In the semi-semi-inclusive case the RMF prediction is compared to those of GENIE using the implemented SuSAv2 model and the factorisation approach. The latter is split depending on whether an FSI cascade was applied and whether the nuclear removal energy is fixed or kinematic dependent (as described in Sec.~\ref{implementation}).}
	\label{fig:ssIncToIncComp_appendix}
\end{figure*}

Fig.~\ref{fig:ssIncToIncComp_appendix} shows the full phase-space equivalent of Fig.~\ref{fig:ssIncToIncComp}, which, as discussed in Sec.~\ref{results-ssinc}, serves as a preliminary first test of the factorisation approach used to extract semi-inclusive predictions from inclusive model implementations in neutrino-nucleus interaction generators. In addition to this, these plots also compare the RMF and SuSAv2 inclusive model predictions. Although the differences were briefly discussed in section~\ref{results-ssinc}, here more detail is provided.


Firstly, a very good agreement between the RMF and SuSAv2 inclusive model predictions can be seen at intermediate angles (0.6-0.94), differing by less than 1\% for the total cross section integrated over this region. The discrepancy in the backward region is due to a correction in SuSAv2 to account for RMF having too strong FSI in the high momentum transfer region. Here the outgoing nucleon carries a large kinetic energy and it would be expected that the FSI effects should be suppressed for such kinematics. However, this does not happen in the RMF theory due to the strong energy-independent scalar and vector potentials included in the model. In order to account for this effect, the SuSAv2 model introduces effects from the Relativistic Plane Wave Impulse Approximation (RPWIA) - where the initial state is described by a mean field but FSI are neglected - at high momentum transfer by using a q-dependent blending function, as detailed in~\cite{Gonzalez-Jimenez:2014eqa,Megias:2016lke}. This effect is fully incorporated into the GENIE implementation. In further works~\cite{Gonzalez-Jimenez:2019qhq}, an improved RMF model with energy-dependent potentials will solve this issue, making the SuSAv2 model more self-contained and avoiding the need of using RPWIA effects to properly describe high kinematics. 

The differences observed at very low kinematics, i.e. very forward angles, are related to the RMF scaling functions employed in the SuSAv2 model. These scaling functions effectively describe the nuclear dynamics of the model and are almost identical for $q\gtrsim400$ MeV/c and for different nuclei (`superscaling')~\cite{super,super1}. However, this scaling behavior is broken at very low q ($<400$ MeV/c) where collective effects which violate superscaling dominate. These effects are indeed accounted for in the RMF theory (producing smaller scaling functions at very low q) but are absent in the SuSAv2 approach (which assumes a general scaling function for all kinematics), producing larger SuSAv2 results at very forward angles. This drawback of the SuSAv2 model will be addressed in further works by considering the $q_3$-dependence of the RMF scaling functions, which will produce more consistent theory-vs-data comparison at these particular kinematics. Accordingly, a full implementation of the upcoming RMF energy-dependent model~\cite{PhysRevC.101.015503,PhysRevC.100.045501} in generators will solve this drawback at very low and high kinematics.

\bibliography{biblio}

\end{document}